%% file: template.tex
\documentclass{article}

\usepackage{arxiv}

\usepackage[T1]{fontenc}    
\usepackage{url}            
\usepackage{booktabs}       
\usepackage{amsfonts}       
\usepackage{nicefrac}       
\usepackage{microtype}      
\usepackage{lipsum}
\usepackage{graphicx}

\usepackage{framed,multirow}
\usepackage{amssymb}
\usepackage{latexsym}

\usepackage{url}
\usepackage{xcolor}
\usepackage[utf8x]{inputenc}
\DeclareUnicodeCharacter{"2068}{}
\DeclareUnicodeCharacter{"2069}{}
\usepackage{amsmath}
\usepackage{amsthm}
\usepackage{hyperref}
\usepackage{cleveref}   
\graphicspath{{images/}}
\usepackage{parskip}
\usepackage{fancyhdr}
\usepackage{xcolor}
\usepackage{pgf}
\usepackage[center,font=small,labelfont=bf]{caption}
\usepackage{subcaption}
\usepackage{pgfplots}
\usepackage{color}
\usepackage{tikz}
\usetikzlibrary{shapes.geometric}
\usepackage{float}
\usepackage{bm}
\usepackage{stmaryrd}
\usepackage{enumitem}
\usepackage{authblk}

\newcommand{\mc}{\mathcal{M}}
\newtheorem{definition}{Definition}
\newtheorem{lemma}{Lemma}
\DeclareMathOperator*{\argmax}{argmax} 
\DeclareMathOperator*{\argmin}{argmin} 
\usetikzlibrary{arrows, arrows.meta}
\definecolor{newcolor}{rgb}{.8,.349,.1}
\usetikzlibrary{calc}
\usepackage{calrsfs}
\pgfplotsset{compat=1.16}

\newtheoremstyle{problemstyle}  
        {3pt}                                               
        {3pt}                                               
        {\normalfont}                               
        {}                                                  
        {\bfseries}                 
        {\normalfont\bfseries:}         
        {.5em}                                          
        {}                                                  
\theoremstyle{problemstyle}

\newtheorem{problem}{Problem}[section]

\newtheoremstyle{Assumptionstyle}  
        {3pt}                                               
        {3pt}                                               
        {\normalfont}                               
        {}                                                  
        {\bfseries}                 
        {\normalfont\bfseries:}         
        {.5em}                                          
        {}                                                  
\theoremstyle{Assumptionstyle}

\newtheorem{assumption}{Assumption}[section]

\title{Optimal Approximation Spaces for Discontinuous Petrov-Galerkin Finite Element Methods}

\author[1]{Ankit Chakarborty}
\author[1]{Ajay Rangarajan}
\author[2]{Georg May}
\affil[1]{AICES, RWTH Aachen University, Schinkelstra\ss e 2, 52062 Aachen, Germany \hspace{2cm} \texttt{chakraborty,rangarajan@aices.rwth-aachen.de}}
\affil[2]{Von Karman Institute For Fluid Dynamics, Waterloosesteenweg 72, 1640 Sint-Genesius-Rode, Belgium \hspace{5cm}
\texttt{georg.may@vki.ac.be} }

\begin{document}
\maketitle
\begin{abstract}
Certain Petrov-Galerkin schemes are inherently stable formulations of variational problems on a given mesh. This stability is primarily obtained by computing an optimal test basis for a given approximation space. Furthermore, these Petrov-Galerkin schemes are equipped with a robust a posteriori error estimate which makes them an ideal candidate for mesh adaptation. One could extend these Petrov-Galerkin schemes not only to have optimal test spaces but also optimal approximation spaces with respect to current estimates of the solution. These extensions are the main focus of this article. In this article, we provide a methodology to drive mesh adaptation to produce optimal meshes for resolving solution features and output functionals which we demonstrate through numerical experiments.

\end{abstract}

\input{Introduction}
\input{discretization}
\input{mesh_metric_duallity}
\input{Anisotropic_error_estimator}
\input{continuous_error_estimate}

\input{Results}

\section{Conclusion and outlook}
We have presented a continuous $h$ mesh model for driving mesh optimization for both solution based and target-based adaptation. Apart from being parameter free, the method directly utilizes the information from the inbuilt error estimator rather being depended on an error model which is not compatible with the numerical scheme. To this end we have been able to produce optimal approximation space in conjunction with optimal test space. We have presented numerical experiments for diffusion and convection-diffusion problems showing the feasibility of the method. In current set up, we have presented two-dimensional results but it is very much possible to extend this approach to three dimensional problem. In order to achieve this, it would be necessary to generalize the definitions of element shape and anisotropy of error estimator. The nature of minimization problem for computing the local anisotropy and mesh density stays the same. Also, a $hp$ extension for the presented methodology is also very much viable and will be presented in future work.
\section{Acknowledgments}
We would like to thank Brendan Keith for fruitful discussions regarding DPG and $\text{DPG}^{\star}$ methods and Vit Dolejsi for insightful discussions on continuous mesh models for goal-oriented adaptations. This work is funded by the Deutsche Forschungsgemeinschaft (DFG, German Research Foundation) – 33849990/ GRK2379 (IRTG Modern Inverse Problems).

\bibliographystyle{unsrt}  
\bibliography{template}
\end{document}

%% file: Introduction.tex
\section{Introduction}
Numerical discretization of Partial differential equations in an efficient and stable manner has been an essential technology in many industrial and academic applications. In terms of stability, over the last decade,  Petrov-Galerkin schemes with optimal test function (Optimal PG schemes) has been a key development  \cite{Demkowicz2010,dem_part2,Zitelli2011,Dahmen2012,Cohen2012,Demkowicz2012a,Keith2016,VaziriAstaneh2018}. Given a weak formulation of the underlying PDE and an approximation space, this framework aims to automatically compute a space of test functions such that the scheme is stable. It can be perceived as adapting the test functions for a given weak formulation to achieve stability which is contrary to the traditional perspective of finite elements where trial and the test functions are selected a priori. One other advantage of this framework is the presence of a residual based inbuilt error estimator. 

Due to the presence of this inbuilt error estimator, PG schemes with optimal test functions are tailor-made for mesh optimization  \cite{Demkowicz2012a,KeithGoal,DEMKOWICZ2020}, such that the optimal test space can be complemented with an optimal approximation space. In recent times, metric-based mesh optimization have emerged as an interesting paradigm for meshes composed of simplexes \cite{inria_a,inria_b,Rangarajan2018,Dolejsi2018}. A significant advantage of these mesh optimization methods over conventional adaptive procedures such as element subdivision and edge swapping is the capability to globally optimize the mesh rather than local modification. It has been long known that anisotropic meshes have substantial advantages, enabling the resolution of anisotropic solution features, such as thin boundary layers and shocks, due to skewed elements aligned along with the solution feature \cite{ceze2013,Leicht2008,LEICHT20107344,VENDITTI200322,Yano2012a,MAVRIPLIS1998141}. 

In the present paper, we propose and analyze a novel anisotropic mesh adaptation/optimization scheme based on the inbuilt error estimator of optimal PG schemes with discontinuous test and approximation spaces \cite{dem_part2}. Our optimization framework uses the idea of analytic optimization \cite{Rangarajan2018,Dolejsi2018,Dolejsi2017} and local anisotropy computation \cite{Dolejsi2019}.  In \cite{Rangarajan2018}, the authors have proposed an analytic optimization technique using a $L^q$-norm error model. In the current paper, we extend the idea of analytic optimization to the inbuilt error estimator accompanying discontinuous Petrov-Galerkin (DPG) schemes, rather than using an extrinsic error model. The problem of discrete mesh optimization can be formulated as
\begin{equation}
\min_{\forall T_h} E_{disc} \quad such \, that\quad  N = const. \label{opt_prob}
\end{equation}

where $E_{disc}$ is the error, obtained from a suitable model, and $N$ is the total degrees of freedom in the mesh. This optimization problem cannot easily be solved using analytic or semi-analytic tools.  The question arises whether one can present a suitable reformulation of the above problem which is continuous in nature. If so, then there are various analytic variational optimization techniques at our disposal. The continuous mesh approach allows us to formulate a suitable problem using a continuous analogue of the discrete error estimator which in turn transforms the discrete problem in~\cref{opt_prob} into a continuous one. While a mesh generated by continuous mesh model may not be provably optimal, it has been shown to be quite effective in terms of good approximation properties for higher order method such as hybridized discontinuous Galerkin schemes (HDG schemes) \cite{Rangarajan2018}, Embedded Discontinuous Galerkin schemes (EDG schemes) \cite{fidkowski2019} and other high order finite element schemes \cite{Formaggia2001}.

In this article, the proposed continuous error model is based on the energy error estimate which accompanies optimal PG schemes i.e $E_{disc} = {\Vert U-U_h \Vert}_{E,\Omega}$ for solution driven adaptations and a duality based error indicator using {DPG}\textsuperscript{$\star$} scheme \cite{DEMKOWICZ2018,Keith_dissert} for goal-oriented adaptations. Though in this paper we have focused on optimal PG schemes with discontinuous trial functions, it can be extended to the optimal PG schemes with continuous trial functions as long as the error estimate can be localized and has a polynomial representation.  

The article is organized as follows. In section 2, we provide a brief introduction to DPG schemes, ultra-weak formulation of a scalar convection-diffusion equation, and {DPG}\textsuperscript{$\star$} scheme. In section 3, we recall mesh-metric duality; in section 4, we introduce the local optimization problem for computing the anisotropy, followed by sections 5 and 6, where we introduce the continuous optimization problem for the element size distribution. Finally, in section 7, we provide numerical results that shows the fidelity of the proposed method, followed by conclusion and outlook.

%% file: discretization.tex
 \section{Discretization} \label{Discretization}
In this section, we provide a brief description of Petrov-Galerkin schemes with optimal test functions and the associated inbuilt error estimator. We will restrict ourselves to two dimensional formulation and scalar problems. First, we present the optimal test function framework in a brief and abstract manner. We consider any inf-sup stable variational problem :
  \begin{equation*}
    u \in (U,{\Vert \cdot \Vert}_U): \qquad \langle  {\mathcal{L}} u,v \rangle = \langle f,v \rangle, \qquad \forall \, v \in (\mathbb{V},{\Vert \cdot \Vert}_\mathbb{V})
  \end{equation*}
  where $ {\mathcal{L}}: U \rightarrow \mathbb{V}^{'}$, $f \in \mathbb{V}^{'}$, $U$ and $\mathbb{V}$ are Hilbert spaces with $U^{'}$ and $\mathbb{V}^{'}$ being the respective dual spaces. One way to characterize Petrov-Galerkin Schemes with optimal test functions~\cite{dem_part2} is the following: Given ${\Vert \cdot \Vert}_{\mathbb{V}}$ and  ${\Vert \cdot \Vert}_U$, they satisfy
\begin{equation*}
{\Vert  {\mathcal{L}} u \Vert}_{\mathbb{V}^{\prime}}={\Vert u \Vert}_E = \sup_{v\in \mathbb{V}} \frac{\vert \langle  {\mathcal{L}} u,v \rangle \vert}{{\Vert v\Vert}_\mathbb{V}} \Rightarrow \vert    \langle  {\mathcal{L}} u,v \rangle \vert \leq {\Vert u \Vert}_E {\Vert v\Vert}_\mathbb{V}
\end{equation*}

In a finite dimensional setting, given $U_h \subset U$, optimal PG schemes construct $\mathbb{V}_h ^{opt} = {{\mathcal{R}}_\mathbb{V}}^{-1} {\mathcal{L}} (U_h)$ such that 
\begin{equation*}
u_h \in U_h :\qquad \langle   {\mathcal{L}} u_h,v_h \rangle= \langle f,v_h \rangle \qquad \forall \, v_h \in \mathbb{V}_h^{opt}
\end{equation*}
inherits the well-posedness, and furthermore 
\begin{equation}
\Vert u-u_h \Vert_E = \min_{v \in U_h} {\Vert u - v \Vert}_E  = {\Vert  {\mathcal{L}} u_h-f \Vert}_{\mathbb{V}^{\prime}} = {\Vert  {{\mathcal{R}}_\mathbb{V}}^{-1}(  {\mathcal{L}}  u_h-f)\Vert}_\mathbb{V} \label{DPGerrorest_a}
\end{equation}

where ${\mathcal{R}}_\mathbb{V}: \mathbb{V} \rightarrow {\mathbb{V}}^{\prime}$ is the Riesz  map defined by the duality as $\langle \mathcal{R}_\mathbb{V} v, v^{\prime} \rangle = {\left( v, v^{\prime} \right)}_{\mathbb{V}} \, \forall \, v^{\prime} \in \mathbb{V}$. 
Since $\mathbb{V}$ is infinite-dimensional, it is not possible to explicitly compute ${{\mathcal{R}}_\mathbb{V}}^{-1}$.  
In order to deal with this issue, a choice of enriched space $\mathbb{V}_r \subset \mathbb{V} \left( M = dim\left(\mathbb{V}_r \right) \geq dim\left( U_h\right) =  N\right)$ is employed over which the Riesz map is discretized and inverted. Since we are approximating the inverse of ${\mathcal{R}}_\mathbb{V}$, we have $\mathbb{V}_h$ which approximates $\mathbb{V}_h^{opt}$. It also implies $\mathbb{V}_h \subset \mathbb{V}_r \subset \mathbb{V}$. The discretization of the Riesz map is represented by a Gram matrix $\mathbb{G}$ which is induced by the inner product on $\mathbb{V}$.
\begin{equation}
\mathbb{G}_{i,j} = {(\psi_i,\psi_j)}_\mathbb{V} \label{Gram_matrix}
\end{equation}
Here $\psi_i$ and $\psi_j$ represents the basis of enriched space $\mathbb{V}_r$. The resulting discrete system on employing a finite dimensional trial basis $U_h$ and $\mathbb{V}_r$ which we call as test search space can be shown to be equivalent to the following linear system \cite{VaziriAstaneh2018} :
\begin{equation}
{B}^T \mathbb{G}^{-1} {B}\hat{x}_h = {B}^TG^{-1}l \label{normal_eq}
\end{equation}
where $u_h =\sum_{i = 1}^{N} {(\hat{x}_h)}_{i}\phi_i $ is the discrete solution with $\phi_i$ being the basis of the finite dimensional trial space $U_h$. $B$ is the enriched stiffness matrix where $B_{ij} = \langle \mathcal{L} \phi_j, \psi_i \rangle$ and $l$ is the enriched load vector with $l_i = \langle f,\psi_i \rangle$. Depending upon the variational formulation employed, $\mathbb{G}$  might need a global assembly, if the test functions are continuous across the element boundaries. In the current work, we have employed an ultra-weak formulation with broken Sobolev spaces which results in block diagonal structure of $\mathbb{G}$ and thus can be inverted locally for each element. However, this results in introduction of additional skeleton variables. Using static condensation, one can solve globally for the skeleton variables, similar to methods such as hybridisable discontinuous Galerkin schemes \cite{Woopen2014}.  The residual based error estimator in~\cref{DPGerrorest_a} can be approximated due to discretization of $\mathcal{R}_{\mathbb{V}}$ and is given by:

\begin{equation}
\hat{y} = {\mathbb{G}}^{-1}\left(B \hat{x}_h - l \right)
\end{equation}

where the dual of residual is approximated as $\varphi = \sum_{i = 1}^{M} {\hat{y}}_i \psi_i$, and hence the error in energy norm is approximated as
\begin{equation}
{\Vert u - u_h \Vert}_E \approx {\Vert \varphi \Vert}_{\mathbb{V}}.
\end{equation}
On employing polynomial basis functions for $\mathbb{V}_r$, we obtain a polynomial representation of $\varphi$ (dual of the residual) which is called the error representation function. Later, in~\cref{anisotropycomp}, we will show how this polynomial representation is paramount to anisotropy computation required for the mesh adaptation.  

Next we present the ultra-weak formulation of a scalar convection-diffusion problem. Let $T_{h}$ be the triangulation of the domain $\Omega$ where $\Omega \subset R^2$ is a bounded connected Lipschitz domain. $ \varepsilon_h^0$ denotes the set of all internal edges and $\varepsilon_h$ denotes the set of all edges. Consider the following function spaces
\begin{align*}
V_h :=&\{ v \in L^2(\Omega) : v\vert_\kappa \in P^p(\kappa), \forall \kappa \in {T}_h\} \\
\Sigma_h :=& \{ \bm{\tau} \in {(L^2(\Omega))}^d :\bm{ \tau}\vert_\kappa \in {(P^p(\kappa))}^d, \forall \kappa \in {T}_h\} \\
\Lambda_h :=& \{ \mu \in H^{\frac{1}{2}}(\varepsilon_h^0): \mu\vert_e \in P^p(e), \forall e \in \varepsilon_h^0 \}\\
\omega_h :=& \{ \mu \in L^2(\varepsilon_h): \mu\vert_e \in P^p(e), \forall e \in \varepsilon_h\}
\end{align*} 

Consider  a scalar second order convection diffusive equation with Dirichlet boundary condition
\begin{align*}
-\epsilon{\nabla}^2  u + \nabla \cdot \bm{\beta} u = s(\mathbf{x}), && \qquad \mathbf{x} \in \Omega \subset \mathbb{R}^d \\
u = g(\mathbf{x}), && \qquad \mathbf{x} \in \partial\Omega
\end{align*}

Re-write as first order system:
\begin{align}
\bm{\sigma} = \epsilon \nabla u, && \qquad \mathbf{x} \in \Omega \label{strng_eq_a}\\
-\nabla \cdot \bm{\sigma} + \nabla \cdot\bm{\beta}u = s(\mathbf{x}), &&  \qquad \mathbf{x} \in \Omega  \label{strng_eq_b}\\
u = g(\mathbf{x}), && \qquad \mathbf{x} \in \partial\Omega \label{strng_eq_c}
\end{align}
We define the space $\mathbb{X}_h = \Sigma_h \times V_h \times \Lambda_h \times \omega_h$ and formulate the DPG method as a discontinuous Galerkin discretization of ~\cref{strng_eq_a} - ~\cref{strng_eq_c}. Find $U_h  = (\bm{\sigma}_h,u_h,\lambda_h,\hat{\sigma}_h) \in \mathbb{X}_h$ such that,
\begin{align*}
\frac{1}{\epsilon}\int_{\Omega_h} \bm{\sigma}_h \cdot \bm{\tau} \, d\mathbf{x} + \int_{\Omega_h} u_h \nabla\cdot \bm{\tau}  \,d\mathbf{x}  -\int_{\varepsilon_h^0} {{\lambda_h}} \llbracket \bm{\tau}\cdot \bm{n}_e \rrbracket \, ds + \int_{\Omega_h} \bm{\sigma}_h \cdot \nabla v \, d\mathbf{x} -  \int_{\Omega_h} \bm{\beta}\cdot \nabla v\, u_h \,d\mathbf{x} - \\  \int_{\varepsilon_h} {\bm{{\hat{\sigma}}}_h} \llbracket v \rrbracket \, ds + \int_{\varepsilon_h^0} \vert \bm{\beta} \cdot \bm{n}_e \vert {{\lambda_h}} \llbracket v \rrbracket   \,ds = \int_{\kappa} s(\mathbf{x}) v \, d\mathbf{x} - \int_{ \partial \Omega_h}  \bm{\beta}\cdot \bm{n}_e  g(\mathbf{x}) \, d\mathbf{x} + \int_{ \partial \Omega_h} \bm{\tau}\cdot \bm{n}_e  g(\mathbf{x}) \, d\mathbf{x}
\end{align*}
for all $(\bm{\tau},v) \in \mathbb{V}_h$. Also to reflect element by element computations, we have used
\begin{align*}
\int_{\Omega_h} \bm{\sigma}_h \cdot \bm{\tau} \, d\mathbf{x} & = \sum_{\kappa \in T_h} \int_{\kappa} \bm{\sigma}_h \cdot \bm{\tau} \, d\mathbf{x}  \\ 
\int_{\varepsilon_h} {\bm{{\hat{\sigma}}}_h} \llbracket v \rrbracket \, ds & =   \sum_{\kappa \in T_h} \int_{\partial k}  {\bm{{\hat{\sigma}}}_h} \, sgn(\bm{n}_k) \,v \, ds \\
 \int_{\varepsilon_h^0} \lambda_h  \llbracket \bm{\tau} \cdot \bm{n}_e \rrbracket \, ds & = \sum_{\kappa \in T_h} \int_{\partial \kappa \setminus \partial \Omega} \lambda_h \bm{\tau} \cdot \bm{n}_k \, ds 
\end{align*}

In above variational formulation, each face has a particular normal $\bm{n}_e$ associated with it. Hence 
\begin{equation}
sgn(\bm{n}_k) = \begin{cases}
1 \qquad \bm{n_k} = \bm{n}_e \\
-1 \qquad \bm{n_k} = -\bm{n}_e
\end{cases}
\end{equation}
where $\bm{n}_k$ is outward facing for an element $k \in T_h$.  In case of ultra-weak variational formulation of diffusion and convection-diffusion equations, we will have two components in error representation function. Thus, in this case we will represent error representation function for convection-diffusion problem and diffusion problem as $\left( \psi_v,\psi_{\tau} \right)$.   

DPG is widely implemented using aforementioned the normal equations,~\cref{normal_eq}, but in certain cases, the normal equation may lead to a very ill-conditioned linear system.  For first order formulations, the normal equations lead to a stiffness matrix whose condition number scales with $O(h^{-2})$. It is a major problem especially for anisotropic adaptations as element size might get very small due to solution features. In order to cope with this situation, a discrete least square formulation (DLS) has been proposed in \cite{KEITHDLS}. We have implemented this DLS formulation for L-shaped domain test case presented in ~\cref{results}.
\subsection{Brief introduction to {DPG}\textsuperscript{$\star$}} \label{DPG_star}
Next, we will provide a brief description of the {DPG}\textsuperscript{$\star$} method \cite{DEMKOWICZ2018,KeithGoal}, which we have implemented for solving the dual problem. {DPG}\textsuperscript{$\star$}  has already been utilized to drive isotropic goal oriented adaptations and an in-depth exposition can be found in \cite{KeithGoal}. Let the target functional be:
\begin{equation}
J(u) = \left( j_{\Omega}, u \right) +  {(j_{\partial \Omega},Cu)}_{\partial \Omega} = \int_{\Omega} j_{\Omega} u \,d\mathbf{x} + \int_{\partial \Omega} j_{\partial \Omega} Cu \,ds \label{target_functional}
\end{equation}
where  $j_{\Omega} \in L^2(\Omega)$ and $j_{\partial \Omega} \in L^2(\partial \Omega)$ and $C$ is a boundary operator on $\partial \Omega$. For our purpose, we will assume that the target functional in~\cref{target_functional} is compatible with the primal problem, i.e. we assume that dual problem satisfies the compatibility condition \cite{Adjcompat} for linear problems given by:
\begin{equation}
\langle \mathcal{L}u,z \rangle + {\langle Au,C^{\star}z \rangle}_{\partial \Omega} = \langle u, {\mathcal{L}}^{\star}z \rangle + {\langle Cu,A^{\star}z \rangle}_{\partial \Omega},
\end{equation}
where $A$ is boundary operator that accompanies the strong form of the primal problem. $\mathcal{L}^{\star},A^{\star}$ and $C^{\star}$ are called the adjoint operators of $\mathcal{L},A$ and $C$. With this compatibility condition, it can be noticed that for given $\mathcal{L}$ and $A$, only certain target functional are compatible. On assuming that the compatibility condition holds, the adjoint problem associated with the primal problem  and the target functional mentioned in ~\cref{target_functional} is given by:
\begin{align}
\mathcal{L}^{\star} z & = j_{\Omega} \qquad  \forall \mathbf{x} \in \Omega  \label{dual_prob_op_eq}\\ 
A^{\star}z  & = j_{\partial \Omega} \qquad \forall \mathbf{x} \in \partial \Omega \label{dual_prob_bnd}
\end{align}

The discrete $\text{DPG}^{\star}$ can be stated as follows:
Find $ \hat{z}_h   = \mathbb{G}^{-1} B \xi_h \in \mathbb{V}_h$ and $\hat{\xi}_h \in \mathbb{X}_h$ such that  
\begin{equation}
{B}^T\mathbb{G}^{-1}{B} \xi_h = J(w_h) \,\qquad \forall w_h \in \mathbb{X}_h 
\end{equation}
where the discrete dual solution is given by $z_h = \sum_{i = 1}^{M} {(\hat{z}_h)}_i \psi_i$ where $\psi_i$ represents the basis of the test search space. The dual solution $z$ of the operator equation and associated boundary condition in ~\cref{dual_prob_op_eq} and ~\cref{dual_prob_bnd} is approximated by $z_h$. 
The error in target functional $J$ is bounded by \cite{KeithGoal}: 
\begin{equation}
\vert J(u- u_h) \vert \leq  {\Vert \mathcal{L} \Vert}^{-1} {\Vert \mathcal{L} u_h - l \Vert}_{\mathbb{V}'} {\Vert {\mathcal{L}}^{\star} z_h - J \Vert}_{U'}  \label{adjointerror_bound_dpg} 
\end{equation}

For diffusion and convection-diffusion problems with ultra-weak formulations, $z_h$ has two components as $z_h \in \mathbb{V}_h$. In ~\cref{adjointerror_bound_dpg}, one can replace the error in residual of the dual problem with an equivalent error estimator mentioned in Theorem 7.14 in  \cite{Keith_dissert} which we will reiterate next in ~\cref{explicit_DPGstar_est}. Since, we are using {DPG}\textsuperscript{$\star$} to solve the dual problem mentioned in  ~\cref{dual_prob_op_eq} and ~\cref{dual_prob_bnd}, we replace $\mathcal{L}$ with $\mathcal{L}^{\star}$ in error estimator mentioned in Theorem 7.14 in  \cite{Keith_dissert}.

\begin{equation}
{(\eta^{\star})}^2 = \sum_{\kappa \in T_h} \left( {\Vert {\mathcal{L}}^{\star} (v,\tau) - j_{\Omega} \Vert }_{L^2(\kappa)}^2 \right) + \sum_{e \in \varepsilon_h^0} h_e {\Vert \llbracket \bm{\tau} \cdot \bm{n} \rrbracket \Vert}_{L^2(e)}^2 + \sum_{e \in \varepsilon_h} {h_e}^{-1}  {\Vert \llbracket v \rrbracket \Vert}_{L^2(e)}^2 \label{explicit_DPGstar_est}
\end{equation}

Above estimator can be decomposed for each element $k \in T_h$ as a refinement indicator \cite{DEMKOWICZ2018} as presented in ~\cref{explicit_DPGstar_element}.
\begin{equation}
\eta^{\star}_k =  {\left({\Vert {\mathcal{L}}^{\star} (v,\tau) - j_{\Omega} \Vert }_{L^2(\kappa)}^2 + \sum_{e \in \partial k \setminus \partial \Omega} h_e {\Vert \llbracket \bm{\tau} \cdot \bm{n} \rrbracket \Vert}_{L^2(e)}^2 + \sum_{e \in \partial k} {h_e}^{-1}  {\Vert \llbracket v \rrbracket \Vert}_{L^2(e)}^2 \right)}^{\frac{1}{2}} \label{explicit_DPGstar_element}
\end{equation}

In ~\cref{explicit_DPGstar_est} and ~\cref{explicit_DPGstar_element}, $\mathcal{L}^{\star}$ is adjoint operator, $h_e$ is measure of size for edges in 2D and faces in 3D and ${\llbracket v \rrbracket}_e = v_{k^{+}} \bm{n}_{k^{+}} +  v_{k^{-}} \bm{n}_{k^{-}} $.
One  crucial point to be noticed here is that global stiffness matrix stays the same for both the primal and dual problem. The adjoint solution exists in approximate optimal test space ($\mathbb{V}_h$) as $  \hat{z}_h  = \mathbb{G}^{-1} B \xi_h $ which is nothing but discrete analogue of $z = {{\mathcal{R}}_\mathbb{V}}^{-1} \mathcal{L} \xi $. Hence, if one needs to compute dual weighted residual (DWR) for error correction using the dual solution $z_h$, they need to project $z_h$ to a richer space in order to overcome Galerkin orthogonality. We execute this by performing a patch-wise reconstruction using test search space.

%% file: mesh_metric_duallity.tex
\section{Mesh - Metric Duality}
\begin{figure}[H]
\begin{center}
\includegraphics[scale=1.5]{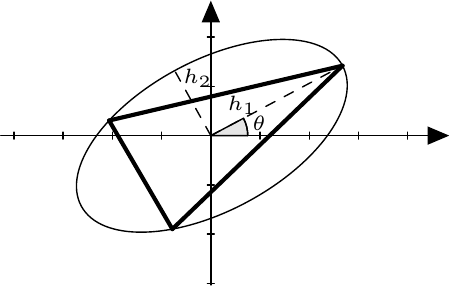} 
\end{center}
\caption{A unit triangle with $C = 3$ and its circumscribing ellipse}\label{e_t_duallity}
\end{figure}

In this section, we present a brief overview of the metric based mesh representation and continuous mesh model.  We have exploited these concepts in our proposed adaptation method. An in-depth exposition to these concepts can be found in \cite{inria_a} and \cite{inria_b}. Given a triangulation, both size and shape of an element can be characterized by a symmetric positive definite matrix of dimensions $2 \times 2$ . Precisely, let $e_k$, for $k =1,2,3$ denote the edges of the triangle, specified as vectors. Then, for any non-degenerate triangle, there exists a unique symmetric positive definite matrix
\begin{equation*}
\mathcal{M} = \begin{pmatrix}
m_{11} && m_{12} \\ m_{21} && m_{22}
\end{pmatrix}
\end{equation*} 
such that, for a given constant $C > 0$, we have
\begin{equation}
e^T\mathcal{M}e = C, \quad k = 1,2,3
\end{equation}
Here we call $\mathcal{M}$ as the metric tensor. Thus, it can inferred that the triangle is equilateral under norm induced by the metric,
\begin{equation*}
{\Vert x \Vert}_{\mathcal{M}} = \sqrt{x^T \mathcal{M} x }
\end{equation*}
An equilateral triangular mesh element is said to be unit with respect to the metric. 
Next, we can associate a unit triangle with $C = 3$ with its circumscribing ellipse provided that the centroid of the triangle and origin of the ellipse coincide. This is illustrated in ~\cref{e_t_duallity}.
The information about the aspect ratio, orientation of ellipse and the associated triangle is encoded within the eigenvectors and eigenvalues of $\mathcal{M}$. This can be illustrated by performing the spectral decomposition of the metric tensor as shown below. 
\begin{align*}
\mathcal{M} = {\begin{bmatrix}
\text{cos}(\theta) && -\text{sin}(\theta) \\
\text{sin}(\theta) && \text{cos}(\theta)
\end{bmatrix}}^T\begin{bmatrix}
\alpha_{1} && 0\\ 0 && \alpha_{2}
\end{bmatrix}\begin{bmatrix}
\text{cos}(\theta) && -\text{sin}(\theta) \\
\text{sin}(\theta) && \text{cos}(\theta)
\end{bmatrix}
\end{align*}

The eigenvectors represents the principal axes of the ellipse whereas eigenvalues($\alpha_{1} $ and $\alpha_{2}$) are inverse of the square of principle lengths $h_1$ and $h_2$. 
\begin{align*}
\alpha_{1}= \frac{1}{h_{1}^2}, \quad \alpha_{2} = \frac{1}{h_{2}^2}
\end{align*}

This relation between eigensystem of metric tensor and the attributes of the ellipse motivates the reason behind encoding information for a desired triangulation in a metric field. From the relation between the area of the ellipse and the triangle, we have 
\begin{align}
\vert \kappa \vert = \frac{3\sqrt{3}}{4} h_{1}h_{2} =  \frac{3\sqrt{3}}{4d}  \label{area_density_relation}
\end{align}

where local mesh density is defined as $d  = \frac{1}{h_1 h_2}$. Furthermore, we define aspect ratio as 
\begin{equation}
\beta = \sqrt{\frac{\alpha_{2}}{\alpha_{1}}} = \frac{h_1}{h_2} 
\end{equation}

Now if we take $\theta$ described above into account, we can describe an ellipse through its discrete metric using only three parameters $d,\beta,\theta$ instead of three elements of the metric tensor. By doing this, we have neatly separated the anisotropy of the triangle and size which is inversely proportional to the local mesh density. Now assume that we have a method through which we can quantify the desirable properties of the triangulation, then we can encode this information in a discrete metric field and produce a metric conforming triangulation. In next section, we will precisely introduce such a method by employing the inbuilt error estimator mentioned in the previous section.  A metric-based mesh generator produces a mesh using a continuous metric field ${(\mathcal{M}(\mathbf{x}))}_{\mathbf{x} \in \Omega}$ which can be generated by a suitable interpolation of the discrete metric field. Such a mesh generator will attempt to produce a triangulation conforming to this metric as closely as possible. The length of a vector under a continuous metric field is computed using the definition of length on Riemannian manifold. In the current work, we have employed freely available metric based mesh generator BAMG \cite{Bamg}. The length of a vector $e := \mathbf{x}_2-\mathbf{x}_1$ under a continuous metric field is computed as 
\begin{equation}
{\Vert e \Vert}_{\mathcal{M}} = \int_0^1 \sqrt{\bm{e}^T \mathcal{M}(\mathbf{x}_1 + t\bm{e})\bm{e}}\, dt
\end{equation}

%% file: Anisotropic_error_estimator.tex
\section{Anisotropic error estimate}   \label{anisotropycomp}
As mentioned in section 2, DPG methods with optimal test functions have an inbuilt residual based error estimator:
 \begin{align*}
 {\Vert U - U_h \Vert}_{E,T_h}^2 \approx {\Vert (\psi_v,\bm{\psi_\tau}) \Vert}^2_{\mathbb{V},T_h} = \sum_{\kappa \in T_h}   {\Vert \psi_v,\bm{\psi_{\tau}}) \Vert}^2_k
 \end{align*}
where $U_h \in \mathbb{X}_h$ represents the solution of convection-diffusion or diffusion problem and $ (\psi_v,\bm{\psi_\tau})$ are the associated error representation functions. It has been already shown in \cite{Demkowicz2011,Jessechan2014,Demkowicz2013} that one can control the equivalence of the energy norm with the norm of their choice by varying the norm associated with the test space. In this section while presenting the anisotropic error estimate, we employ the natural inner product associated with $H^1(T_h) \times H(div;T_h)$ which we will call standard V norm given by ~\cref{norm_MN} and a scaled version of V norm  \cite{Demkowicz2012a} given by ~\cref{scaled_norm_MN}. For a scalar convection diffusion problem and the diffusion problem with ultra weak formulation, $\psi_v \in H^1(T_h)$ and $\bm{\psi}_{\tau} \in H(div;T_h)$, the standard V norm for $k \in T_h$ is given by:  
\begin{align}
{\Vert (\psi_v,\bm{\psi}_{\tau} ) \Vert}^2_{V,k} & = {\Vert \psi_{v} \Vert}^2 + {\Vert \nabla \psi_v \Vert}^2  + {\Vert \bm{\psi}_{\tau}\Vert}^2 + {\Vert \nabla \cdot \bm{\psi}_{\tau}\Vert}^2 \notag \\ &= \int_k\underbrace{ {(\psi_{v}(\mathbf{x}))}^2+ \bm{\psi_\tau}(\mathbf{x}) \cdot\bm{\psi}_{\tau}(\mathbf{x}) +  ({\nabla \psi_{v}(\mathbf{x}) \cdot \nabla \psi_{v}(\mathbf{x})}  + {(\nabla \cdot\bm{\psi}_{\tau}(\mathbf{x}))}^2))}_{e_k(\mathbf{x})} \,d\mathbf{x}  \label{norm_MN}
\end{align}

and the scaled V norm as
\begin{align}
{\Vert (\psi_v,\psi_{\tau}) \Vert}^2_{V,k} & = {\Vert \psi_{v} \Vert}^2 +  \sqrt{\vert K \vert} \,  {\Vert \nabla \psi_v \Vert}^2 + {\Vert \bm{\psi}_{\tau}\Vert}^2 +  \sqrt{\vert K \vert} \, {\Vert \nabla \cdot \bm{\psi}_{\tau}\Vert}^2 \notag \\ &= \int_k\underbrace{ {(\psi_{v}(\mathbf{x}))}^2+ \bm{\psi_\tau}(\mathbf{x}) \cdot\bm{\psi}_{\tau}(\mathbf{x}) + \sqrt{\vert K \vert} ({\nabla \psi_{v}(\mathbf{x}) \cdot \nabla \psi_{v}(\mathbf{x})}  + {(\nabla \cdot\bm{\psi}_{\tau}(\mathbf{x}))}^2))}_{e_k(\mathbf{x})} \,d\mathbf{x}  \label{scaled_norm_MN}
\end{align}
 Here ${\vert K \vert}$ represents the volume of the element $k \, \in \, T_h$. 
\subsection{Anisotropy of Homogeneous polynomials}

\begin{definition}
Let $\bar{\mathbf{x}} =( {\bar{x},\bar{y}} )\in \Omega$ and $i \in \mathbb{N}$. We say that a polynomial $P_i(\mathbf{x}): \Omega \rightarrow \mathbb{R}$ is a homogeneous polynomial of order $i$ located at $\bar{\mathbf{x}}$ if 
\begin{equation}
P(\mathbf{x}) = \sum_{l=0}^{i} c_l {(x-\bar{x})}^l {(y-\bar{y})}^{i-l}
\end{equation}
\end{definition}

In [\cite{Dolejsi2014}, Lemma 3.12 ], the following result on the boundedness of a homogeneous polynomial can be found. We mention this in the next lemma as it is one of the fundamental results which we will need in our anisotropic error estimator.
\begin{lemma}
Let $P_i(\mathbf{x}):\Omega\rightarrow\mathbb{R}$ be a $i$ th order homogeneous polynomial  located at $\bar{\mathbf{x}}$, $i \geq 2$.  Then there exist values $A_i,\rho_i$ and $\phi_i \in [0,2\pi)$ such that
\begin{align}
\vert P_{i,\bar{\mathbf{x}}}(\mathbf{x}) \vert  \lesssim A_{i}\left((\mathbf{x}-\bar{\mathbf{x}})^TQ_{\phi_i}D_{\rho_i}{Q_{\phi_i}}^T(\mathbf{x}-\bar{\mathbf{x}})\right)^{\frac{i}{2}}
\end{align} \label{homogenousbndeq}
\end{lemma} 

In particular one can choose
\begin{align*}
Q_{\phi_i} = \begin{bmatrix}
\text{cos}(\phi_i) && -\text{sin}(\phi_i) \\ \text{sin}(\phi_i) && \text{cos}(\phi_i) 
\end{bmatrix}, \qquad D_{\rho_i} = \begin{bmatrix}
1 && 0 \\ 0 && \rho_i^{\frac{-2}{i}}
\end{bmatrix}
\end{align*}
where 
\begin{equation*}
\phi_i = \argmax_{\phi}  \vert P_{i,\bar{x}}(\xi(\phi)) \vert 
\end{equation*}

\begin{align*}
A_{i} = \max_{\xi \in B} \vert P_{i,\bar{\mathbf{x}}}(\xi(\phi_i)) \vert \qquad A_{i,\perp} =  { \vert P_{i,\bar{\mathbf{x}}}(\xi(\phi_i-\frac{\pi}{2})) \vert}
\end{align*}

\begin{align*}
\rho_i = \frac{A_{i}}{A_{i.\perp}}
\end{align*}

\begin{figure}[H]
\begin{center}
\includegraphics[scale=0.4]{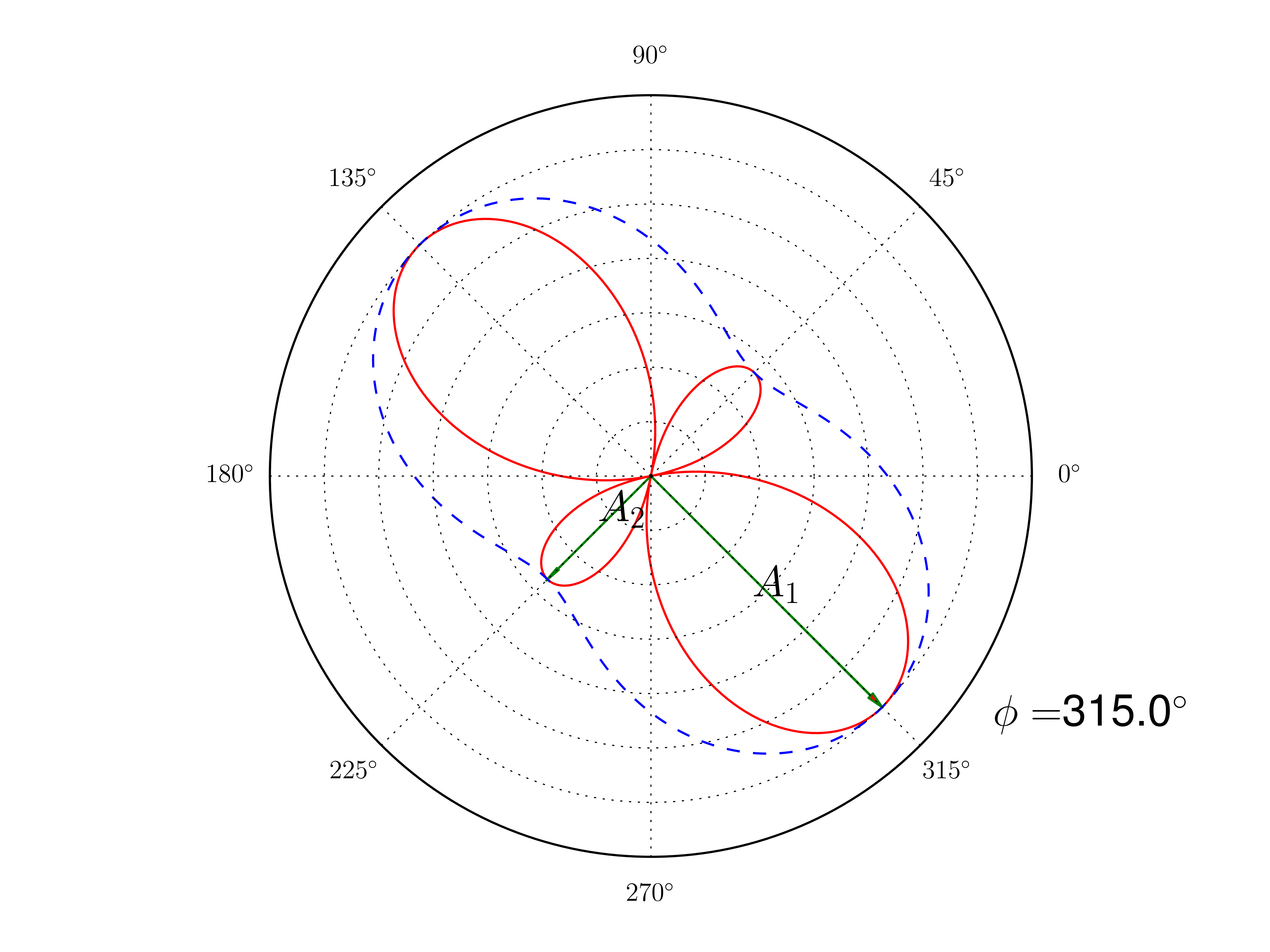}  
\end{center}
\caption{Bound on an arbitrary fourth order polynomial.}\label{homogenousboundfig}
\end{figure}
In ~\cref{homogenousboundfig}, the graph denoted by the red continuous line represents fourth order polynomial whereas the graph denoted by the blue dotted line represents the bound mentioned in ~\cref{homogenousbndeq}.
In order to obtain the anisotropic error estimator, we will be dealing
with the polynomial representation of the error representation function which is computed locally due to broken test spaces as mentioned in ~\cref{Discretization}. Let the test space and the trial space be chosen as mentioned in section 2, then error representation functions $\psi_v$ and ${(\bm{\psi}_{\tau})}_{i,j} \forall i,j \in {1,2}$ are polynomial of order $p+\delta p$ and hence, in ~\cref{norm_MN} $e_k(\mathbf{x})$ is a polynomial of $2(p+\delta p)$. Also let $E_k$ be the circumscribing ellipse of the element $K$ as shown in ~\cref{e_t_duallity}.

\begin{align}
e_k(\mathbf{x}) = \sum_{i = 0}^{2(P +\delta P)} \underbrace{P_{i,\bar{\mathbf{x}}_k}(\mathbf{x})}_ {\sum_{l=0}^{i} c_l {(x-\bar{x}_k)}^l {(y-\bar{y}_k)}^{i-l}} \label{net_energy_function}
\end{align}
~\Cref{net_energy_function} represents the decomposition of $e_k(\mathbf{x})$ into homogeneous polynomials centred at the centroid $\bar{\mathbf{x}}_k$ of $k \in T_h$. Since integrals of monomials whose cumulative power is odd over a symmetric interval about origin is zero, half of the monomials in \cref{net_energy_function} gets dropped while integrating which can noticed in \cref{integral_energy_func}.

\begin{align}
\int_k e_k(\mathbf{x}) \, d\mathbf{x} \leq \int_{E_k} e_k(\mathbf{x}) \, d\mathbf{x} \lesssim \sum_{i = 2,i \in Z^{+}_{E}}^{2(P +\delta P)} \int_{E_k}A_{i}\left((\mathbf{x}-\bar{\mathbf{x}}_k)^TQ_{\phi_i}D_{\rho_i}{Q_{\phi_i}}^T(\mathbf{x}-\bar{\mathbf{x}}_k)\right)^{\frac{i}{2}} \, d\mathbf{x}   \label{integral_energy_func}
\end{align}
In above expression, we have ignored $i = 0$ as it represents a constant which acts as an offset to the bound and has no anisotropy. $Z^{+}_{E}$ represents the set of positive even integers. We denote by $h_1$ and $h_{2} = \frac{h_1}{\beta_{\mathcal{M}}}$ the size of semi-axes of sought ellipse $E_k$,$\beta_{\mathcal{M}}$ is the aspect ratio and the angle between the major axis of $E_k$ and x-axis is denoted by $\phi_{\mathcal{M}}$. Let $\hat{E} := \{ \zeta \in R^2;\vert \zeta \vert \leq 1 \}$ be a  closed unit ball. Then a mapping $F: \hat{E} \rightarrow R^2 $ can be defined as
\begin{align}
F(\hat{\mathbf{x}}) = Q_eS_e\hat{\mathbf{x}} + \bar{\mathbf{x}}_k
\end{align}
where $Q_e$ is the rotation matrix for rotation of angle $\phi_\mathcal{M}$ and
\begin{align*}
S_e = \begin{bmatrix}
h_1 & 0 \\ 0 & h_{2}
\end{bmatrix} = h_1 \begin{bmatrix}
1 & 0 \\ 0 & \frac{1}{\beta_\mathcal{M}}
\end{bmatrix}
\end{align*}
If we treat $\bar{\mathbf{x}}_k$ as origin and introduce polar coordinates $\hat{\mathbf{x}} = {\begin{bmatrix}
cos(\theta) & sin(\theta)
\end{bmatrix}}^T$, we have $\mathbf{x}-\bar{\mathbf{x}}_k = Q_eS_e\hat{\mathbf{x}} = Q_eS_e{\begin{bmatrix}
cos(\theta) & sin(\theta)
\end{bmatrix}}^T $. On substituting $\mathbf{x}-\bar{\mathbf{x}}_k $ in ~\cref{integral_energy_func},
\begin{align*}
\int_k e_k(\mathbf{x}) \, d\mathbf{x} \leq \int_{E_k} e_k(\mathbf{x}) \, d\mathbf{x} \lesssim {\sum_{i = 2,i \in Z^{+}_{E}}^{2(P +\delta P)}} \frac{A_{i} {\lambda_k}^{\frac{i+2}{2}}}{i+2}\int_{0}^{2\pi} {(g_i(\theta;{\beta_{\mc}},{\phi_\mc}-\phi_i))}^{\frac{i}{2}} \, d\theta
\end{align*}
where $\lambda_k = \frac{1}{d_k}$ and $d_k$ is computed using the implied metric (computed using the vertices of $k \in T_h$ where $T_h$ is the current triangulation). The anisotropic properties $\{\beta_\mc,\phi_\mc\}$ are set by minimizing the error bound for $k \in T_h$.

\begin{align}
{\beta_\mc,\phi_\mc} = \argmin_{\beta,\phi} \underbrace{{ \sum_{i = 2,i \in Z^{+}_{E}}^{2(P +\delta P)} \frac{A_{i} {\lambda_k}^{\frac{i+2}{2}}}{i+2}\int_{0}^{2\pi} {(g_i(\theta;{\beta},{\phi}-\phi_i))}^{\frac{i}{2}} \, d\theta }}_{\bar{G}}\label{min_beta_phi}
\end{align}
 where
 \begin{eqnarray}
&  g_i(\theta,\beta,\phi-\phi_i) =  G_{11,i} cos^2(\theta) + G_{22,i} sin^2(\theta) +2G_{12,i}sin(\theta)cos(\theta) \label{g_i}\\
& with \quad G_{11,i} = \beta(cos^2({\phi}-\phi_i) + {\rho_i}^{\frac{-2}{i}} sin^2({\phi}-\phi_i)) \nonumber  \\
& \qquad  \quad  G_{12,i} = -sin(\phi-\phi_i)cos(\phi-\phi_i)(1-{\rho_i}^{\frac{-2}{i}})\nonumber \\ 
& \qquad \quad   G_{22,i} = \beta(sin^2({\phi}-\phi_i) + {\rho_i}^{\frac{-2}{i}} cos^2({\phi}-\phi_i))  \nonumber
 \end{eqnarray}
  
Next, we will be briefly mentioning the procedure implemented for solving the minimization problem stated by ~\cref{min_beta_phi}. From ~\cref{g_i}, it can inferred that $g_i(\theta,\beta,\phi-\phi_i) $ (consequently $\bar{G}$) is $\pi$ periodic in $\phi$ and $\bar{G} \rightarrow \infty$ for $\beta \rightarrow \infty$ as $g_i  \rightarrow \infty$ for $\beta \rightarrow \infty$. Hence, the continuity of $\bar{G}$ with respect to $\beta$ and $\phi$ implies the existence of atleast one minima. The minimum of ~\cref{min_beta_phi} is sought iteratively by performing an alternate search in $\beta$ and $\phi$. More details can be found in [section 7.2 \cite{Dolejsi2019}].

%% file: continuous_error_estimate.tex
\section{Continuous model of energy error estimate for H adaptation} 	\label{cont_model_h_adap}

When $U_h = (u,\bm{\sigma},\hat{u}, \hat{\sigma}_h)$ is approximated with $\mathbb{X}_h = \Sigma_h \times V_h \times \Lambda_h \times \omega_h$  using the DPG framework with optimal test functions, it has already been shown that $\Vert U-U_h\Vert_E$ is equivalent to the best approximation error under various test norms. In order to achieve this equivalence, the primary ingredient is the equivalence of prescribed test norm with optimal test norm. These salient results regarding equivalence of standard V norm and the optimal test norm are rigorously proved in  \cite{Demkowicz2011a}. In the present work, we have prescribed discontinuous trace approximations. It has already shown in \cite{Norbert2014} that this is a valid functional setting and is quasi optimal when compared to conforming trace approximations. Hence, we assume that ${\Vert U-U_h\Vert}_E = o(h^{p+1})$ which is essential for the proposed continuous error estimate. We have also used scaled V norm. The reason behind this scaling is the fact that this keeps the Gram matrix $\mathbb{G}$ well behaved as the elements size starts reducing locally which in turn keeps the condition number of Global stiffness matrix tractable. This ill conditioning of the gram matrix is already mentioned in \cite{Jessechan2014}. Other option  to overcome this issue is to implement the DLS framework but it comes with added cost of solving a least squares problem.

\begin{figure}[H]
\hspace*{-2cm}  
\minipage{0.4\textwidth}
\includegraphics[width=\linewidth]{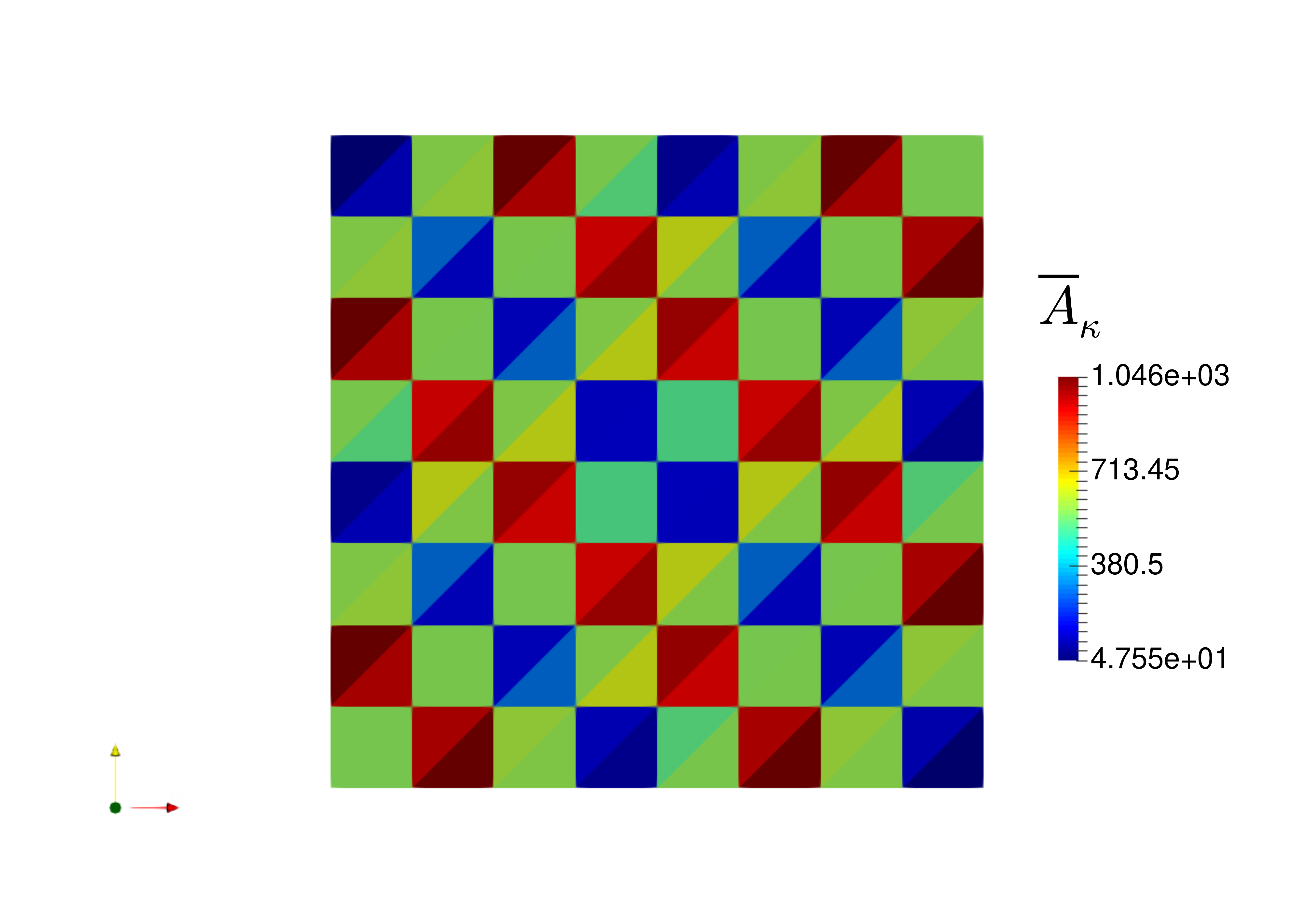}
\captionsetup{labelformat=empty}
\caption*{$ne  = 128$}
\endminipage
\minipage{0.4\textwidth}
\includegraphics[width=\linewidth]{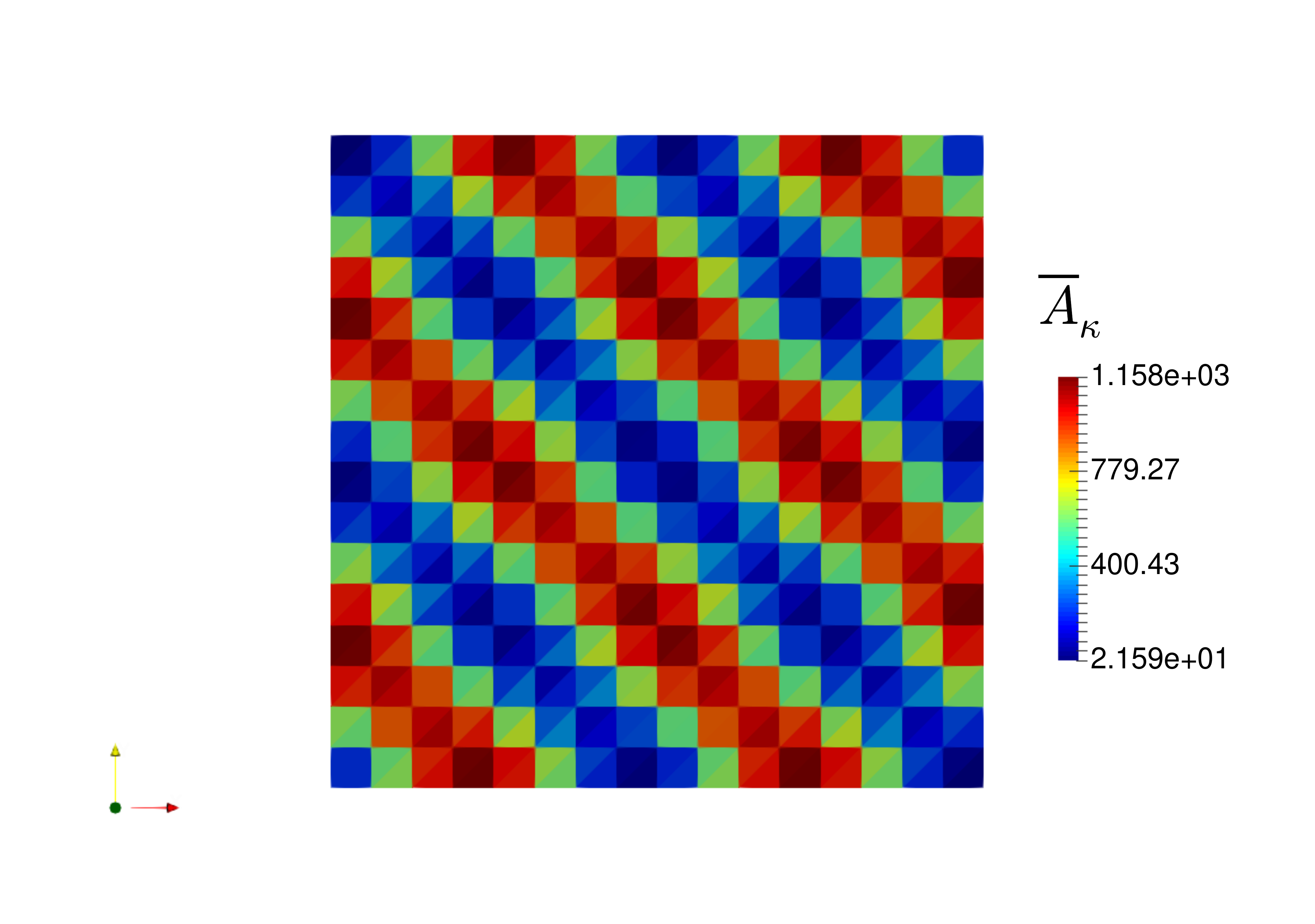}
\captionsetup{labelformat=empty}
\caption*{$ne  = 512$}
\endminipage
\minipage{0.4\textwidth}
\includegraphics[width=\linewidth]{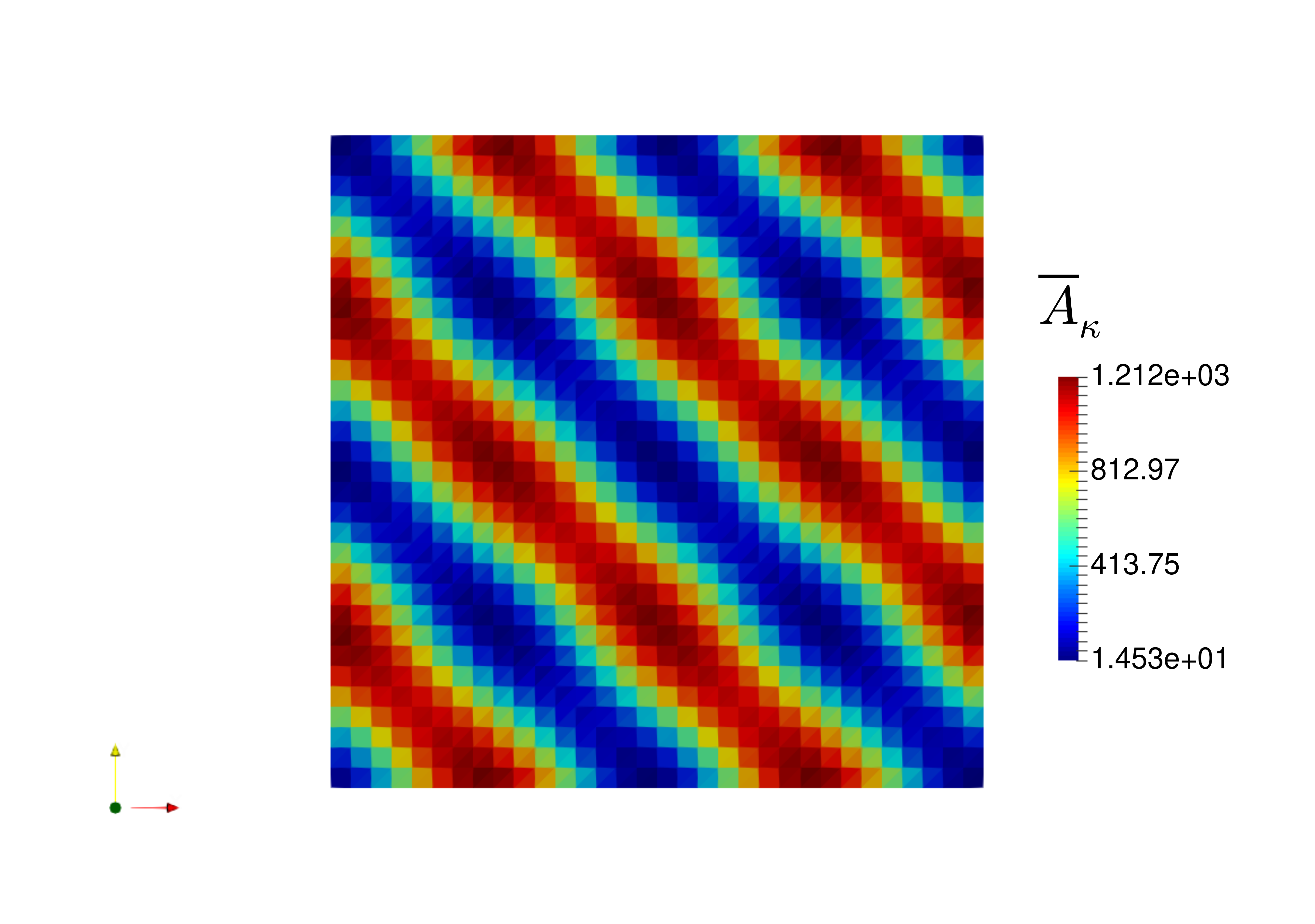}
\captionsetup{labelformat=empty}
\caption*{$ne  = 2048$}
\endminipage\hfill
\caption{Numerical Justification of ~\cref{assump1} using $P = 1$} \label{verf_1}
\end{figure}

\begin{figure}[H]
\hspace*{-2cm}  
\minipage{0.4\textwidth}
\includegraphics[width=\linewidth]{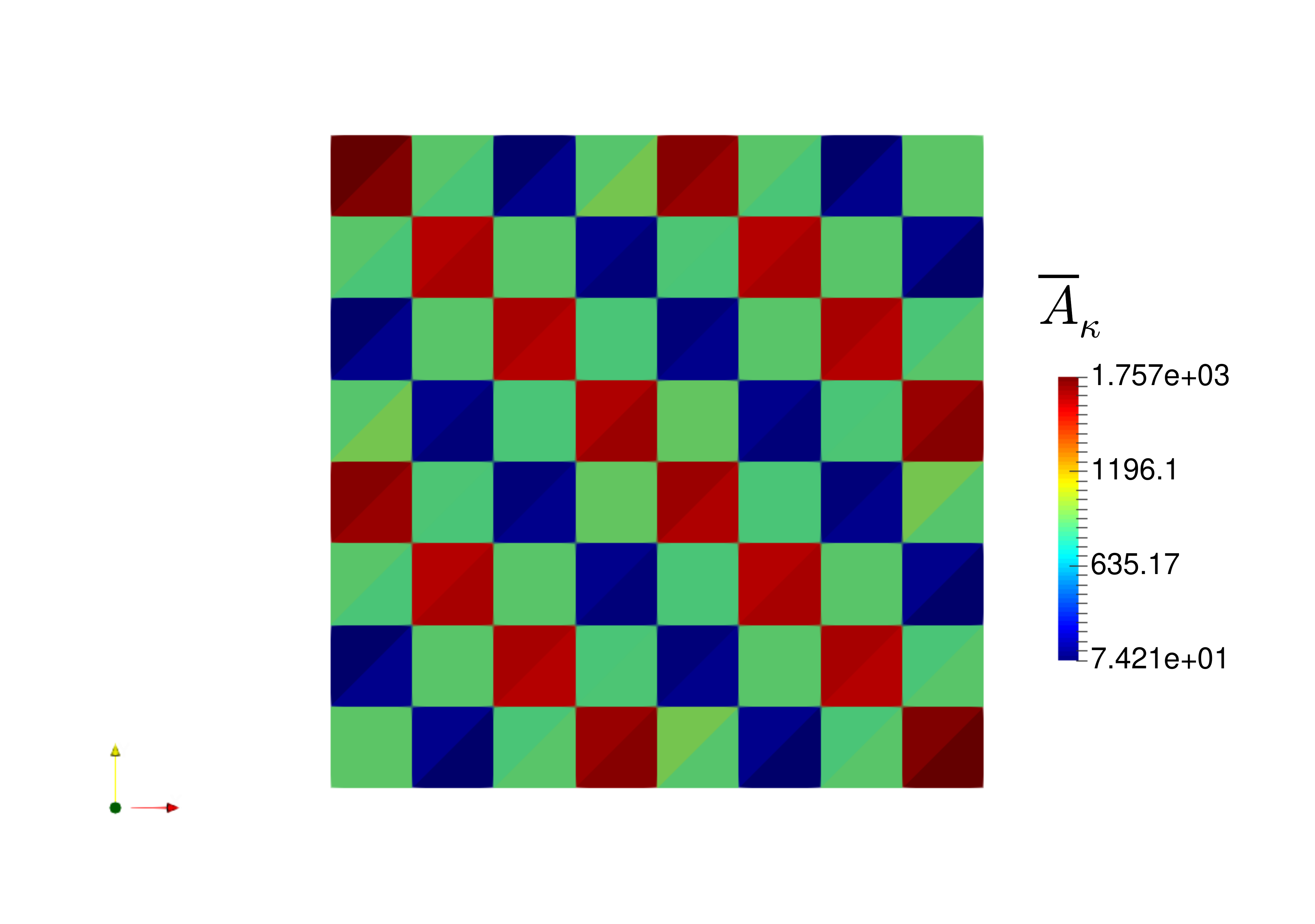}
\captionsetup{labelformat=empty}
\caption*{$ne  = 128$}
\endminipage
\minipage{0.4\textwidth}
\includegraphics[width=\linewidth]{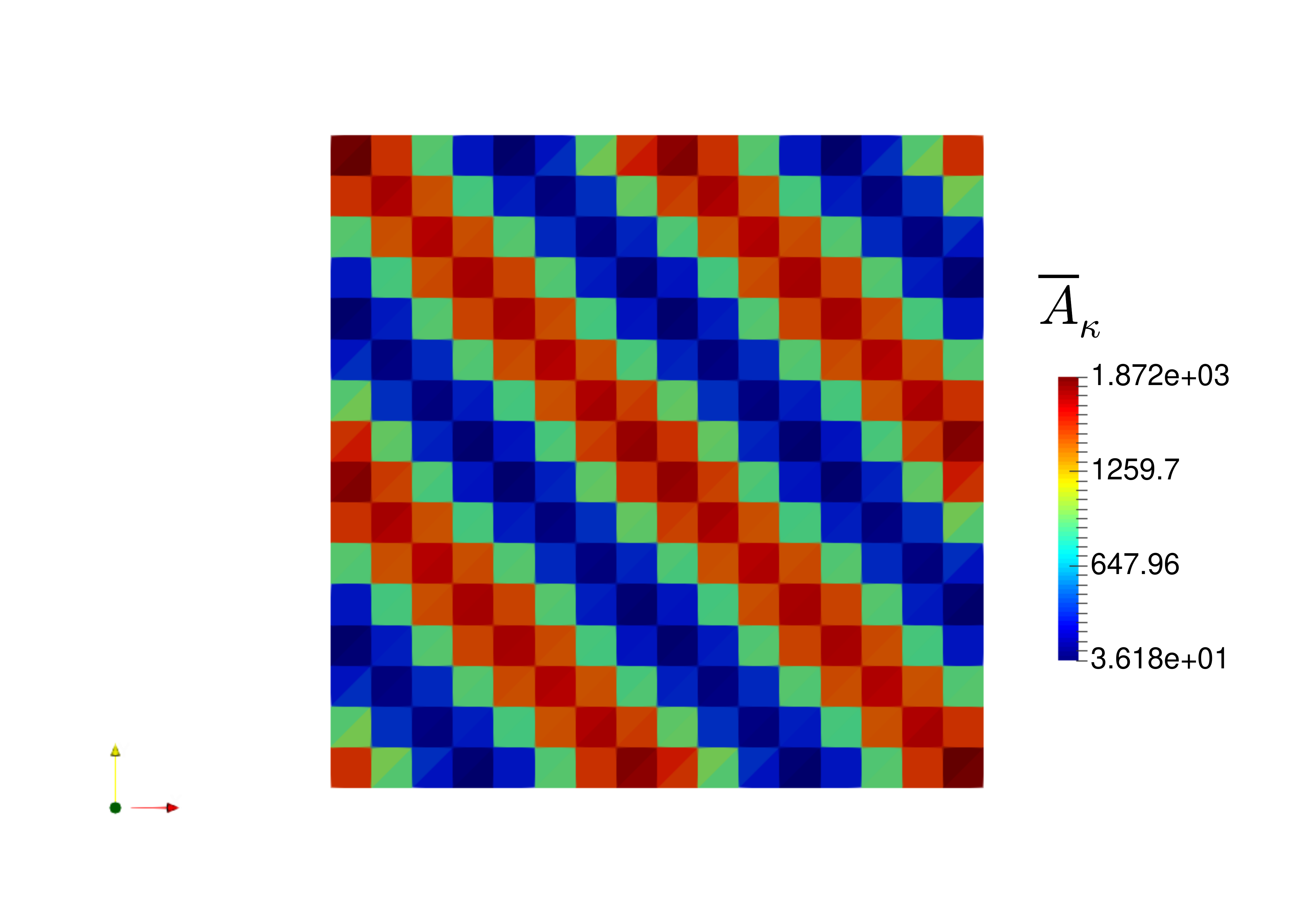}
\captionsetup{labelformat=empty}
\caption*{$ne  = 512$}
\endminipage
\minipage{0.4\textwidth}
\includegraphics[width=\linewidth]{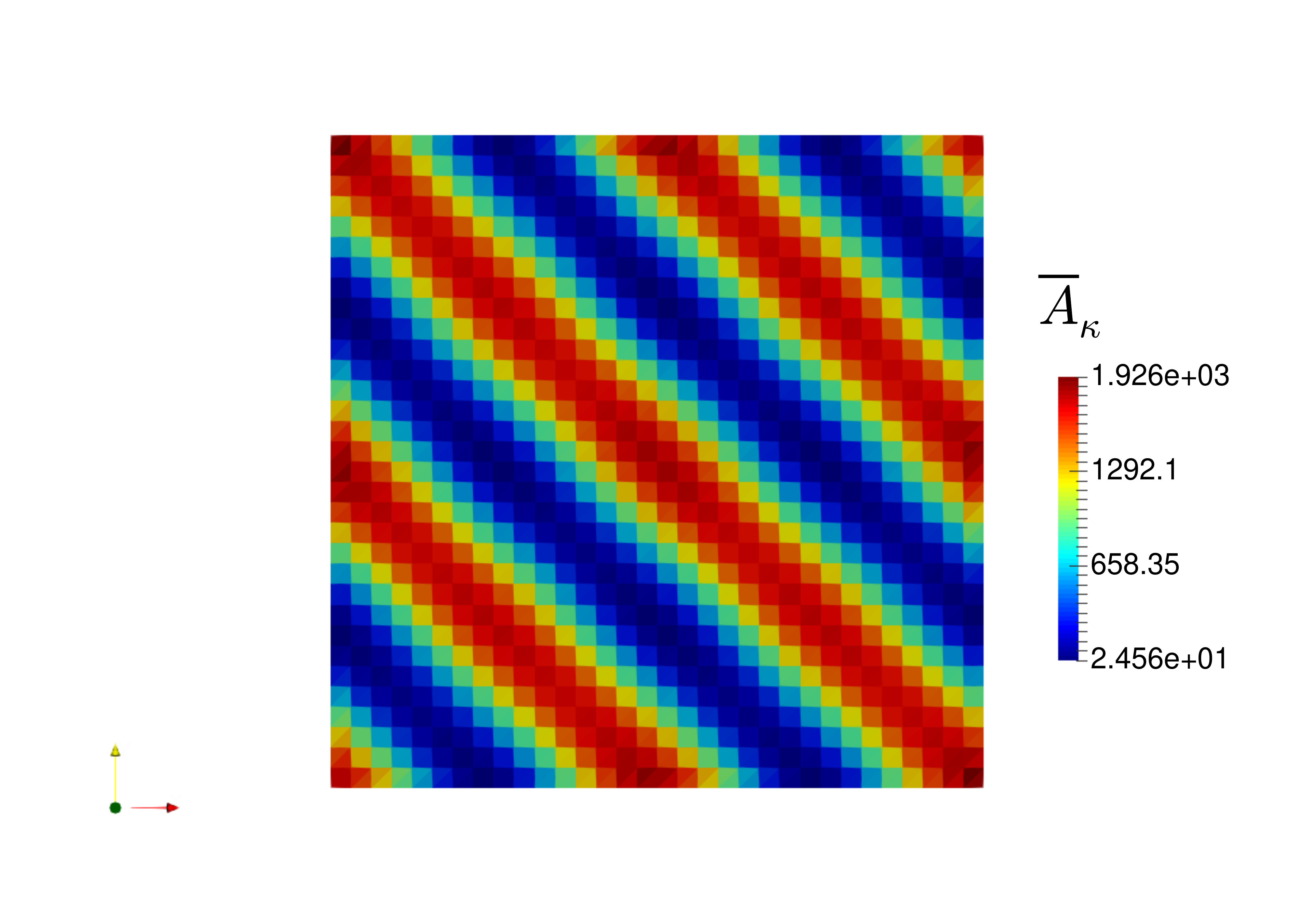}
\captionsetup{labelformat=empty}
\caption*{$ne  = 2048$}
\endminipage\hfill
\caption{Numerical Justification of ~\cref{assump1}  using $P = 2$} \label{verf_2}
\end{figure}

\begin{figure}[H]
\hspace*{-2cm}  
\minipage{0.4\textwidth}
\includegraphics[width=\linewidth]{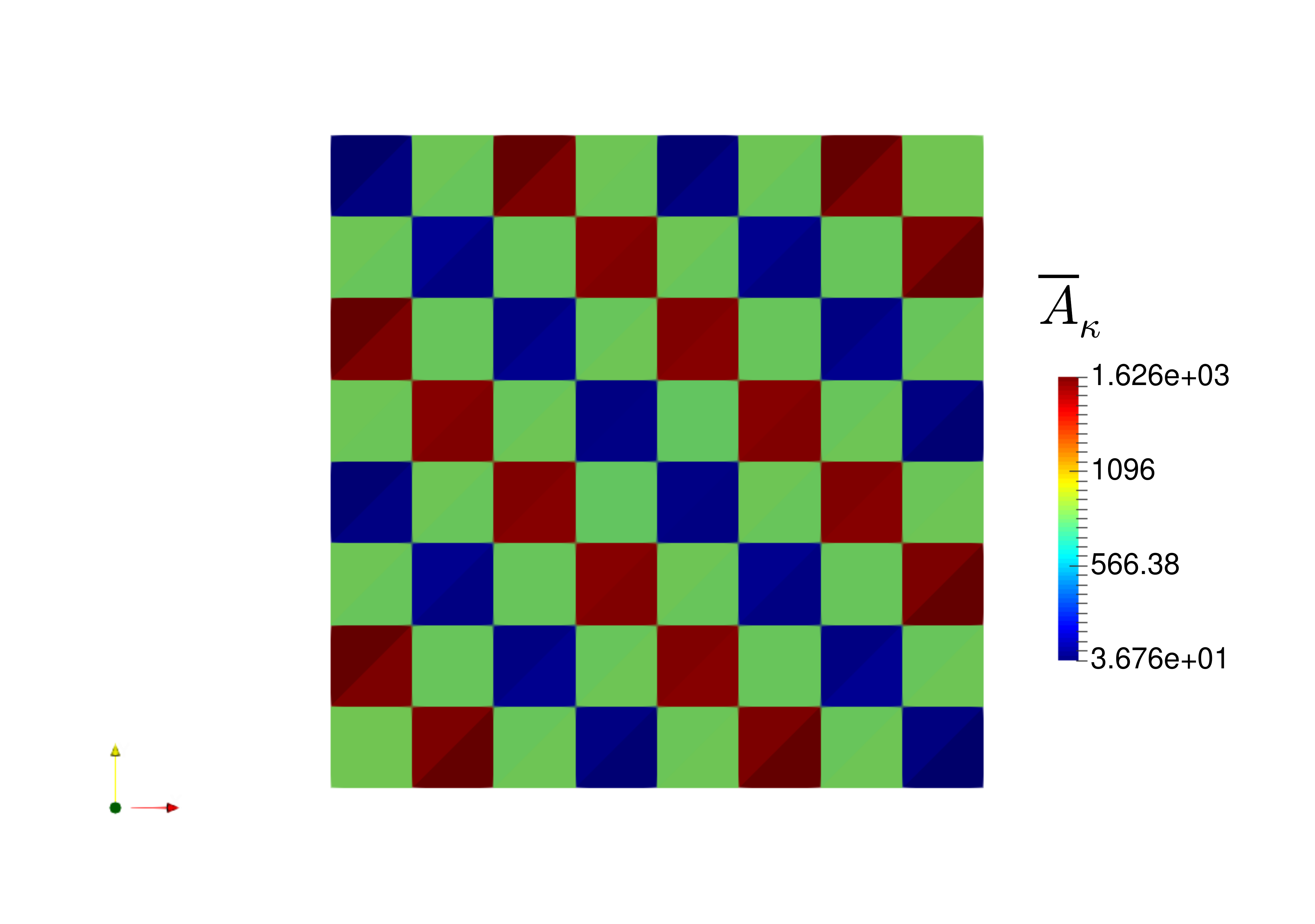}
\captionsetup{labelformat=empty}
\caption*{$ne  = 128$}
\endminipage
\minipage{0.4\textwidth}
\includegraphics[width=\linewidth]{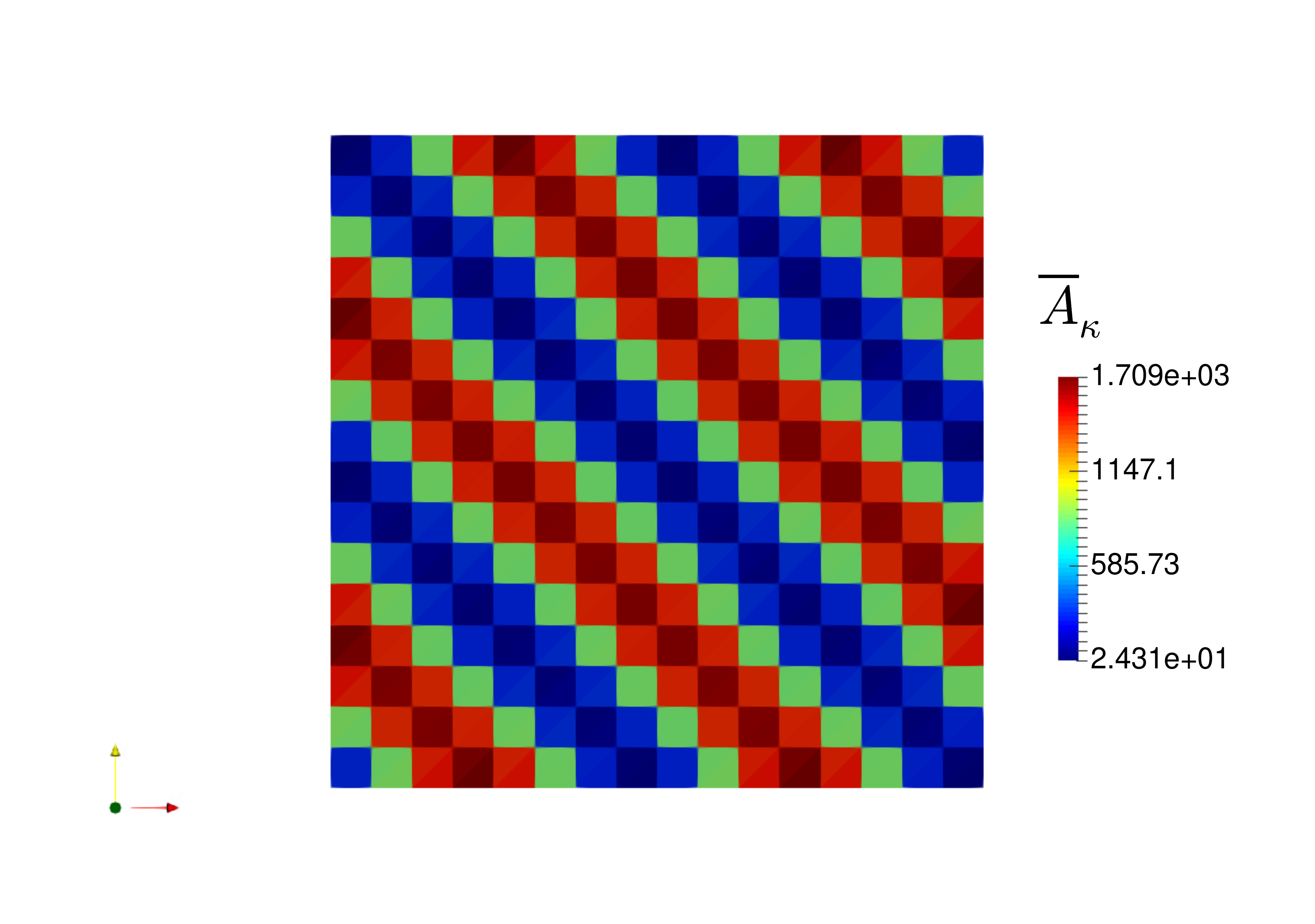}
\captionsetup{labelformat=empty}
\caption*{$ne  = 512$}
\endminipage
\minipage{0.4\textwidth}
\includegraphics[width=\linewidth]{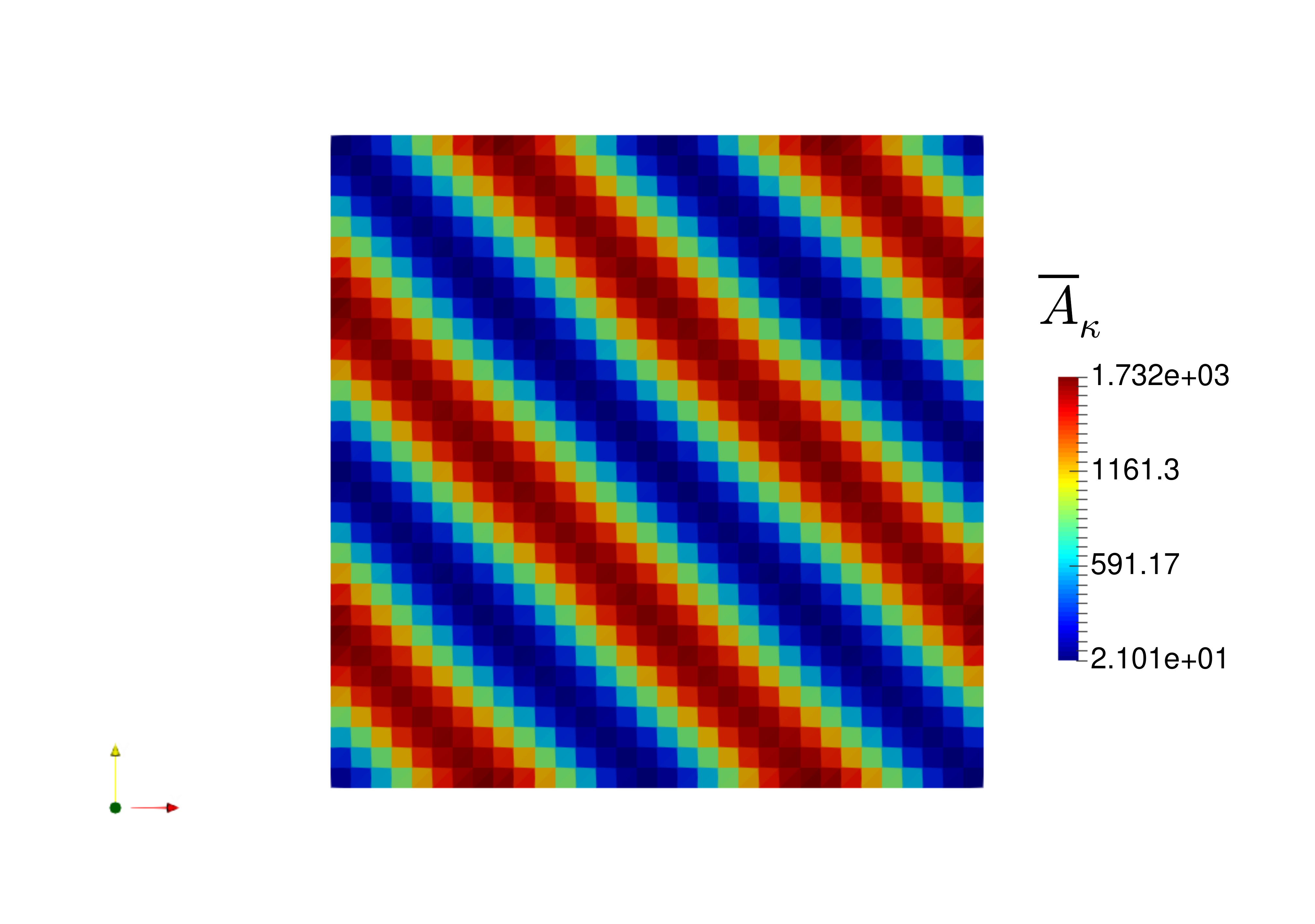}
\captionsetup{labelformat=empty}
\caption*{$ne  = 2048$}
\endminipage\hfill
\caption{Numerical Justification of ~\cref{assump1}  using $P = 3$} \label{verf_3}
\end{figure}

\begin{assumption}
Let ${T}_h$ be a triangulation and  $\eta_k$ be a local error estimate such that ${\eta}^2 = \sum_{k \in {T}_h} {\eta}^2_k$ and $\eta = O(h^s)$ i.e error estimate converges at $s^{th}$ order \cite{Dolejsi2017}. We assume that local error estimate $\eta_k$ scales \footnote{To motivate the scaling $\eta_k \propto {\vert K \vert}^{(s+1)}$, recall that the global error estimate $\eta^2$ scales as $h^{2s} \propto {\vert K \vert}^s$. (we have a $s^{th}$ order method.) At the same time, the contribution from each individual sub-element scales with an additional factor of K, because the local domain of integration becomes smaller.} as $\eta_k = \overline{A}_k {\vert K \vert}^{(s+1)}$ where $A_k$ depends only upon the anisotropy of the element. Furthermore, the order of the method $s$ directly depends upon the polynomial degree of approximation $p$. \label{assump1}
\end{assumption}
Verification of ~\cref{assump1} can be found in [section 8.1, \cite{Dolejsi2017}] where the authors have demonstrated this with a Laplace problem and in [section 3.2, \cite{VENDITTI200240}] where similar properties are studied for compressible flow problems. In present work, we have shown a similar demonstration solving the same Laplace problem as in [section 8.1, \cite{Dolejsi2017}] where the exact solution  u has the form $u = sin(2 \pi x) sin(2 \pi y)$  in ~\cref{verf_1}, ~\cref{verf_2} and ~\cref{verf_3} for $P = 1,2,3$ using the standard V norm.
Similar to the observation made in \cite{Dolejsi2017}, there is very small dependence of maximal value of $\overline{A}_k$ on element size whereas dependence of minimal value is not. However as already mentioned in  \cite{Dolejsi2017}, these elements with smallest error usually do not contribute too much to the total error.

Let $\left\{{T}_h\right\}_n$ be the sequence of triangulation employed, we would like to construct an error estimate which achieves global equality with the inbuilt energy error estimate asymptotically.
\begin{align}
\Vert U - U_h \Vert^2_{E,k} \approx e_d(\mathbf{x}_k) \vert K \vert \quad \text{for} \, \, \mathbf{x}_k \in k, \, k \in T_h\label{origin}
\end{align}
where $e_d(\mathbf{x}_k)$ is the density of the error estimate. Thus  Asymptotically ($h \rightarrow 0$), we would like to have
\begin{align*}
\Vert U-U_h \Vert^2_{E,\Omega} = \sum_{k \in {T}_h} e_d(\mathbf{x}_k) \vert K \vert \rightarrow \int_{\Omega}  e_d(\mathbf{x}) \, d \mathbf{x}
\end{align*}
 Next, we define the error density function as following
 \begin{equation}
 e_d(\mathbf{x}) = \overline{A}(\mathbf{x}){d(\mathbf{x})}^{-s}\alpha^s \label{error_density}
 \end{equation}
where $\alpha = \frac{3\sqrt{3}}{4}$. Using ~\cref{assump1}  and apriori established order of convergence of energy estimate in  \cite{Demkowicz2012a}, we can set $ s = p+1 $.

Now we introduce the global continuous estimate $E =\int_{\Omega} e_d(\mathbf{x}) \, d\mathbf{x}$ which will be the primary target of the optimization ~\cref{opt_prob_a}. In problem stated below, we will be treating $\overline{A}(\mathbf{x})$ as a continuous function though it can only be computed in a discrete way on a triangulation. Also, we would like to introduce the notion of mesh complexity. We define the mesh complexity as 
\begin{equation*}
N = \int_{\Omega} d(\mathbf{x})\, d\mathbf{x}
\end{equation*}
In the case of triangulation, we will have a discrete density field i.e. $d(\mathbf{x})$ is piecewise constant. Hence, we have  
\begin{equation}
\int_{\Omega} d(\mathbf{x})\, d\mathbf{x} \approx \sum_{k \in {T}_h} d(\mathbf{x}_k) \vert k \vert = \frac{3\sqrt{3}}{4}  Ne  \quad \text{where} \, \, \mathbf{x}_k \in k
\end{equation}

In above relation, we have used ~\cref{area_density_relation} , $Ne$ is the total number of elements in the mesh. Thus, if $N$ is prescribed as a constraint, we can formulate a problem of minimizing continuous error at fixed cost i.e desired number of elements in the mesh.
\begin{problem}
Let $N$ be the desired complexity and $e_d(\mathbf{x})$ be the error density function. We seek a mesh density distribution $d(\mathbf{x}): \Omega \rightarrow \mathbb{R}^{+}$ on the next triangulation  such that: \label{opt_prob_a}

(a) $N = \int_{\Omega} d(\mathbf{x})\, d\mathbf{x}$. 

(b) $E =\int_{\Omega} e_d(\mathbf{x}) \, d\mathbf{x}$ is minimized. 
\end{problem}
In order to compute the minima of the mesh density distribution $d(\mathbf{x})$ under the constraint of constant complexity $N$, we will be employing calculus of variations. We begin by taking variation of the constraint in \cref{opt_prob_a} with respect to $d(\mathbf{x})$ which results in the following integral:
\begin{equation}
\int_{\Omega} \delta d(\mathbf{x}) \, d\mathbf{x} = 0 \label{const_var}
\end{equation}
Next, on substituting error density function into global continuous error estimate and taking variation with respect to $d(\mathbf{x})$,
\begin{equation}
\delta E = -(p+1) {\alpha}^{(p+1)} \int_{\Omega} {\overline{A}(\mathbf{x})}{d}^{-(p+2)} \delta d(\mathbf{x}) \, d\mathbf{x} \label{glob_err_var}
\end{equation}
From ~\cref{const_var} and ~\cref{glob_err_var}, it can  be implied that 
\begin{equation}
\overline{A}(\mathbf{x}) {d}^{-(p+2)} = const.
\end{equation}
Hence again using the constraint, we obtain 
\begin{equation}
d^{\star} = C {\overline{A}(\mathbf{x})}^{\frac{1}{p+2}}, \qquad C = \frac{N}{\int_{\Omega}{\overline{A}(\mathbf{x})}^{\frac{1}{(p+2)}} \, d\mathbf{x}}
\end{equation}

On substituting  the expression for $C$, we obtain the expression for the optimal density.
\begin{equation}
d^{\star}(\mathbf{x}) = \frac{N\overline{A}(\mathbf{x})^{\frac{1}{p+2}}}{\int_{\Omega}{\overline{A}(\mathbf{x})}^{\frac{1}{(p+2)}} \, d\mathbf{x}} \label{optdensity_a}
\end{equation}

Since, we are in a discrete setting in terms of triangulation, we will compute this desired density for each element $k \in {T}_h$. This allows us to replace $\overline{A}(\mathbf{x})$ with $\overline{A}(\mathbf{x}_k)$ and thus, we obtain
\begin{align}
\overline{A}(\mathbf{x}_k) = \frac{{\Vert U - U_h \Vert}^2_{E,k}}{{\vert K \vert}^{p+2}} \label{abar}
\end{align}

\begin{equation}
d^{\star}(\mathbf{x}_k) = \frac{N{\Vert U-U_h\Vert}_{E,k}^{\frac{2}{p+2}}}{\vert K \vert \left(\sum \limits_{k \in {T}_h} {\Vert U-U_h \Vert}^\frac{2}{(p+2)}_{E,k}  \right)} \label{optdensity_b}
\end{equation}
\\
\begin{equation}
 \sqrt{e(\mathbf{x}_k,d^{\star}){\vert K^{\star} \vert}} = {\alpha}^{\frac{(p+2)}{2}}{N}^{-\frac{(p+2)}{2}} {\left( \sum \limits_{k \in {T}_h} {\Vert U-U_h \Vert}^\frac{2}{(p+2)}_{E,k}\right)}^\frac{(p+2)}{2}\label{equidist}
\end{equation}

\begin{equation}
E^{\star} = \sqrt{ \sum\limits_{k \in {T}_h} {e(\mathbf{x}_k,d^{\star})} \vert K^{\star} \vert
} = {\left(\frac{\alpha}{N}\right)}^{\frac{p+1}{2}}{\left( \sum \limits_{k \in {T}_h} {\Vert U-U_h\Vert}_{E,k}^{\frac{2}{(p+2)}}\right)}^{\frac{(p+2)}{2}} \label{cont_error_hadap}
\end{equation}

Finally, after  computing  $(\beta_{\mathcal{M}}, \theta_{\mathcal{M}}, d^{\star})$ for every element in $T_h$, the discrete metric field can generated and can be passed on to the metric-based mesh generator to produce the desired triangulation.
\subsection{Goal Oriented Adaptation}
For Goal oriented adaptation, we use $ \tilde{\eta}_k = \eta^{\star}_k \eta_k$ where $\eta_k = {\Vert U-U_h \Vert}_{E,k}$ and $\eta^{\star}_k$ from ~\cref{explicit_DPGstar_element} in place of just ${\Vert U-U_h\Vert}_{E,k} $ in above mentioned algorithm for global size computation whereas the anisotropy computation stays the same. We kept the value of $s$ in this case also $p+1$. In ~\cref{assump1}, we require sufficient regularity but in many cases the regularity of both the dual and primal solution is questionable. Hence, we prefer to use $s = p+1$ which is a pessimistic approach in place of $s = 2p+1$. This results in overestimation of optimal area and hence might not be sufficiently refined initially. But this gets improved with subsequent adaptations. In \cite{Dolejsi2017}, this has already been highlighted for non-regular solutions where the authors have used $s = p+1$ which can be higher than the expected regularity of the solution. In that case, optimal area will be underestimated and with subsequent adaptation computed area will converge to the optimal one. In order to get the global density distribution, we have employed the {DPG}\textsuperscript{$\star$} based error estimator from  \cite{Keith_dissert}  but this can be replaced with DWR error estimator using the reconstructed adjoint solutions.
While reconstructing adjoint, we employ patchwise reconstruction, the details are documented in previous work\cite{Balan2015,Dolejs2014}.
\begin{figure}[H]
\begin{center}
\includegraphics[scale=0.3]{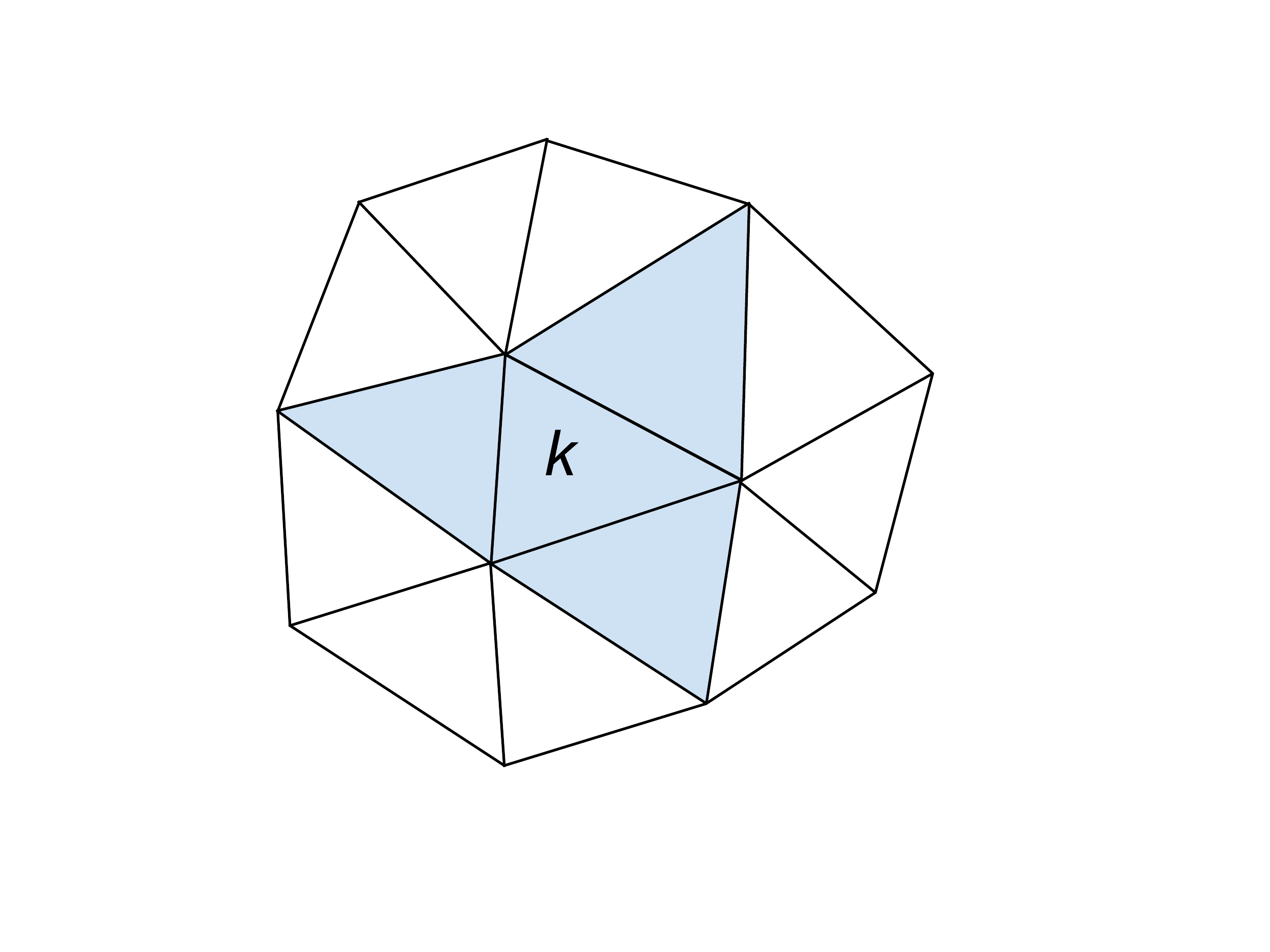}
\caption{Stencil representing the first level neighbours for patchwise reconstruction.}
\end{center}
\end{figure}
\subsection{Regularization} \label{regularization}
In the equation for optimal density ~\cref{abar}, $\overline{A}$ can drastically reduce for an element if a posteriori error estimate or indicator has very low magnitude for the particular element.  From ~\cref{optdensity_a}, it can be easily inferred that this can result in a excessively small value of density. Even though mesh generators might still produce a valid mesh but this could lead to excessive coarsening in certain parts of the domain. This might lead to a large value of a posteriori estimate in the next adaptation cycle which in turn can lead to refinement in the same region of the domain. Thus, it might lead to a cycle of alternative refinement and coarsening. One can overcome this by a simple regularization of parameter $\overline{A}(\mathbf{x}_k) $. The fundamental idea behind regularization is to determine a new local mesh size in such a way that the local error contribution is approximately same as that of the optimized mesh described this far. We attain this by employing the local optimal error$(\sqrt{e(\mathbf{x}_k,d^{\star}){\vert K^{\star} \vert}})$ mentioned in ~\cref{equidist} rather than the local error estimate on the current triangulation. An in-depth exposition to the idea of regularization can be found in \cite{RANGARAJAN2020109321}.

%% file: Results.tex
\section{Results} \label{results}
\subsection{Boundary Layer}
\begin{figure}[H]
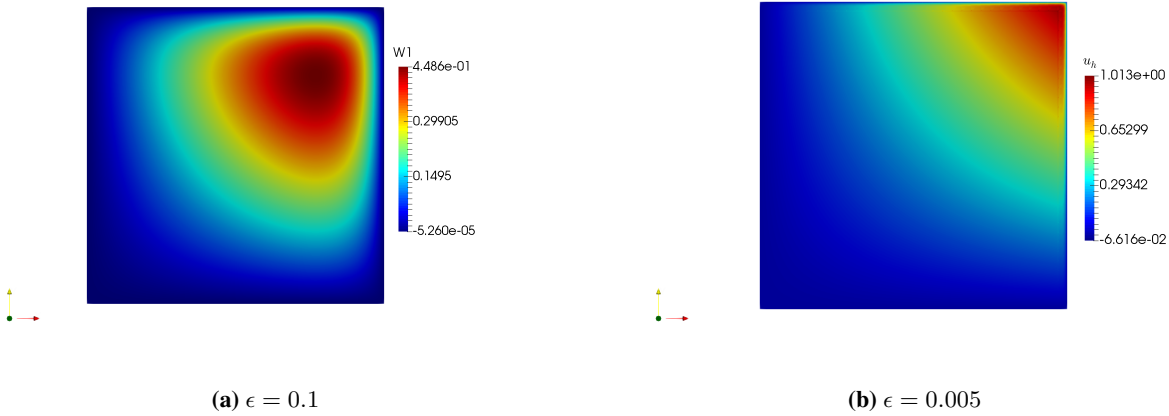

\hspace{-1cm}
\begin{subfigure}[]{0.5\textwidth}
\includegraphics[scale=0.2]{Data/Boundary_layer⁩/MNS_0p5/uniform_sol_0p1_2048-eps-converted-to.pdf}
\caption{$\epsilon = 0.1$}
\end{subfigure}
\hspace{0.2cm}
\begin{subfigure}[$\epsilon = 0.005$]{0.5\textwidth}
\includegraphics[scale=0.2]{Data/Boundary_layer⁩/MNS_0p5/uniform_sol_0p005_2048-eps-converted-to.pdf}
\caption{$\epsilon = 0.005$}
\end{subfigure}
\caption{\label{fig:bdyContour}Boundary Layer: Solution contour for two values of $\epsilon$ }
\end{figure}

Sharp boundary layers are one of the most encountered features in flow fields. This test problem is selected in order to validate the fidelity of the proposed algorithm in presence of a boundary layer. In particular, we solve,
\begin{equation}
\begin{aligned}
\beta \cdot {\nabla}u-\epsilon{\nabla}^2u &= s(\mathbf{x}) \qquad&& \mathbf{x} \in \Omega = {(0,1)}^2 \\
u &= 0 && \mathbf{x} \in \partial \Omega 
\end{aligned}
\end{equation}
where $\beta = {[1,1]}^T$. The source term $s(\mathbf{x})$ is selected in such a way that the exact solution is given by
\begin{equation}
u(\mathbf{x}) = \left( x + \frac{e^{\frac{x}{\epsilon}}-1}{1-e^{\frac{1}{\epsilon}}} \right) \left( y + \frac{e^{\frac{y}{\epsilon}}-1}{1-e^{\frac{1}{\epsilon}}} \right). 
\end{equation}
The solution exhibits a sharp boundary layer near $x \approx 1$,$y \approx 1$  and its sharpness is determined by $\epsilon$. The smaller the value of $\epsilon$, the sharper will be the boundary layer (see \cref{fig:bdyContour}).

\subsubsection{h-adaptations}
\subsubsection*{Solution Based Refinement}
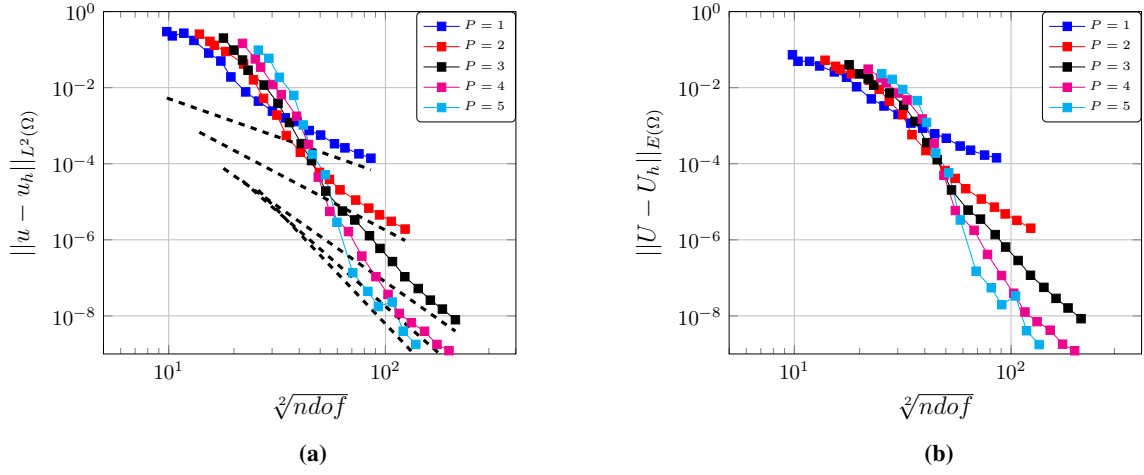
\begin{figure}[H]
\begin{subfigure}[b]{0.5\textwidth}
\begin{tikzpicture}[scale=0.8]
		\begin{loglogaxis}[xmin=5,xmax=400, ymin=1e-9,ymax=1,xlabel=\large{$\sqrt[2]{ndof}$},ylabel=\large{$||u-u_h||_{L^{2}(\Omega)}$},grid=major,legend style={at={(1,1)},anchor=north east,font=\tiny,rounded corners=2pt}]
		\addplot[color = blue,mark=square*] table[x= ndof, y=err_l2, col sep = comma] {Data/Boundary_layer⁩/Vnorm/L2_error_DPG_BL_op005_VN_p1.txt};
		\addplot [color = red,mark=square*] table[x= ndof, y=err_l2, col sep = comma] {Data/Boundary_layer⁩/Vnorm/L2_error_DPG_BL_op005_VN_p2.txt};
		\addplot [color = black,mark=square*] table[x= ndof, y=err_l2, col sep = comma] {Data/Boundary_layer⁩/Vnorm/L2_error_DPG_BL_op005_VN_p3.txt};
		\addplot [color = magenta,mark=square*] table[x= ndof, y=err_l2, col sep = comma] {Data/Boundary_layer⁩/Vnorm/L2_error_DPG_BL_op005_VN_p4.txt};
			\addplot [color = cyan,mark=square*] table[x= ndof, y=err_l2, col sep = comma] {Data/Boundary_layer⁩/Vnorm/L2_error_DPG_BL_op005_VN_p5.txt};					
	    \addplot  [dashed,line width=1.5pt,mark=none, black,forget plot] table[x= ndof, y=exslp, col sep = comma] {Data/Boundary_layer⁩/Vnorm/L2_error_DPG_BL_op005_VN_p1.txt};
	    	    \addplot  [dashed,line width=1.5pt,mark=none, black,forget plot] table[x= ndof, y=exslp, col sep = comma] {Data/Boundary_layer⁩/Vnorm/L2_error_DPG_BL_op005_VN_p2.txt};
	    	    	    \addplot  [dashed,line width=1.5pt,mark=none, black,forget plot] table[x= ndof, y=exslp, col sep = comma] {Data/Boundary_layer⁩/Vnorm/L2_error_DPG_BL_op005_VN_p3.txt};
	    	    	    	    \addplot  [dashed,line width=1.5pt,mark=none, black,forget plot] table[x= ndof, y=exslp, col sep = comma] {Data/Boundary_layer⁩/Vnorm/L2_error_DPG_BL_op005_VN_p4.txt};
	    	    	    \addplot  [dashed,line width=1.5pt,mark=none, black,forget plot] table[x= ndof, y=exslp, col sep = comma] {Data/Boundary_layer⁩/Vnorm/L2_error_DPG_BL_op005_VN_p5.txt};
		\legend{$P =1$,$P =2$,$P =3$,$P =4$,$P =5$}
		\end{loglogaxis}
	\end{tikzpicture}
	\caption{}
\end{subfigure}
\begin{subfigure}[b]{0.5\textwidth}
\begin{tikzpicture}[scale=0.8]
		\begin{loglogaxis}[xmin=5,xmax=400, ymin=1e-9,ymax=1,xlabel=\large{$\sqrt[2]{ndof}$},ylabel=\large{$||U-U_h||_{E(\Omega)}$},grid=major,legend style={at={(1,1)},anchor=north east,font=\tiny,rounded corners=2pt}]
		\addplot[color = blue,mark=square*]  table[x= ndof, y=EE, col sep = comma] {Data/Boundary_layer⁩/Vnorm/EE_error_DPG_BL_op005_VN_p1.txt};
		\addplot [color = red,mark=square*] table[x= ndof, y=EE, col sep = comma] {Data/Boundary_layer⁩/Vnorm/EE_error_DPG_BL_op005_VN_p2.txt};
		\addplot [color = black,mark=square*] table[x= ndof, y=EE, col sep = comma] {Data/Boundary_layer⁩/Vnorm/EE_error_DPG_BL_op005_VN_p3.txt};
	    \addplot [color = magenta,mark=square*] table[x= ndof, y=EE, col sep = comma] {Data/Boundary_layer⁩/Vnorm/EE_error_DPG_BL_op005_VN_p4.txt};
			\addplot [color = cyan,mark=square*] table[x= ndof, y=EE, col sep = comma] {Data/Boundary_layer⁩/Vnorm/EE_error_DPG_BL_op005_VN_p5.txt};
		\legend{$P =1$,$P =2$,$P =3$,$P =4$,$P =5$}
		\end{loglogaxis}
\end{tikzpicture}
\caption{}
\end{subfigure}	
\caption{Convergence plots of (a) $L^2$ error in $u_h$ and (b) Energy norm using standard V norm.} \label{convergence_stdVnorm}
\end{figure}

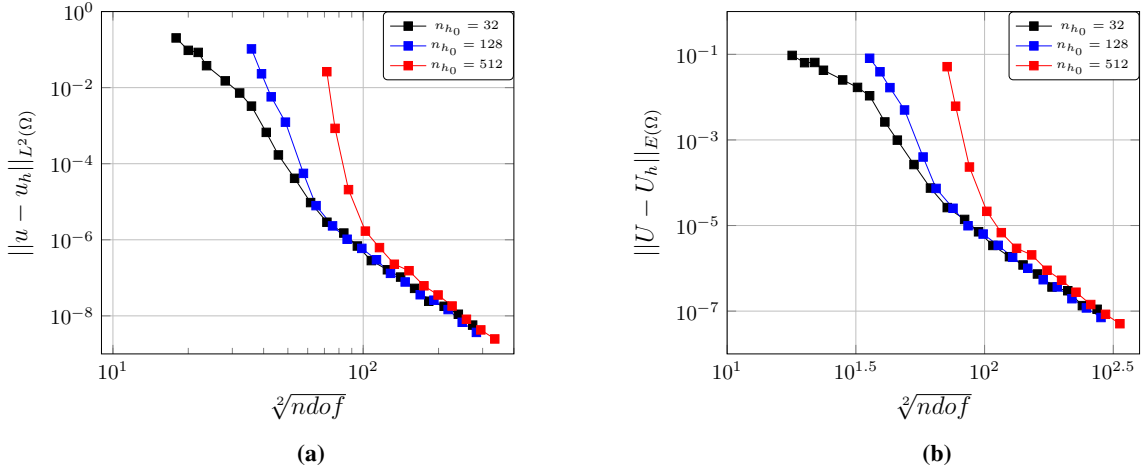
\begin{figure}[H]
\begin{subfigure}[b]{0.5\textwidth}
\begin{tikzpicture}[scale=0.8]
		\begin{loglogaxis}[xmin=9,xmax=400, ymin=1e-9,ymax=1,xlabel=\large{$\sqrt[2]{ndof}$},ylabel=\large{$||u-u_h||_{L^{2}(\Omega)}$},grid=major,legend style={at={(1,1)},anchor=north east,font=\tiny,rounded corners=2pt}]
		\addplot [color = black,mark=square*] table[x= ndof, y=err_l2, col sep = comma] {Data/Boundary_layer⁩/MNS_0p5/L2_error_DPG_BL_0p005_p3_MN_0p5.txt};	
		\addplot[color = blue,mark=square*] table[x= ndof, y=err_l2, col sep = comma] {Data/Boundary_layer⁩/MNS_0p5/L2_error_DPG_BL_p3_start_mesh_128_MN_0p5.txt};
		\addplot [color = red,mark=square*] table[x= ndof, y=err_l2, col sep = comma] {Data/Boundary_layer⁩/MNS_0p5/L2_error_DPG_BL_p3_start_mesh_512_MN_0p5.txt};

		\legend{$n_{h_0} = 32$,$n_{h_0} = 128$,$n_{h_0} = 512$}
		\end{loglogaxis}
	\end{tikzpicture}
	\caption{}
\end{subfigure}
\begin{subfigure}[b]{0.5\textwidth}
\begin{tikzpicture}[scale=0.8]
		\begin{loglogaxis}[xmin=10,xmax=400, ymin=1e-8,ymax=1,xlabel=\large{$\sqrt[2]{ndof}$},ylabel=\large{$||U-U_h||_{E(\Omega)}$},grid=major,legend style={at={(1,1)},anchor=north east,font=\tiny,rounded corners=2pt}]
			\addplot [color = black,mark=square*] table[x= ndof, y=EE, col sep = comma] {Data/Boundary_layer⁩/MNS_0p5/EE_error_DPG_BL_0p005_p3_MN_0p5.txt};		
		\addplot[color = blue,mark=square*]  table[x= ndof, y=EE, col sep = comma] {Data/Boundary_layer⁩/MNS_0p5/EE_DPG_BL_p3_0p005_startmesh_128.txt};
		\addplot [color = red,mark=square*] table[x= ndof, y=EE, col sep = comma] {Data/Boundary_layer⁩/MNS_0p5/EE_DPG_BL_p3_0p005_startmesh_512.txt};
		\legend{$n_{h_0} = 32$,$n_{h_0} = 128$,$n_{h_0} = 512$}
		\end{loglogaxis}
\end{tikzpicture}
\caption{}
\end{subfigure}	
\caption{Convergence plots of (a) $L^2$ error in $u_h$ and (b) Energy norm with different initial meshes using scaled V norm.($\epsilon = 0.005, P = 3$).} \label{convergence_BL_diff_ini_mesh}
\end{figure}

In ~\cref{convergence_stdVnorm}  and ~\cref{convergence_BL_scaled_math_norm}, we have presented the convergence results for the proposed $h$ adaptation algorithm for the standard V norm and scaled V norm associated with the test space. In $h$ adaptations, we start with an initial mesh of 32 elements. Between each adaptation cycle, $N$ is being increased by $30 \%$. The growth in $N$ is arbitrarily chosen, a different choice in growth may result in different pre-asymptotic behaviour but should produce similar asymptotic result. In $h$ adaptations, it can be observed that there is an initial drop in the error at a rate more than the asymptotic rate. This pre-asymptotic drop is due to the fact that redistribution of dofs contributes significantly to the error reduction. The algorithm identifies the anisotropic features present in the solution field during these initial adaptations. Once, this identification is done in the first few adaptations, we reach the asymptotic region in convergence plot. The extent of this pre-asymptotic region also depends upon the resolution of the initial mesh. From~\cref{convergence_BL_diff_ini_mesh}, it can be easily inferred that an increase in the resolution of the initial mesh results in fewer adaptations required to reach the asymptotic convergence ($n_{h_0}$ denotes the number of elements in the initial mesh). The rationale behind this is the fact that information for local anisotropy minimization and analytic optimization is sampled at higher number of points which in turn enhances the performance of the algorithm. Though in pre-asymptotic region the convergence may depend on the initial mesh, asymptotically each of convergence plots in ~\cref{convergence_BL_diff_ini_mesh} are very similar. 

\begin{figure}[H]
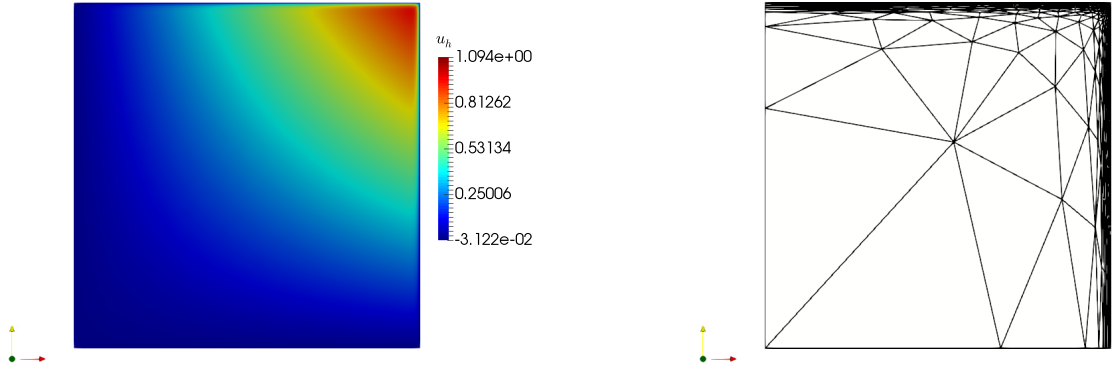

\begin{subfigure}[b]{0.5\textwidth}
\includegraphics[scale=0.25]{Data/Boundary_layer⁩/MNS_0p5/Solution_BL_484_10adap_0p005_p3_MNS_0p5-eps-converted-to.pdf}
\end{subfigure}
\hspace{0.8cm}
\begin{subfigure}[b]{0.5\textwidth}
\includegraphics[scale=0.25]{Data/Boundary_layer⁩/MNS_0p5/Mesh_BL_484_10adap_0p005_p3_MNS_0p5-eps-converted-to.pdf}
\end{subfigure}
\caption{Solution and Mesh after 10 adaptations $(\epsilon = 0.005,Ne = 484, P = 3)$ using scaled V norm} \label{484_adap_mesh_result}
\end{figure}

Also, one major advantage of the proposed $h$ adaptation algorithm is the redistribution of the degrees of freedom when done at a fixed cost. In ~\cref{fixed_cost_scaled_math_norm}, we have provided the results showing the drop in error in both $L^2$ and energy norm. One can think of this drop in error as an effect of reallocating the degrees of freedom from regions with smoother solution to regions with anisotropic features in the solution. It only takes very few adaptations to reach the saturation in terms of drop in error. The saturation is seen due to the fact that we have kept the DOFs fixed during this numerical experiment. After a few adaptations  the algorithm identifies the regions where DOFs need to be reallocated along with local anisotropic modifications to the metric tensor. Once this optimal distribution is achieved, subsequent adaptations will not result in any further decrease in error.
In ~\cref{484_adap_mesh_result}, it can be seen that the algorithm prescribes larger elements away form the boundary layer with most the degrees of freedom being allocated to the boundary layer.
\begin{figure}[H]
\begin{subfigure}[b]{0.5\textwidth}
\begin{tikzpicture}[scale=0.8]
		\begin{loglogaxis}[xmin=5,xmax=400, ymin=1e-11,ymax=1,xlabel=\large{$\sqrt[2]{ndof}$},ylabel=\large{$||u-u_h||_{L^{2}(\Omega)}$},grid=major,legend style={at={(1,1)},anchor=north east,font=\tiny,rounded corners=2pt}]
		\addplot[color = blue,mark=square*] table[x= ndof, y=err_l2, col sep = comma] {Data/Boundary_layer⁩/MNS_0p5/L2_error_DPG_BL_0p005_p1_MN_0p5.txt};
		\addplot [color = red,mark=square*] table[x= ndof, y=err_l2, col sep = comma] {Data/Boundary_layer⁩/MNS_0p5/L2_error_DPG_BL_0p005_p2_MN_0p5.txt};
		\addplot [color = black,mark=square*] table[x= ndof, y=err_l2, col sep = comma] {Data/Boundary_layer⁩/MNS_0p5/L2_error_DPG_BL_0p005_p3_MN_0p5.txt};
		\addplot [color = magenta,mark=square*] table[x= ndof, y=err_l2, col sep = comma] {Data/Boundary_layer⁩/MNS_0p5/L2_error_DPG_BL_0p005_p4_MN_0p5.txt};
			\addplot [color = cyan,mark=square*] table[x= ndof, y=err_l2, col sep = comma] {Data/Boundary_layer⁩/MNS_0p5/L2_error_DPG_BL_0p005_p5_MN_0p5.txt};					
	    \addplot  [dashed,line width=1.5pt,mark=none, black,forget plot] table[x= ndof, y=exslp, col sep = comma] {Data/Boundary_layer⁩/MNS_0p5/L2_error_DPG_BL_0p005_p1_MN_0p5.txt};
	    	    \addplot  [dashed,line width=1.5pt,mark=none, black,forget plot] table[x= ndof, y=exslp, col sep = comma] {Data/Boundary_layer⁩/MNS_0p5/L2_error_DPG_BL_0p005_p2_MN_0p5.txt};
	    	    	    \addplot  [dashed,line width=1.5pt,mark=none, black,forget plot] table[x= ndof, y=exslp, col sep = comma] {Data/Boundary_layer⁩/MNS_0p5/L2_error_DPG_BL_0p005_p3_MN_0p5.txt};
	    	    	    	    \addplot  [dashed,line width=1.5pt,mark=none, black,forget plot] table[x= ndof, y=exslp, col sep = comma] {Data/Boundary_layer⁩/MNS_0p5/L2_error_DPG_BL_0p005_p4_MN_0p5.txt};
	    	    	    \addplot  [dashed,line width=1.5pt,mark=none, black,forget plot] table[x= ndof, y=exslp, col sep = comma] {Data/Boundary_layer⁩/MNS_0p5/L2_error_DPG_BL_0p005_p5_MN_0p5.txt};
		\legend{$P =1$,$P =2$,$P =3$,$P =4$,$P =5$}
		\end{loglogaxis}
	\end{tikzpicture}
	\caption{}
\end{subfigure}
\begin{subfigure}[b]{0.5\textwidth}
\begin{tikzpicture}[scale=0.8]
		\begin{loglogaxis}[xmin=5,xmax=400, ymin=1e-11,ymax=1,xlabel=\large{$\sqrt[2]{ndof}$},ylabel=\large{$||U-U_h||_{E(\Omega)}$},grid=major,legend style={at={(1,1)},anchor=north east,font=\tiny,rounded corners=2pt}]
		\addplot[color = blue,mark=square*]  table[x= ndof, y=EE, col sep = comma] {Data/Boundary_layer⁩/MNS_0p5/EE_error_DPG_BL_0p005_p1_MN_0p5.txt};
		\addplot [color = red,mark=square*] table[x= ndof, y=EE, col sep = comma] {Data/Boundary_layer⁩/MNS_0p5/EE_error_DPG_BL_0p005_p2_MN_0p5.txt};
			\addplot [color = black,mark=square*] table[x= ndof, y=EE, col sep = comma] {Data/Boundary_layer⁩/MNS_0p5/EE_error_DPG_BL_0p005_p3_MN_0p5.txt};
					\addplot [color = magenta,mark=square*] table[x= ndof, y=EE, col sep = comma] {Data/Boundary_layer⁩/MNS_0p5/EE_error_DPG_BL_0p005_p4_MN_0p5.txt};
			\addplot [color = cyan,mark=square*] table[x= ndof, y=EE, col sep = comma] {Data/Boundary_layer⁩/MNS_0p5/EE_error_DPG_BL_0p005_p5_MN_0p5.txt};
		\legend{$P =1$,$P =2$,$P =3$,$P =4$,$P =5$}
		\end{loglogaxis}
\end{tikzpicture}
\caption{}
\end{subfigure}	
\caption{Convergence plots of (a) $L^2$ error in $u_h$ and (b) Energy norm using scaled V norm.} \label{convergence_BL_scaled_math_norm}
\end{figure}

\begin{table}[H]
\begin{center}
\begin{tabular}{|c|c|c|c|}
\hline
Adaptation & Ne  & ${\Vert u - u_h \Vert}_{2,\Omega}$ & ${\Vert U - U_h \Vert}_{E,\Omega}$ \\ \hline
 0          & 512 & 0.0262192                          & 0.0514999                          \\ \hline
2          & 437 & 0.000289966                        & 0.00218748                         \\ \hline
4          & 463 & 6.05824e-06                        & 4.77721e-05                        \\ \hline
6          & 484 & 3.65609e-06                        & 3.09974e-05                        \\ \hline
8          & 489 & 2.02129e-06                        & 1.86422e-05                        \\ \hline
\end{tabular}
\end{center}
\vspace{0.2cm}
\caption{Adaptation Vs. Error for constant complexity using scaled V norm. ($P =3, N = \int_{\Omega} d(\mathbf{x}) \, d\mathbf{x} = 512.0$)} \label{fixed_cost_scaled_math_norm}
\end{table}

\begin{figure}[H]
\begin{subfigure}[b]{0.5\textwidth}
\begin{tikzpicture}[scale=0.7]
		\begin{loglogaxis}[xmin=9,xmax=200, ymin=1e-4,ymax=1,xlabel=\large{$\sqrt[2]{ndof}$},grid=major,legend style={at={(1,1)},anchor=north east,font=\tiny,rounded corners=2pt}]
		\addplot[color = black,mark=square*]  table[x= ndof, y=EE, col sep = comma] {Data/Boundary_layer⁩/MNS_0p5/EE_error_DPG_BL_0p005_p1_MN_0p5.txt};
		\addplot [color = red,mark=square*] table[x= ndof, y=pred, col sep = comma] {Data/Boundary_layer⁩/MNS_0p5/EE_errorpred_DPG_BL_0p005_p1_MN_0p5.txt};
		\legend{$||U-U_h||_{E(\Omega)}$,$E^{\star}$}
		\end{loglogaxis}
	\end{tikzpicture}
	\caption{}
\end{subfigure}
\begin{subfigure}[b]{0.5\textwidth}
\begin{tikzpicture}[scale=0.7]
		\begin{loglogaxis}[xmin=10,xmax=300, ymin=1e-6,ymax=1,xlabel=\large{$\sqrt[2]{ndof}$},grid=major,legend style={at={(1,1)},anchor=north east,font=\tiny,rounded corners=2pt}]
		\addplot[color = black,mark=square*]  table[x= ndof, y=EE, col sep = comma] {Data/Boundary_layer⁩/MNS_0p5/EE_error_DPG_BL_0p005_p2_MN_0p5.txt};
		\addplot [color = red,mark=square*] table[x= ndof, y=pred, col sep = comma] {Data/Boundary_layer⁩/MNS_0p5/EE_errorpred_DPG_BL_0p005_p2_MN_0p5.txt};
		\legend{$||U-U_h||_{E(\Omega)}$,$E^{\star}$}
		\end{loglogaxis}
\end{tikzpicture}
\caption{}
\end{subfigure}	

\begin{center}
\begin{subfigure}[b]{0.5\textwidth}
\begin{tikzpicture}[scale=0.7]
		\begin{loglogaxis}[xmin=10,xmax=350, ymin=1e-8,ymax=1,xlabel=\large{$\sqrt[2]{ndof}$},grid=major,legend style={at={(1,1)},anchor=north east,font=\tiny,rounded corners=2pt}]
		\addplot[color = black,mark=square*]  table[x= ndof, y=EE, col sep = comma] {Data/Boundary_layer⁩/MNS_0p5/EE_error_DPG_BL_0p005_p3_MN_0p5.txt};
		\addplot [color = red,mark=square*] table[x= ndof, y=pred, col sep = comma] {Data/Boundary_layer⁩/MNS_0p5/EE_errorpred_DPG_BL_0p005_p3_MN_0p5.txt};
		\legend{$||U-U_h||_{E(\Omega)}$,$E^{\star}$}
		\end{loglogaxis}
\end{tikzpicture}
\caption{}
\end{subfigure}	
\end{center}

\caption{Comparison of $||U-U_h||_{E(\Omega)}$ and $E^{\star}$ for (a) $P =1$, (b) $P =2$ and (c) $P =3$ at $\epsilon = 0.005$ using scaled V norm.} \label{EE_pred_BL}
\end{figure}
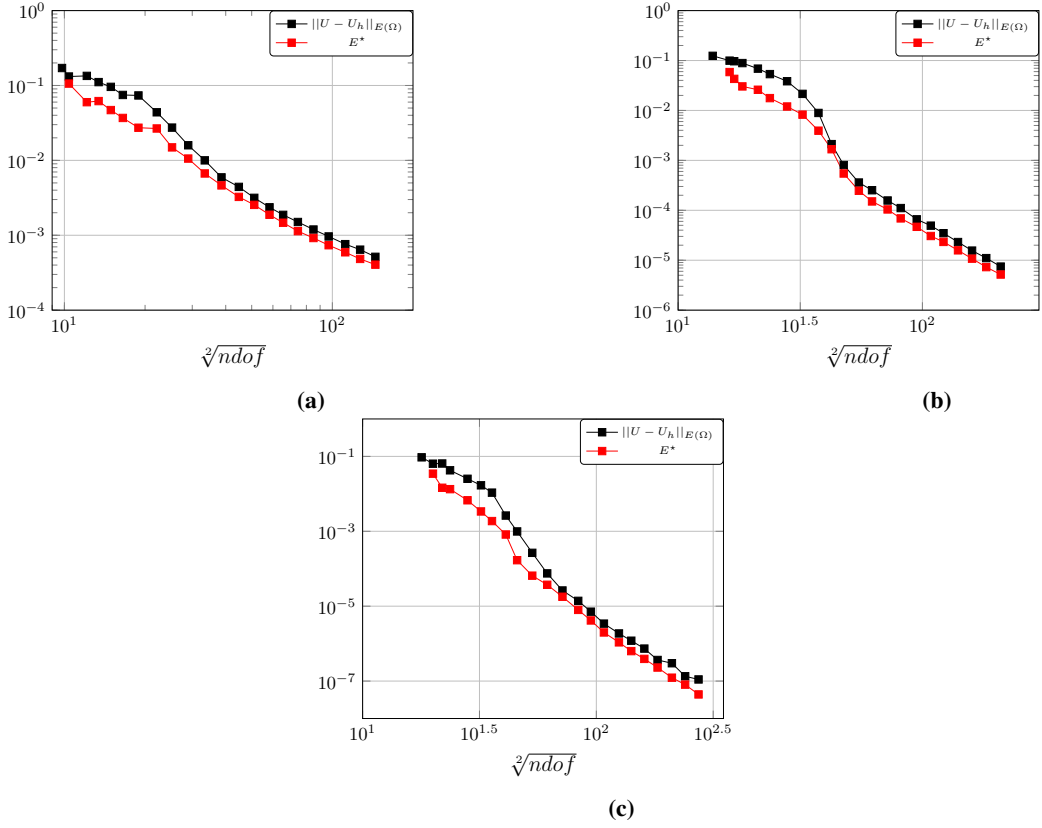

Finally, in ~\cref{EE_pred_BL} we present the comparison of the continuous error estimate in ~\cref{cont_error_hadap} with the in-built error estimate of the DPG  scheme for various polynomial order for $h$ adaptations. The purpose of this comparison is to show that the continuous estimate which drives the adaptation follows the inbuilt error estimate which is the very purpose of the presented  adaptation algorithm. It can be observed that the convergence curve of the continuous error estimate follows the same trend as of the inbuilt error estimate for different polynomial orders.

\subsubsection{Goal Oriented Adaptation - Opposite Boundary Layer}
Similar to the primal problem, the dual problem is formulated by suitably choosing the target $J_1$ such that the exact solution $\eta_1$ to the dual problem is known. The target functional chosen here is a weighted volume integral of $u(\mathbf{x})$ over the domain.

\begin{equation}
J_1(u) = \int_\Omega j_{\Omega}(\mathbf{x}) u(\mathbf{x}) \, d\mathbf{x} \label{volumetarget}
\end{equation}

\begin{figure}[H]
\begin{subfigure}[b]{0.5\textwidth}
\includegraphics[scale=0.25]{Data/Boundary_layer⁩/MNS_0p5/Solution_BL_484_10adap_0p005_p3_MNS_0p5-eps-converted-to.pdf}
\caption{Primal Solution ($u$)}
\end{subfigure}
\hspace{0.3cm}
\begin{subfigure}[b]{0.5\textwidth}
\includegraphics[scale=0.225]{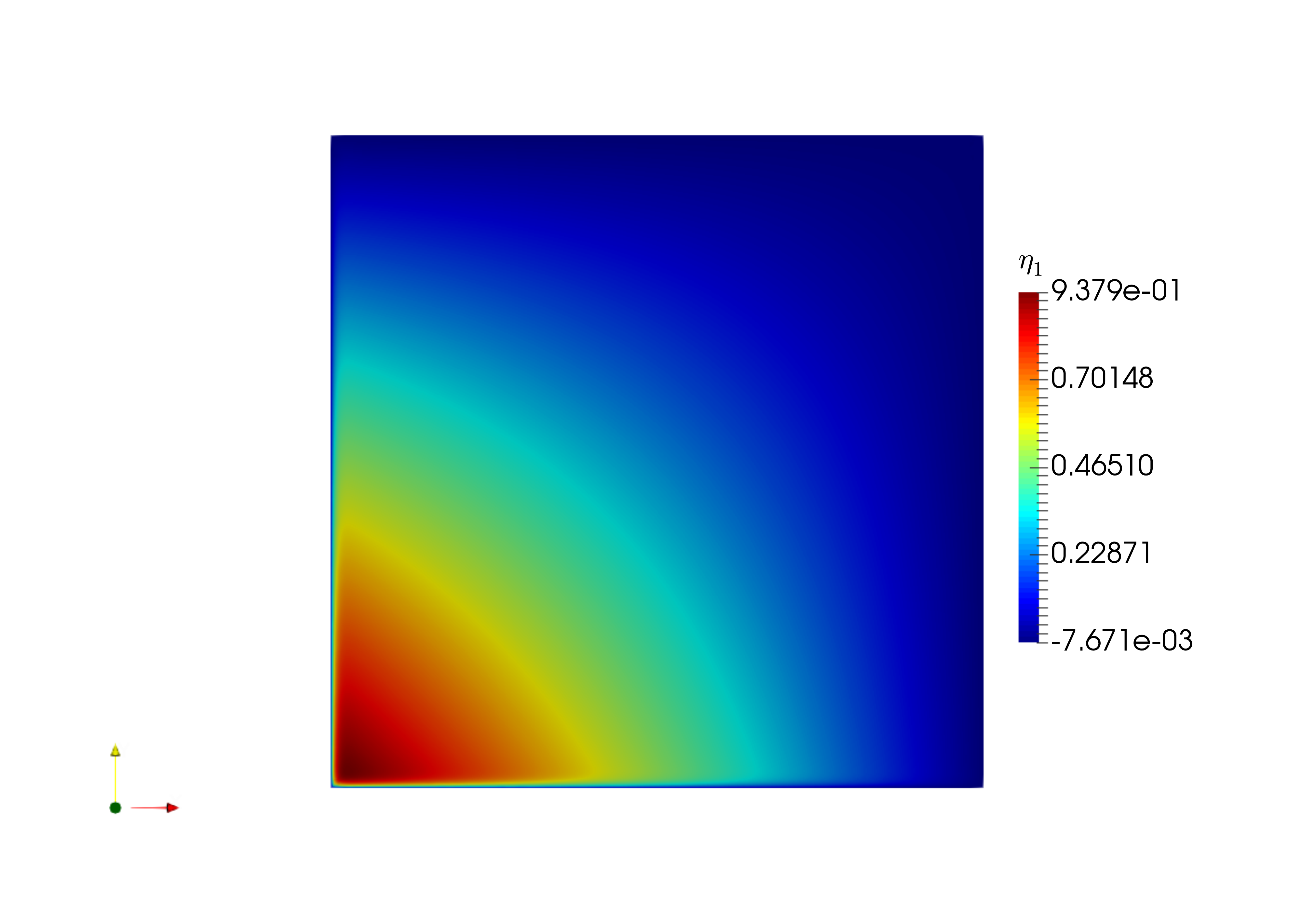}
\caption{Dual Solution ($\eta_1$)}
\end{subfigure}
\caption{\label{fig:bdyControurDual}Boundary Layer test case: Contour plot for the primal $u$ and adjoint $\eta_1$}
\end{figure}

In the above target, we can change the functional by prescribing $j_{\Omega}$. For the current linear set-up, we can write the corresponding dual problem as

\begin{align}
-\beta \cdot {\nabla} \eta_1-\epsilon{\nabla}^2 \eta_1 &= j_{\Omega}(\mathbf{x}) \qquad && \mathbf{x} \in \Omega = {(0,1)}^2  \label{adjstrngeqa}\\
\eta_1 &= 0 && \mathbf{x} \in \partial \Omega  \label{adjbnda}
\end{align}

where $\beta = {[1, \,1]}^T$. With appropriately chosen $j_{\Omega}$, we can have
\begin{equation}
\eta_1 (\mathbf{x}) = \left( -x + 1+\frac{e^{\frac{1-x}{\epsilon}}-1}{1-e^{\frac{1}{\epsilon}}} \right) \left( -y + 1+\frac{e^{\frac{1-y}{\epsilon}}-1}{1-e^{\frac{1}{\epsilon}}} \right)
\end{equation}

which will solve~\cref{adjstrngeqa} and~\cref{adjbnda}. A contour plot of  both the primal and the dual solution is shown in \cref{fig:bdyControurDual}.

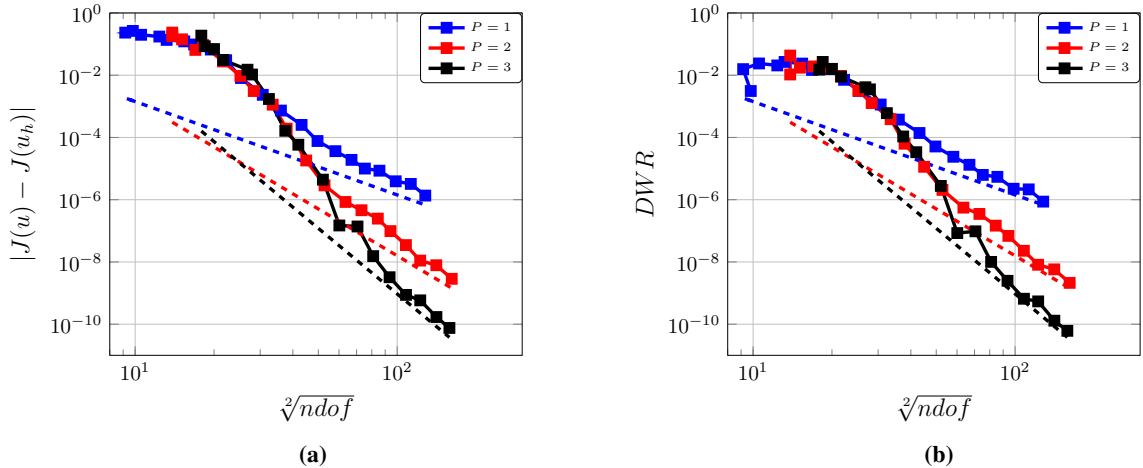
\begin{figure}[H]
\begin{subfigure}[b]{0.5\textwidth}
	\begin{tikzpicture}[scale = 0.8]
		\begin{loglogaxis}[xmin=8,xmax=300, ymin=1e-11,ymax=1,xlabel=\large{$\sqrt[2]{ndof}$},ylabel=\large{$\vert J(u) - J(u_h) \vert$},grid=major,legend style={at={(1,1)},anchor=north east,font=\tiny,rounded corners=2pt}]
       \addplot [color = blue,mark=square*,ultra thick] table[x=ndof, y=target_err,col sep = comma]{⁨Data⁩/TargetBasedAdaptation⁩/targetbasedadaptation_s_p_plus_1_reverseboundarylayer_BL_p1.txt⁩};
       \addplot [color = red,mark=square*,ultra thick] table[x=ndof, y=target_err,col sep = comma]{⁨Data⁩/TargetBasedAdaptation⁩/targetbasedadaptation_s_p_plus_1_reverseboundarylayer_BL_p2⁩.txt};
       \addplot [color = black,mark=square*,ultra thick] table[x=ndof, y=target_err,col sep = comma]{⁨Data⁩/TargetBasedAdaptation⁩/targetbasedadaptation_s_p_plus_1_reverseboundarylayer_BL_p3⁩.txt};

              \addplot [dashed,color = blue,mark=none,ultra thick] table[x=ndof, y=extslp,col sep = comma]{⁨Data⁩/TargetBasedAdaptation⁩/targetbasedadaptation_s_p_plus_1_reverseboundarylayer_BL_p1.txt⁩};
       \addplot [dashed,color = red,mark=none,ultra thick] table[x=ndof, y=extslp,col sep = comma]{⁨⁨Data⁩/TargetBasedAdaptation⁩/targetbasedadaptation_s_p_plus_1_reverseboundarylayer_BL_p2.txt⁩};
       \addplot [dashed,color = black,mark=none,ultra thick] table[x=ndof, y=extslp,col sep = comma]{⁨Data⁩/TargetBasedAdaptation⁩/targetbasedadaptation_s_p_plus_1_reverseboundarylayer_BL_p3.txt};
		\legend{$P =1$,$P =2$,$P =3$}
		\end{loglogaxis}
	\end{tikzpicture}
	\caption{}\label{gmr_conv_reverse_bnd_layer_error}
\end{subfigure} 
\begin{subfigure}[b]{0.5\textwidth}
	\begin{tikzpicture}[scale = 0.8]
		\begin{loglogaxis}[xmin=8,xmax=300, ymin=1e-11,ymax=1,xlabel=\large{$\sqrt[2]{ndof}$},ylabel=\large{$DWR$},grid=major,legend style={at={(1,1)},anchor=north east,font=\tiny,rounded corners=2pt}]       
              \addplot [color = blue,mark=square*,ultra thick] table[x=ndof, y=DWR_est,col sep = comma]{⁨Data⁩/TargetBasedAdaptation⁩/targetbasedadaptation_s_p_plus_1_reverseboundarylayer_BL_p1.txt⁩};
       \addplot [color = red,mark=square*,ultra thick] table[x=ndof, y=DWR_est,col sep = comma]{⁨Data⁩/TargetBasedAdaptation⁩/targetbasedadaptation_s_p_plus_1_reverseboundarylayer_BL_p2.txt⁩};
       \addplot [color = black,mark=square*,ultra thick] table[x=ndof, y=DWR_est,col sep = comma]{⁨Data⁩/TargetBasedAdaptation⁩/targetbasedadaptation_s_p_plus_1_reverseboundarylayer_BL_p3.txt};

              \addplot [dashed,color = blue,mark=none,ultra thick] table[x=ndof, y=extslp,col sep = comma]{⁨Data⁩/TargetBasedAdaptation⁩/targetbasedadaptation_s_p_plus_1_reverseboundarylayer_BL_p1.txt⁩};
       \addplot [dashed,color = red,mark=none,ultra thick] table[x=ndof, y=extslp,col sep = comma]{⁨⁨Data⁩/TargetBasedAdaptation⁩/targetbasedadaptation_s_p_plus_1_reverseboundarylayer_BL_p2.txt⁩};
       \addplot [dashed,color = black,mark=none,ultra thick] table[x=ndof, y=extslp,col sep = comma]{⁨Data⁩/TargetBasedAdaptation⁩/targetbasedadaptation_s_p_plus_1_reverseboundarylayer_BL_p3.txt};
		\legend{$P =1$,$P =2$,$P =3$}
		\end{loglogaxis}
	\end{tikzpicture}
\caption{}\label{gmr_conv_reverse_bnd_layer_dwr}
\end{subfigure}
\caption{Convergence plots for (a) error in target functional  and (b) Dual Weighted Residual using scaled V norm.} 
\end{figure}

\begin{figure}[H]
\hspace{-2cm}
\begin{subfigure}[b]{0.5\textwidth}
\includegraphics[scale=0.25]{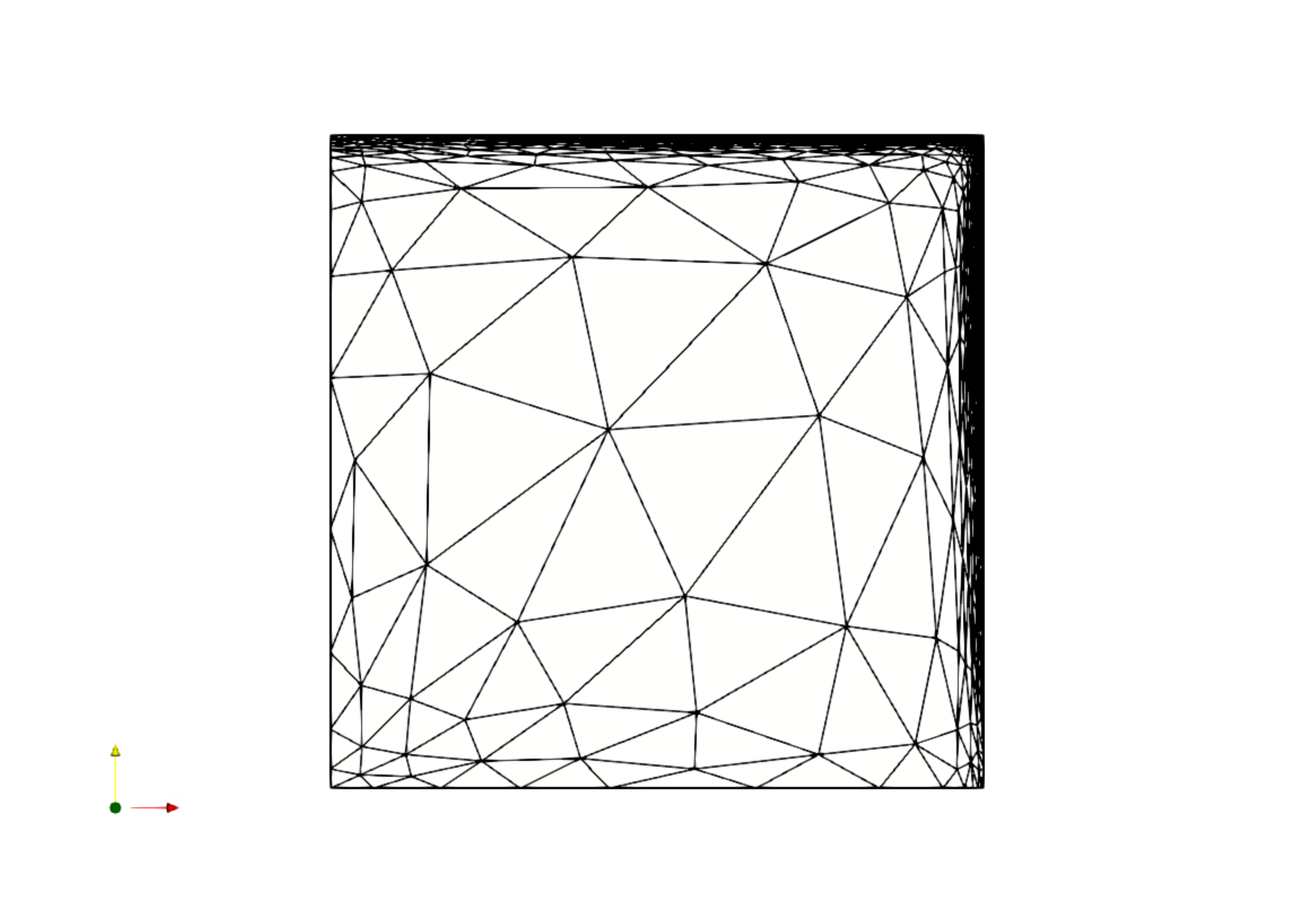}
\caption{$Ne  = 1133$}
\end{subfigure}
\hspace{0.2cm}
\begin{subfigure}[b]{0.5\textwidth}
\includegraphics[scale=0.25]{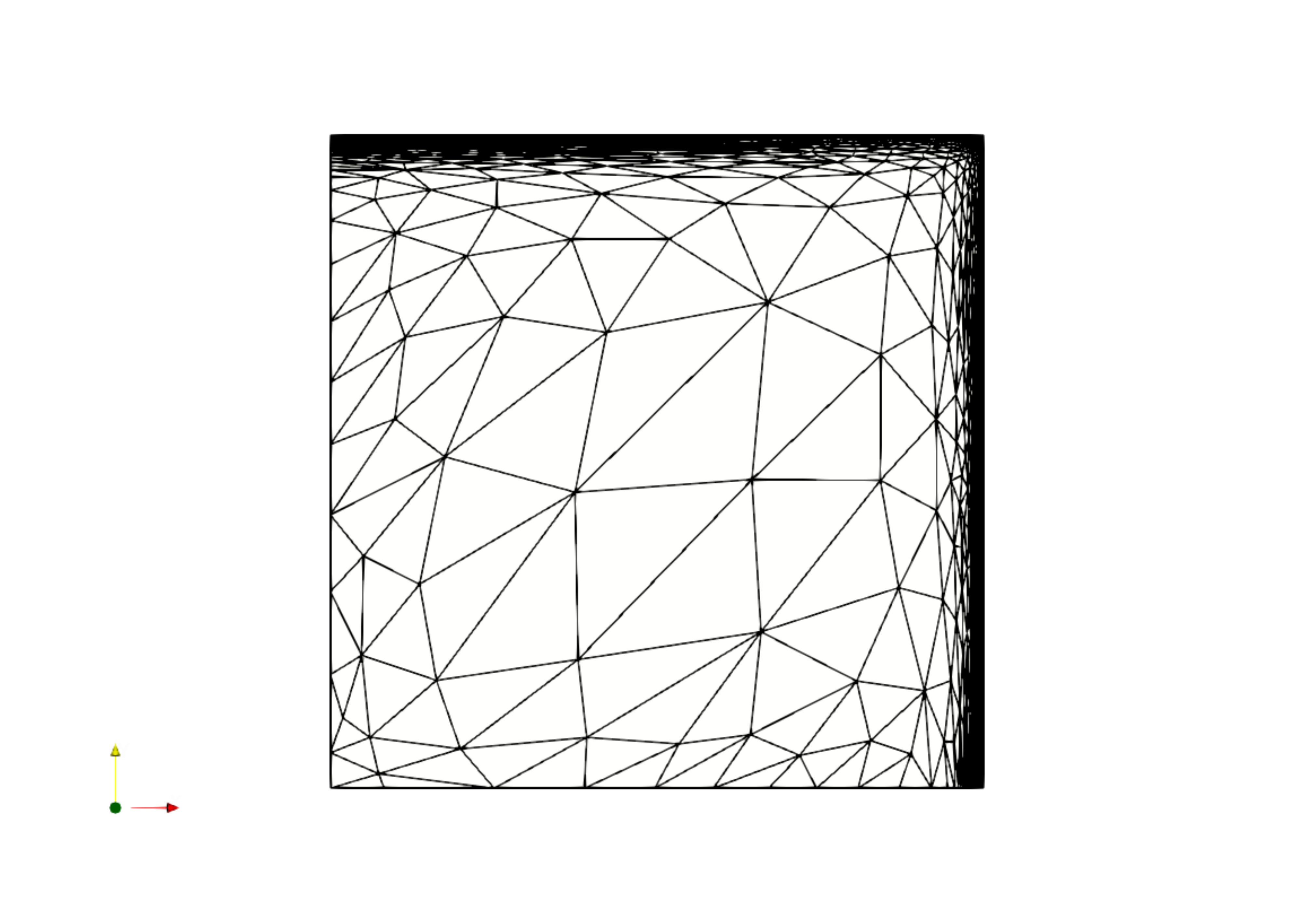}
\caption{$Ne = 1177$}
\end{subfigure}
\caption{Adapated mesh for (a)$P =2$ and (b) $P =3$ after $15^{th}$ adaptation using scaled V norm.} \label{adapted_mesh_reversebnd_layer}
\end{figure}

The anisotropy and the global size distribution are computed as mentioned in~\cref{anisotropycomp} and~\cref{cont_model_h_adap}. In each adaptation cycle, we prescribe an increase of $30\%$ in the mesh complexity ($N$). In~\cref{gmr_conv_reverse_bnd_layer_error}, we show the convergence of the target functional by computing the actual error as we have both exact adjoint solution needed for computing $j_{\Omega}$ and exact solution $u$. In ~\cref{gmr_conv_reverse_bnd_layer_dwr}, we have shown the convergence of dual weighted residual (DWR) error estimator for target functional. The rate of convergence for both error in the target function and DWR estimator is $2p + 1$ where $p$ is the polynomial order of approximation for the primal $u$. This super convergence is in accordance  with the previous results in \cite{RANGARAJAN2020109321}. In ~\cref{adapted_mesh_reversebnd_layer}, we have shown the adapted mesh for $P =2$ and $P =3$. One important point to notice  in these meshes are the presence of smaller elements near the bottom left corner. Since the dual variable has a boundary layer, the lower value of ${\Vert U - U_h \Vert}_{E,k}$ is being compensated by a larger value of $\eta^{\star}_k$ mentioned in ~\cref{explicit_DPGstar_est} for the elements present near the bottom left corner. This difference can be noticed by comparing the meshes in ~\cref{adapted_mesh_reversebnd_layer} and the solution adapted mesh shown in ~\cref{484_adap_mesh_result}.

\subsubsection{Goal Oriented Adaptation - Gaussian Peak }
Next, we consider the same primal problem and the target is given by ~\cref{volumetarget} where
\begin{equation}
j_{\Omega}(\mathbf{x}) = e^{-\alpha \left({(x - x_c)}^2 + {(y - y_c)}^2\right)}
\end{equation}

\begin{figure}[H]
\begin{subfigure}[b]{0.5\textwidth}
\includegraphics[scale=0.25]{Data/Boundary_layer⁩/MNS_0p5/Solution_BL_484_10adap_0p005_p3_MNS_0p5-eps-converted-to.pdf}
\caption{Primal Solution ($u$)} \label{primalsol_gaussianpeak}
\end{subfigure}
\hspace{0.2cm}
\begin{subfigure}[b]{0.5\textwidth}
\includegraphics[scale=0.225]{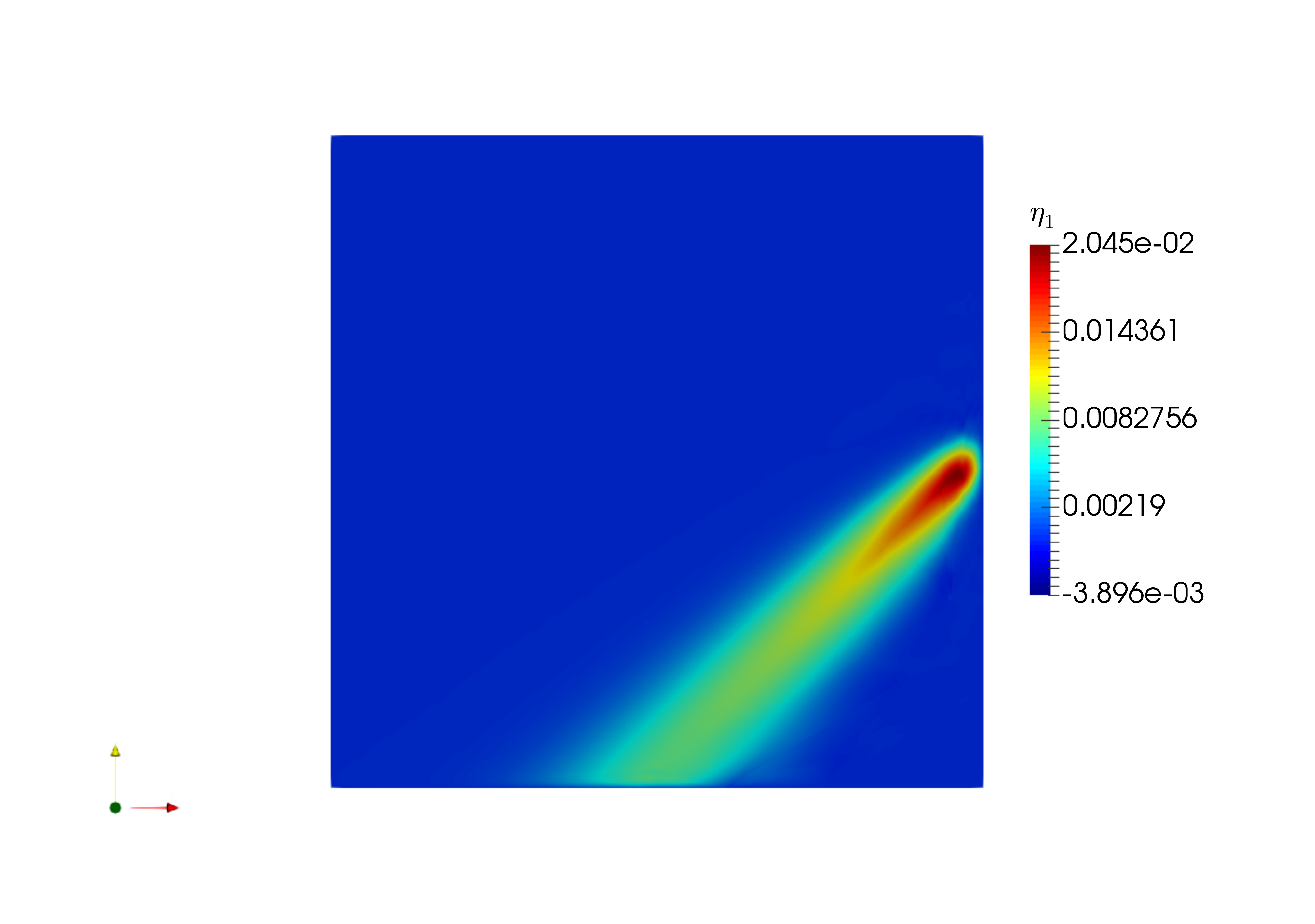}
\caption{Dual Solution ($\eta_1$)} \label{dualsol_gaussianpeak}
\end{subfigure}
\caption{Boundary Layer test case: Contour plot for the primal $u$ and adjoint $\eta_1$}
\end{figure}

\begin{figure}[h]
\begin{subfigure}[b]{0.5\textwidth}
	\begin{tikzpicture}[scale = 0.8]
		\begin{loglogaxis}[xmin=5,xmax=200, ymin=1e-14,ymax=1,xlabel=\large{$\sqrt[2]{ndof}$},ylabel=\large{$\vert J(u) - J(u_h) \vert$},grid=major,legend style={at={(1,1)},anchor=north east,font=\tiny,rounded corners=2pt}]
       \addplot [color = blue,mark=square*,ultra thick] table[x=ndof, y=target_err,col sep = comma]{⁨Data⁩/TargetBasedAdaptation⁩/targetbasedadaptation_s_p_plus_1_gaussian_peak_alpha_1000_p_BL1.txt⁩};
       \addplot [color = red,mark=square*,ultra thick] table[x=ndof, y=target_err,col sep = comma]{⁨Data⁩/TargetBasedAdaptation⁩/targetbasedadaptation_s_p_plus_1_gaussian_peak_alpha_1000_p_BL2.txt⁩};
       \addplot [color = black,mark=square*,ultra thick] table[x=ndof, y=target_err,col sep = comma]{⁨⁨Data⁩/TargetBasedAdaptation⁩/targetbasedadaptation_s_p_plus_1_gaussian_peak_alpha_1000_p_BL3.txt⁩};

              \addplot [dashed,color = blue,mark=none,ultra thick] table[x=ndof, y=extslp,col sep = comma]{⁨Data⁩/TargetBasedAdaptation⁩/targetbasedadaptation_s_p_plus_1_gaussian_peak_alpha_1000_p_BL1.txt⁩};
       \addplot [dashed,color = red,mark=none,ultra thick] table[x=ndof, y=extslp,col sep = comma]{⁨Data⁩/TargetBasedAdaptation⁩/targetbasedadaptation_s_p_plus_1_gaussian_peak_alpha_1000_p_BL2.txt⁩};
       \addplot [dashed,color = black,mark=none,ultra thick] table[x=ndof, y=extslp,col sep = comma]{⁨Data⁩/TargetBasedAdaptation⁩/targetbasedadaptation_s_p_plus_1_gaussian_peak_alpha_1000_p_BL3.txt⁩};
		\legend{$P =1$,$P =2$,$P =3$}
		\end{loglogaxis}
	\end{tikzpicture}
	\caption{}\label{convergence_gaussian_peak_error}
\end{subfigure}
\begin{subfigure}[b]{0.5\textwidth}
	\begin{tikzpicture}[scale = 0.8]
		\begin{loglogaxis}[xmin=5,xmax=200, ymin=1e-14,ymax=1,xlabel=\large{$\sqrt[2]{ndof}$},ylabel=\large{$DWR$},grid=major,legend style={at={(1,1)},anchor=north east,font=\tiny,rounded corners=2pt}]       
              \addplot [color = blue,mark=square*,ultra thick] table[x=ndof, y=DWR_est,col sep = comma]{⁨Data⁩/TargetBasedAdaptation⁩/targetbasedadaptation_s_p_plus_1_gaussian_peak_alpha_1000_p_BL1.txt⁩};
       \addplot [color = red,mark=square*,ultra thick] table[x=ndof, y=DWR_est,col sep = comma]{⁨Data⁩/TargetBasedAdaptation⁩/targetbasedadaptation_s_p_plus_1_gaussian_peak_alpha_1000_p_BL2.txt⁩};
       \addplot [color = black,mark=square*,ultra thick] table[x=ndof, y=DWR_est,col sep = comma]{⁨Data⁩/TargetBasedAdaptation⁩/targetbasedadaptation_s_p_plus_1_gaussian_peak_alpha_1000_p_BL3.txt⁩};
		\legend{$P =1$,$P =2$,$P =3$}
		\end{loglogaxis}
	\end{tikzpicture}
	\caption{}\label{convergence_gaussian_peak_dwr}
\end{subfigure}
\caption{Convergence plots for (a) error in target functional  and (b) Dual Weighted Residual using scaled V norm.} 
\end{figure}
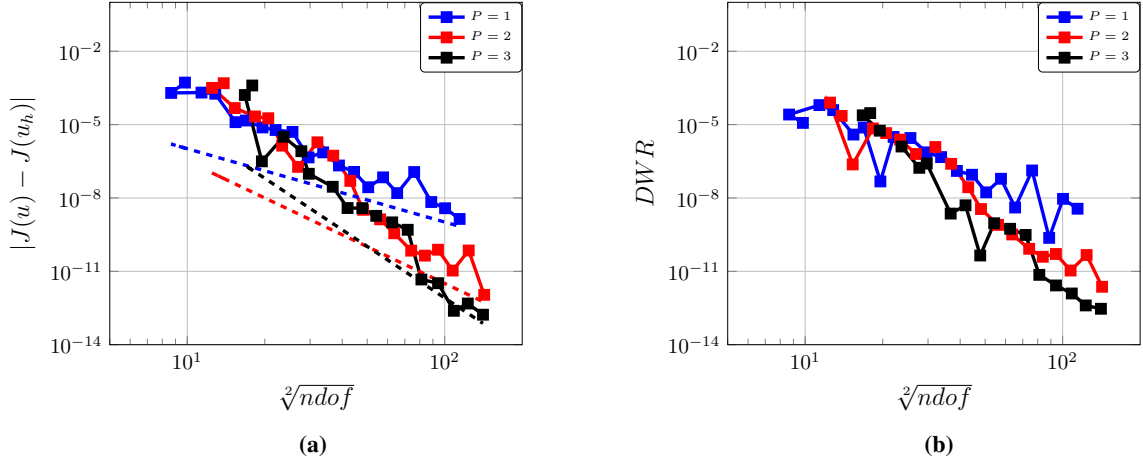

Here $(x_c,y_c) \in \Omega$ determines the location of the Gaussian peak and $\alpha$ controls the steepness of the Gaussian peak.  A similar problem have been mentioned in \cite{MITCHELL2013350}. In the present iteration of the test case, we have considered $(x_c,y_c) = (0.99,0.5)$. This point is located in the middle of one of the boundary layers. In ~\cref{primalsol_gaussianpeak} and ~\cref{dualsol_gaussianpeak}, we have presented the primal solution and the dual solution.  In ~\cref{convergence_gaussian_peak_error} and ~\cref{convergence_gaussian_peak_dwr}, we have shown the convergence of the error and dual weighted residual (DWR). Here also we observe that the rate of convergence for both error in the target function and DWR estimator is $2p + 1$ where $p$ is the polynomial order of approximation for the primal $u$.  In this test case also, we prescribe an increase of $30\%$ in the mesh complexity between each adaptation cycle.

\begin{figure}[H]
\hspace{-1cm}
\begin{subfigure}[b]{0.5\textwidth}
\includegraphics[scale=0.25]{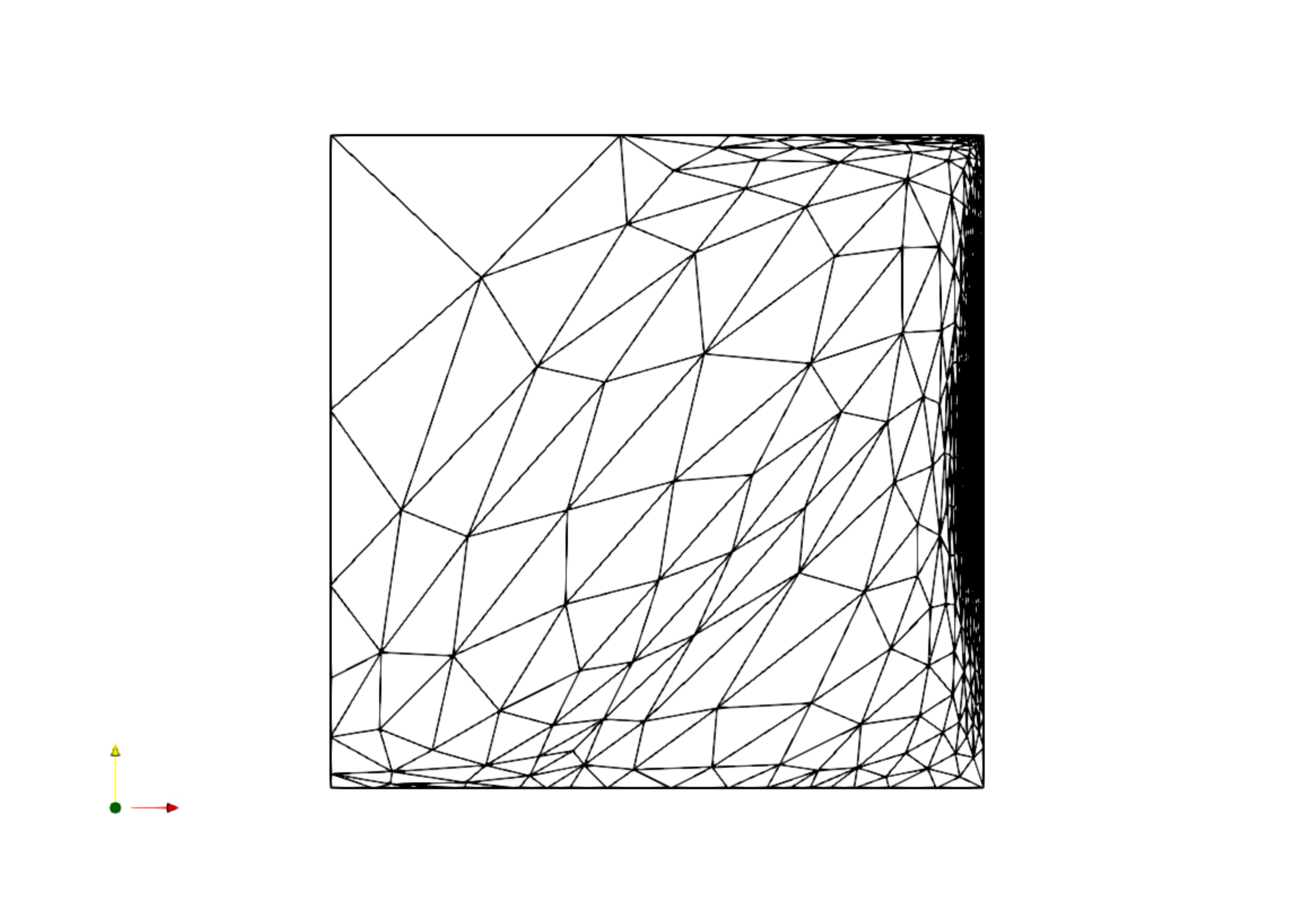}
\caption{$Ne  = 1169$}
\end{subfigure}
\hspace{0.2cm}
\begin{subfigure}[b]{0.5\textwidth}
\includegraphics[scale=0.25]{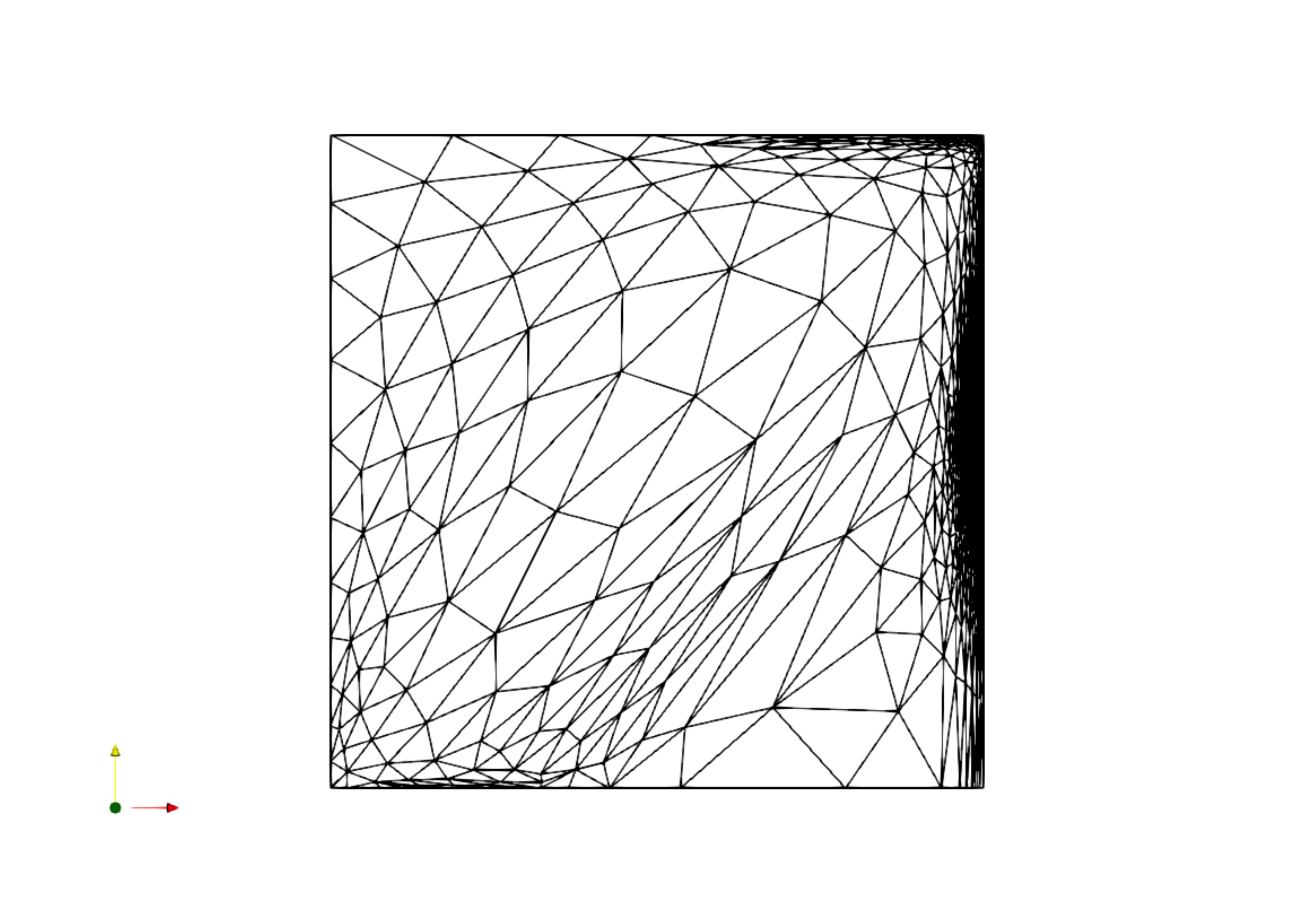}
\caption{$Ne = 1178$}
\end{subfigure}
\caption{Adapated mesh for (a)$P =2$ and (b) $P =3$ after $15^{th}$ adaptation using scaled V norm.} \label{adaptedmeshgaussianpeak}
\end{figure}

In ~\cref{adaptedmeshgaussianpeak}, we have presented the adapted mesh for resolving the Gaussian peak. It can be easily inferred that only the right boundary layer near $x = 1.0$ is resolved as the center of the peak lies in this boundary layer.

Next, we repeat the numerical experiment where we keep the complexity $N$ constant and then adapt the mesh. In this case also, we can notice that it just takes only few adaptation to achieve a drop of $5-6$ order of magnitude in the error of the target functional and the DWR estimator. In ~\cref{fixed_cost_targetfunctional}, we have tabulated the error and the DWR with adaptation.
\begin{table}[H]
\captionsetup{justification=centering,margin=2cm}
\centering
\begin{tabular}{|c|c|c|c|}
\hline
Adaptation & Ne  & ${\vert J(u) - J(u_h) \vert}$ & $DWR$ \\ \hline
 0          & 512 & 0.000136149                         & 3.55914e-05                         \\ \hline
2          & 436 &7.47054e-09                       & 1.03836e-08                     \\ \hline
4          & 469 & 9.99534e-10                      &    1.75062e-09                  \\ \hline
6          & 470 & 3.56406e-10                       & 5.92587e-10                        \\ \hline
8          & 478 & 2.35767e-10                       & 8.58174e-10                       \\ \hline
\end{tabular}
\vspace{0.2cm}
\caption{Adaptation Vs. Error for constant complexity using scaled V norm. ($P =2, N = \int_{\Omega} d(x) \, dx = 512.0$)} \label{fixed_cost_targetfunctional}
\end{table}
\subsection{Inverse Tangent Test Case -Flux Target}
Here, the governing primal PDE stays the same as in previous cases.
\begin{equation}
\begin{aligned}
\beta \cdot {\nabla}u-\epsilon{\nabla}^2u &= s(x) \qquad&& \mathbf{x} \in \Omega = {(0,1)}^2 \\
u &= 0 && \mathbf{x} \in \partial \Omega 
\end{aligned}
\end{equation}
where $\beta = {[1,1]}^T$. The source term $s(\mathbf{x})$ is selected in such a way that the exact solution is given by
\begin{equation}
u(\mathbf{x}) = \left( tan^{-1}(\alpha(x - x_1)) + tan^{-1}(\alpha (x_2 - x)) \right)  \left( tan^{-1}(\alpha(y -y_1)) + tan^{-1}(\alpha (y_2 - y)) \right) \notag
\end{equation}

where $x_1 = y_1 = \frac{1.0}{3.0}$, $x_2 = y_2 = \frac{2.0}{3.0}$, $\alpha = 50.0 $ and $\epsilon = 0.01$. 
Here we focus on a flux target for the dual problem. The target for this problem is 
\begin{equation}
J({\sigma}) = \int_{\partial \Omega} j_{\partial \Omega}(\mathbf{x}) \bm{\sigma} \cdot \mathbf{n} ds
\end{equation}
where we have
\begin{equation}
j_{\partial \Omega} = \begin{cases}
1 \qquad \, \mathbf{n} = \left( 1,0 \right) \\
0 \qquad otherwise
\end{cases}
\end{equation}

Thus, we have taken the flux on the right most boundary as the target function. In this case, the dual problem has no source, while the weighting function $j_{\partial \Omega}(\mathbf{x})$ appears as a boundary condition. Hence, the dual problem is as follows

\begin{align}
-\beta \cdot {\nabla} \eta_1-\epsilon{\nabla}^2 \eta_1 &= 0 \qquad && \mathbf{x} \in \Omega = {(0,1)}^2  \label{adjstrngeqb}\\
\eta_1 &= j_{\partial \Omega} && \mathbf{x} \in \partial \Omega  \label{adjbndb}
\end{align}

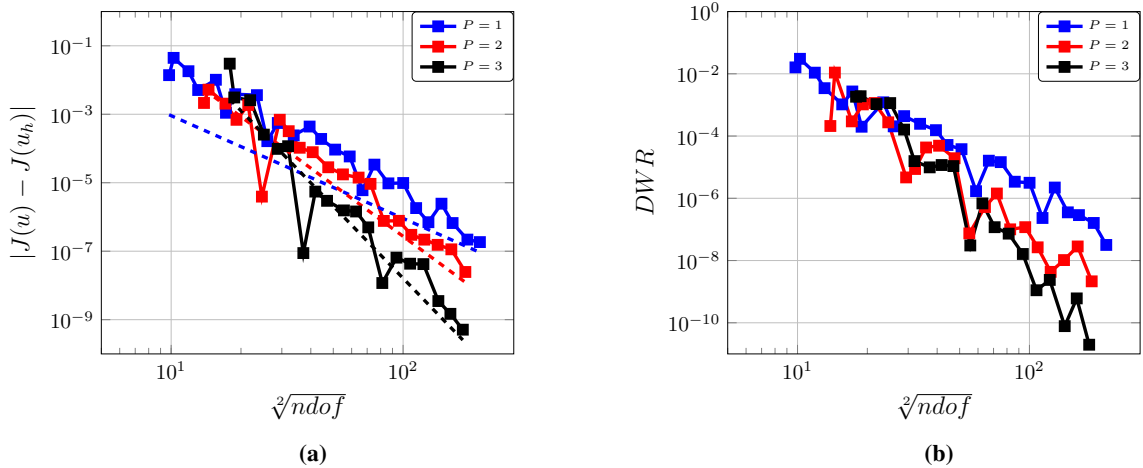
\begin{figure}[h]
\begin{subfigure}[b]{0.5\textwidth}
	\begin{tikzpicture}[scale = 0.8]
		\begin{loglogaxis}[xmin=5,xmax=300, ymin=1e-10,ymax=1,xlabel=\large{$\sqrt[2]{ndof}$},ylabel=\large{$\vert J(u) - J(u_h) \vert$},grid=major,legend style={at={(1,1)},anchor=north east,font=\tiny,rounded corners=2pt}]
       \addplot [color = blue,mark=square*,ultra thick] table[x=ndof, y=target_err,col sep = comma]{⁨Data⁩/TargetBasedAdaptation⁩/targetbasedadaptation_s_p_plus_1_squarejump_Nadarajah_regularized_alpha_50_rightbndflux_eps_0p01_p1.txt⁩};
       \addplot [color = red,mark=square*,ultra thick] table[x=ndof, y=target_err,col sep = comma]{⁨Data⁩/TargetBasedAdaptation⁩/targetbasedadaptation_s_p_plus_1_squarejump_Nadarajah_regularized_alpha_50_rightbndflux_eps_0p01_p2.txt⁩};
       \addplot [color = black,mark=square*,ultra thick] table[x=ndof, y=target_err,col sep = comma]{⁨⁨Data⁩/TargetBasedAdaptation⁩/targetbasedadaptation_s_p_plus_1_squarejump_Nadarajah_regularized_alpha_50_rightbndflux_eps_0p01_p3.txt⁩};       
              \addplot [dashed,color = blue,mark=none,ultra thick] table[x=ndof, y=extslp,col sep = comma]{⁨Data⁩/TargetBasedAdaptation⁩/targetbasedadaptation_s_p_plus_1_squarejump_Nadarajah_regularized_alpha_50_rightbndflux_eps_0p01_p1.txt⁩};
       \addplot [dashed,color = red,mark=none,ultra thick] table[x=ndof, y=extslp,col sep = comma]{⁨Data⁩/TargetBasedAdaptation⁩/targetbasedadaptation_s_p_plus_1_squarejump_Nadarajah_regularized_alpha_50_rightbndflux_eps_0p01_p2.txt⁩};
       \addplot [dashed,color = black,mark=none,ultra thick] table[x=ndof, y=extslp,col sep = comma]{⁨Data⁩/TargetBasedAdaptation⁩/targetbasedadaptation_s_p_plus_1_squarejump_Nadarajah_regularized_alpha_50_rightbndflux_eps_0p01_p3.txt⁩};
		\legend{$P =1$,$P =2$,$P =3$}
		\end{loglogaxis}
	\end{tikzpicture}
	\caption{}
\end{subfigure}
\begin{subfigure}[b]{0.5\textwidth}
	\begin{tikzpicture}[scale = 0.8]
		\begin{loglogaxis}[xmin=5,xmax=300, ymin=1e-11,ymax=1,xlabel=\large{$\sqrt[2]{ndof}$},ylabel=\large{$DWR$},grid=major,legend style={at={(1,1)},anchor=north east,font=\tiny,rounded corners=2pt}]       
              \addplot [color = blue,mark=square*,ultra thick] table[x=ndof, y=DWR_est,col sep = comma]{⁨Data⁩/TargetBasedAdaptation⁩/targetbasedadaptation_s_p_plus_1_squarejump_Nadarajah_regularized_alpha_50_rightbndflux_eps_0p01_p1.txt};
       \addplot [color = red,mark=square*,ultra thick] table[x=ndof, y=DWR_est,col sep = comma]{⁨Data⁩/TargetBasedAdaptation⁩/targetbasedadaptation_s_p_plus_1_squarejump_Nadarajah_regularized_alpha_50_rightbndflux_eps_0p01_p2.txt⁩};
       \addplot [color = black,mark=square*,ultra thick] table[x=ndof, y=DWR_est,col sep = comma]{⁨Data⁩/TargetBasedAdaptation⁩/targetbasedadaptation_s_p_plus_1_squarejump_Nadarajah_regularized_alpha_50_rightbndflux_eps_0p01_p3.txt⁩};
		\legend{$P =1$,$P =2$,$P =3$}
		\end{loglogaxis}
	\end{tikzpicture}
	\caption{}
\end{subfigure}
\caption{Convergence plots for (a) error in target functional  and (b) Dual Weighted Residual with regularization (~\cref{regularization} ) using scaled V norm.} \label{convfluxtarget}
\end{figure}

\begin{figure}[H]
\hspace{-1cm}
\begin{subfigure}[b]{0.5\textwidth}
\includegraphics[scale=0.25]{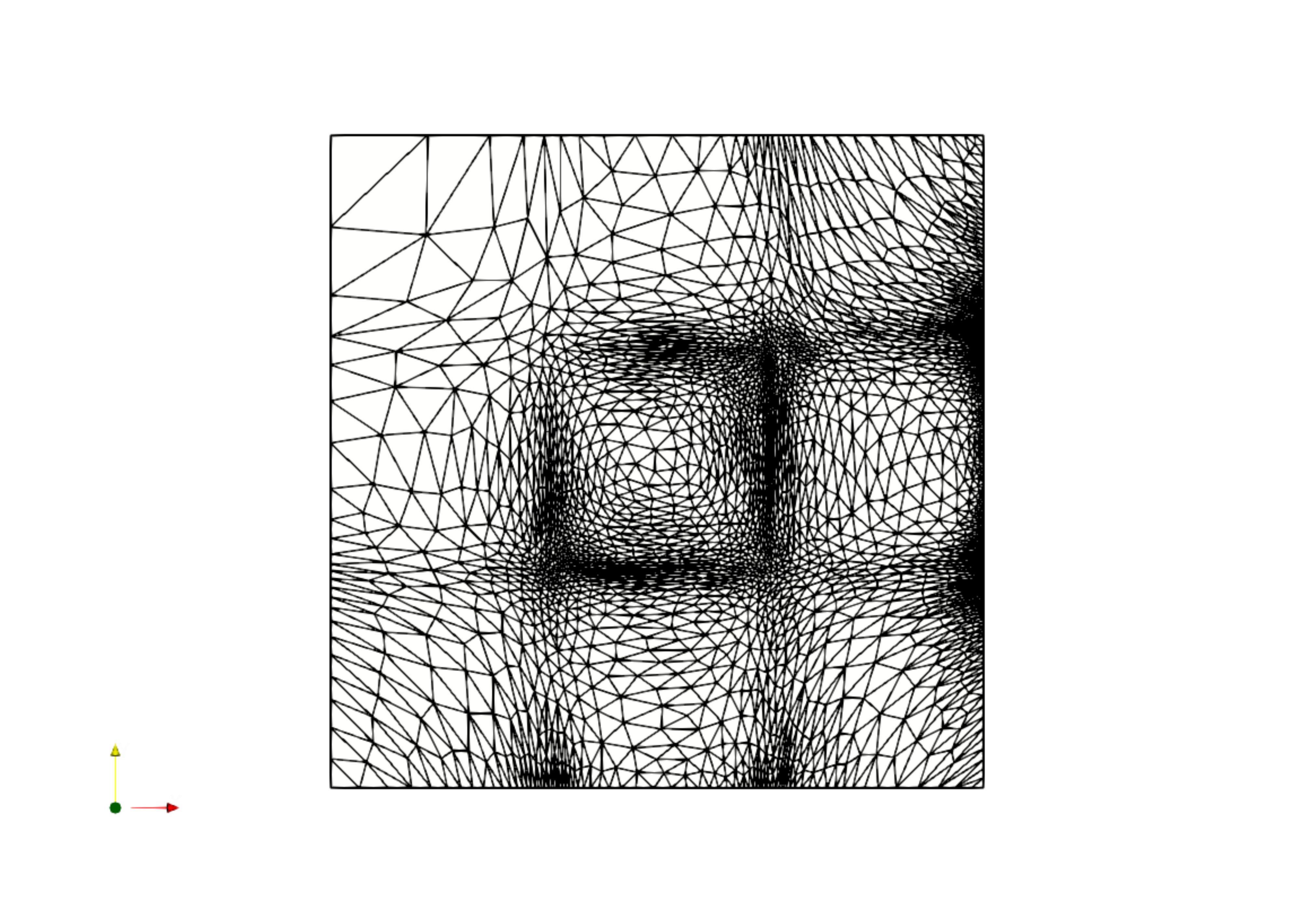}
\caption{$Ne  = 4344$}
\end{subfigure}
\hspace{0.2cm}
\begin{subfigure}[b]{0.5\textwidth}
\includegraphics[scale=0.25]{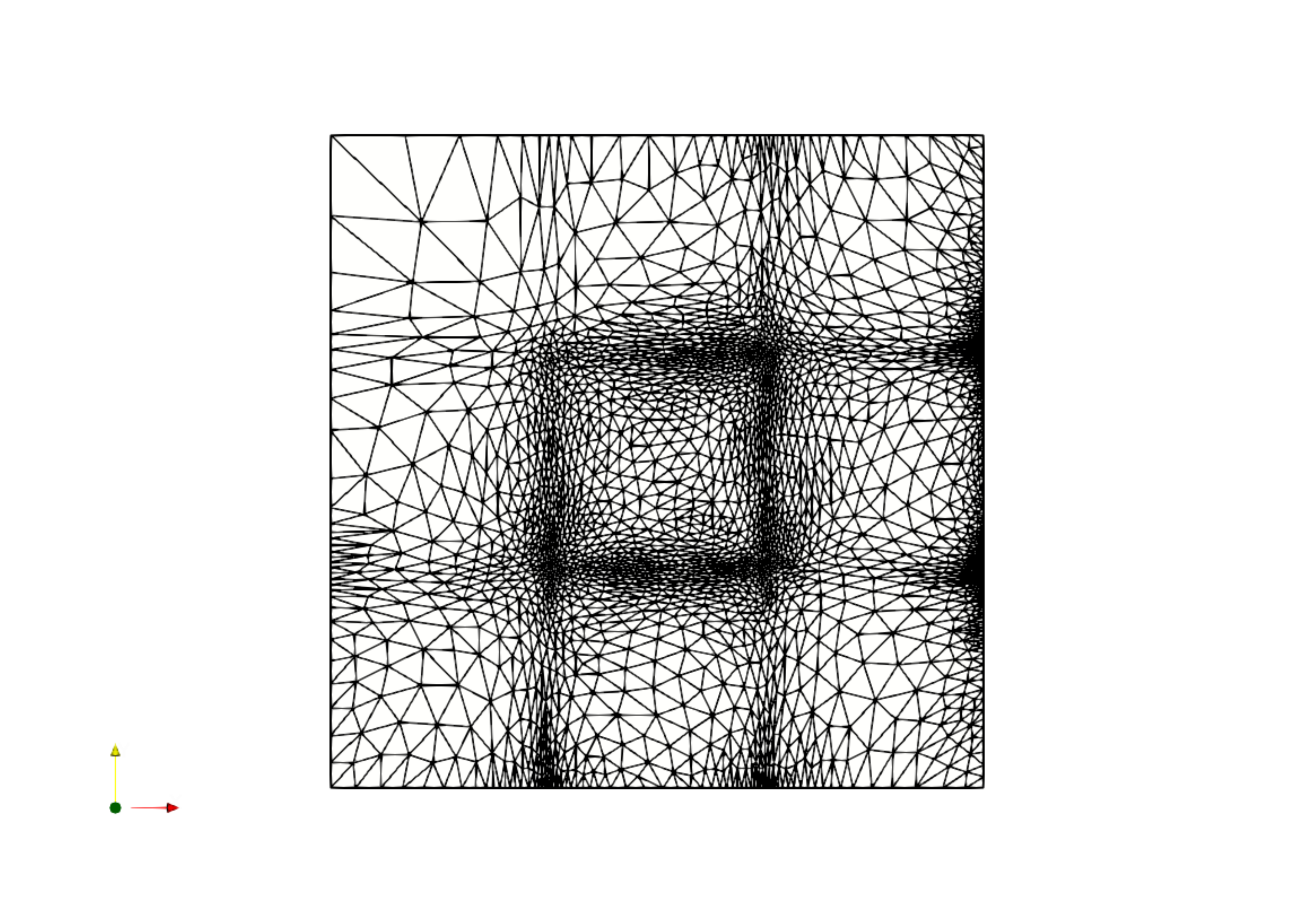}
\caption{$Ne = 4276$}
\end{subfigure}
\caption{Adapated mesh for (a)$P =2$ and (b) $P =3$ after $15^{th}$ adaptation using scaled V norm.}\label{adaptedmeshfluxtarget}
\end{figure}

\begin{figure}[H]
\begin{subfigure}[b]{0.3\textwidth}
\includegraphics[scale=0.16]{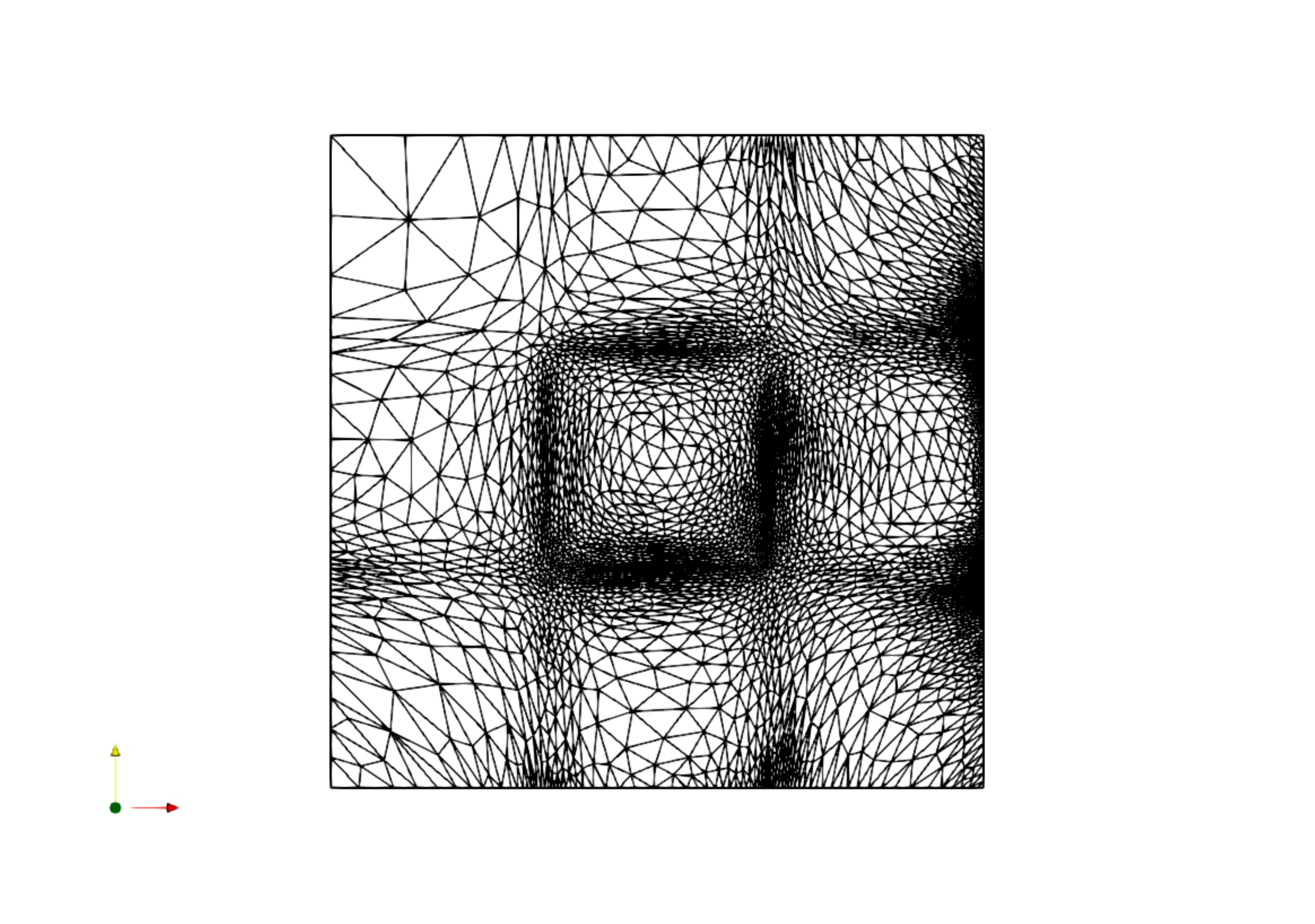}
\caption{$Ne  = 5785$}
\end{subfigure}
\hspace{0.25cm}
\begin{subfigure}[b]{0.3\textwidth}
\includegraphics[scale=0.16]{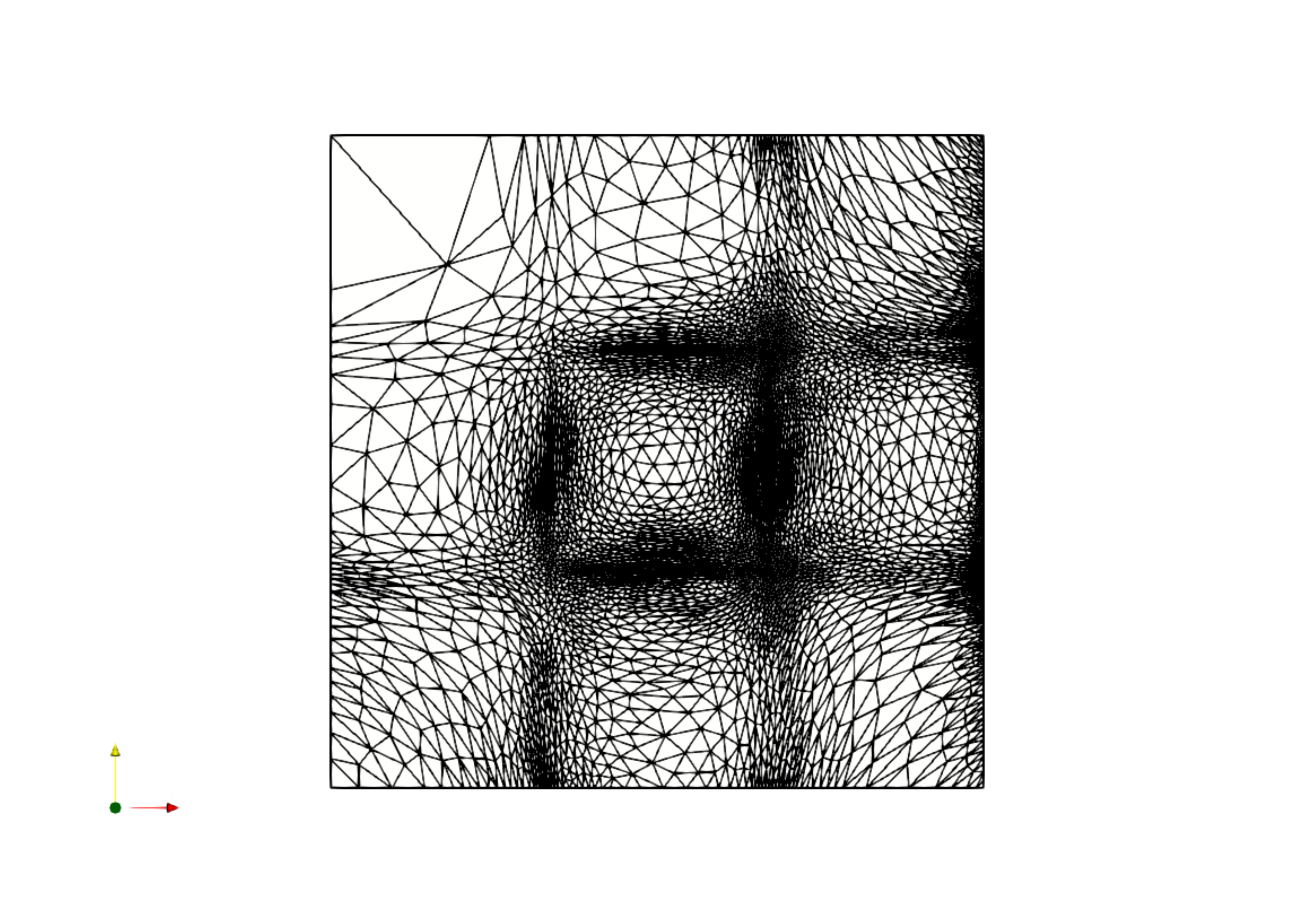}
\caption{$Ne = 7386$}
\end{subfigure}
\hspace{0.25cm}
\begin{subfigure}[b]{0.3\textwidth}
\includegraphics[scale=0.16]{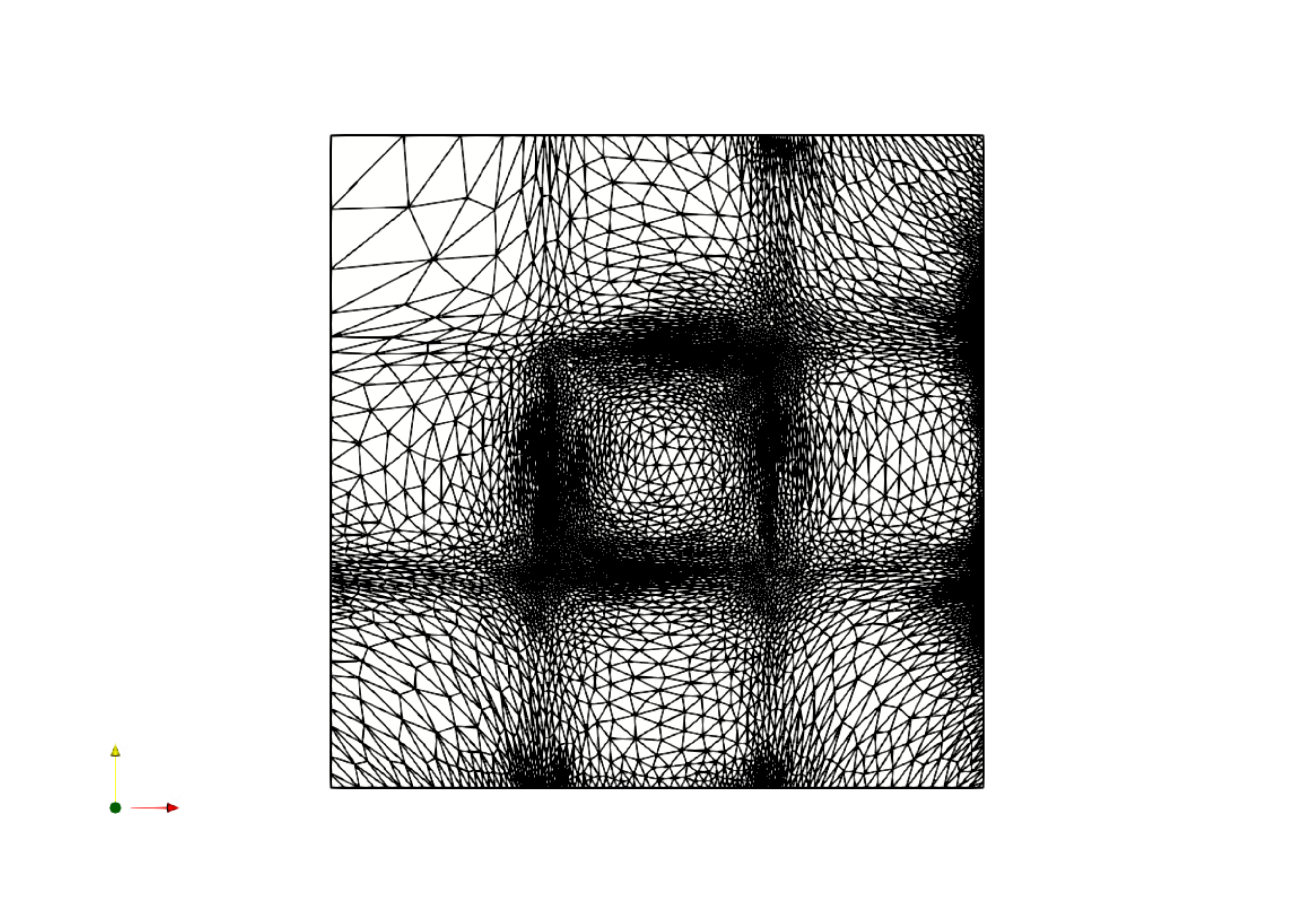}
\caption{$Ne = 9742$}
\end{subfigure}
\caption{Oscillation in mesh during adaptation cycle without regularization using scaled V norm.} \label{oscillation_mesh}
\end{figure}

\begin{figure}[H]
\begin{subfigure}[b]{0.3\textwidth}
\includegraphics[scale=0.16]{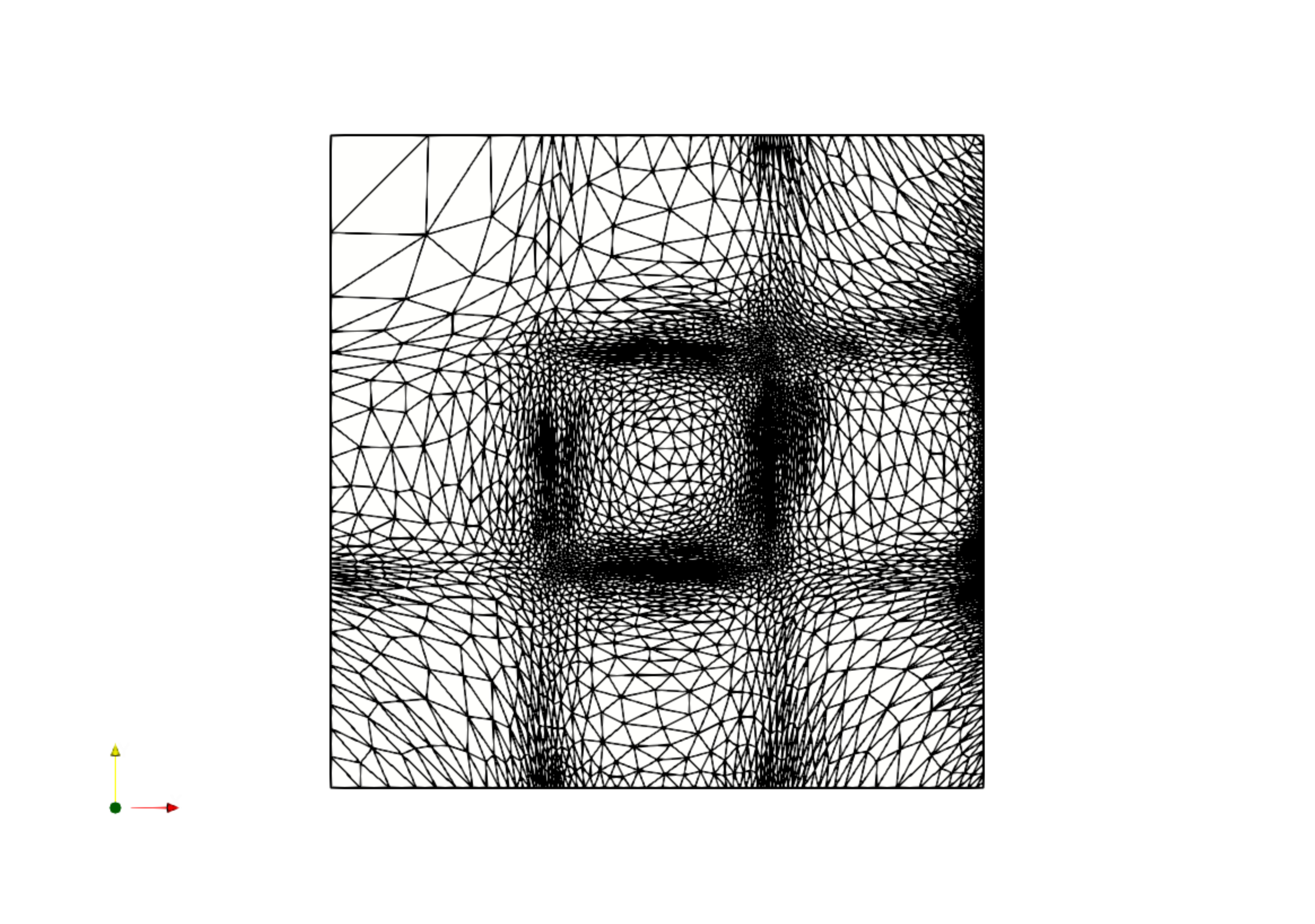}
\caption{$Ne  = 5737$}
\end{subfigure}
\hspace{0.25cm}
\begin{subfigure}[b]{0.3\textwidth}
\includegraphics[scale=0.16]{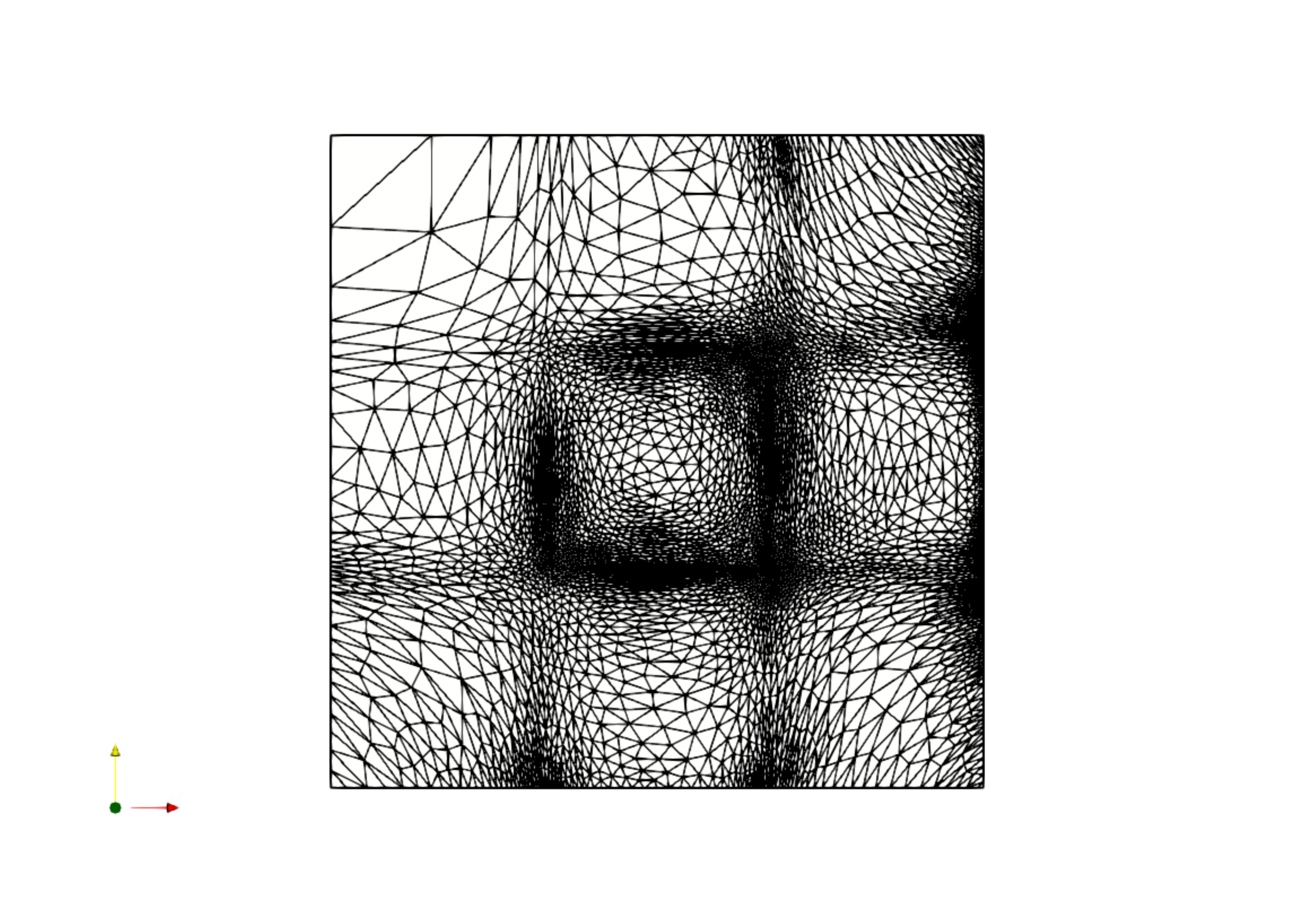}
\caption{$Ne = 7412$}
\end{subfigure}
\hspace{0.25cm}
\begin{subfigure}[b]{0.3\textwidth}
\includegraphics[scale=0.16]{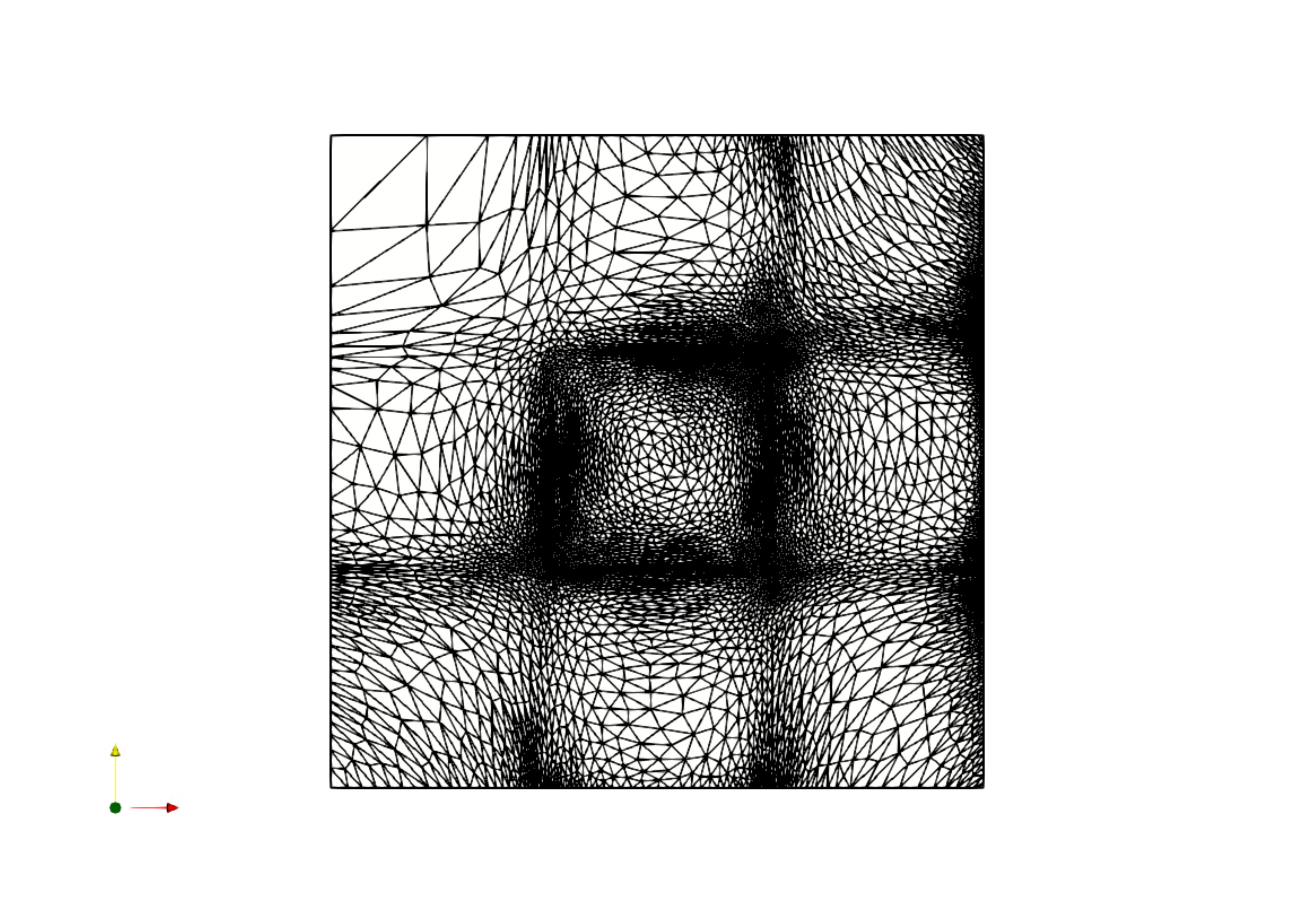}
\caption{$Ne = 9751$}
\end{subfigure}
\caption{Adaptation cycle with regularization using scaled V norm.} \label{non_oscillation_mesh}
\end{figure}

\begin{figure}[h]
\centering
	\begin{tikzpicture}[scale = 0.8]
		\begin{loglogaxis}[xmin=5,xmax=300, ymin=1e-8,ymax=1,xlabel=\large{$\sqrt[2]{ndof}$},ylabel=\large{$\vert J(u) - J(u_h) \vert$},grid=major,legend style={at={(1,1)},anchor=north east,font=\tiny,rounded corners=2pt}]
       \addplot [color = blue,mark=square*,ultra thick] table[x=ndof, y=target_err,col sep = comma]{⁨Data⁩/TargetBasedAdaptation⁩/targetbasedadaptation_s_p_plus_1_squarejump_Nadarajah_regularized_alpha_50_rightbndflux_eps_0p01_p2.txt⁩};
       \addplot [color = red,mark=square*,ultra thick] table[x=ndof, y=target_err,col sep = comma]{⁨Data⁩/TargetBasedAdaptation⁩/targetbasedadaptation_s_p_plus_1_squarejump_Nadarajah_non_regularized_alpha_50_rightbndflux_eps_0p01_p2.txt⁩};
		\legend{$Regularized$,$Non-Regularized$}
		\end{loglogaxis}
	\end{tikzpicture}
\caption{Comparison of Convergence plots for error in $J(u)$ between regularized and non regularized adaptation cycles using scaled V norm and $P = 2$.} \label{comparison_regularization}
\end{figure}
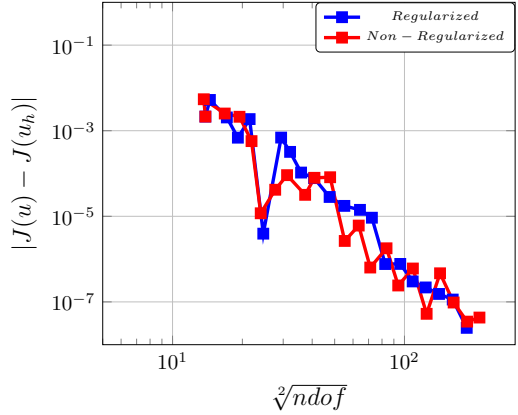

In ~\cref{convfluxtarget}, we have presented the convergence in the target functional and DWR estimator which we followed by presenting the adapted mesh in ~\cref{adaptedmeshfluxtarget}.
Through this numerical example, we also wanted to show the need of regularization mentioned in ~\cref{regularization}. In ~\cref{oscillation_mesh}, we have presented the adapted meshes from three subsequent adaptation and it can be quite easily noticed that alternate refinement and coarsening of mesh elements is  taking place at the top left corner of the domain.  This causes spurious oscillation in the convergence of the  target function even though we are not in the pre-asymptotic region.  If regularization is employed, these oscillation tend to disappear which can be seen in ~\cref{non_oscillation_mesh}. In ~\cref{comparison_regularization}, we have presented the comparison of convergence in target functional with and without regularization employed. The convergence with regularization employed is much more smoother compared to oscillatory convergence without regularization. In current setup, the oscillations are 
not overwhelming but this can't be guaranteed and will depend upon the nature of the problem. Hence, implementation of regularization is subjective and needs to be treated on a case by case manner.

\subsection{Interior Line Singularity}
\begin{figure}[H]
\begin{subfigure}[b]{0.5\textwidth}
\begin{tikzpicture}[scale=0.8]
		\begin{loglogaxis}[xmin=9,xmax=300, ymin=1e-9,ymax=1,xlabel=\large{$\sqrt[2]{ndof}$},ylabel=\large{$||u-u_h||_{L^{2}(\Omega)}$},grid=major,legend style={at={(1,1)},anchor=north east,font=\tiny,rounded corners=2pt}]
		\addplot[color = blue,mark=square*]  table[x= ndof, y=err_l2, col sep = comma] {Data/⁨Interior_line_singularity⁩/MNS_0p5/L2_error_DPG_ILS_gamma3p5_p1_MN_0p5.txt};
		\addplot[color = red,mark=square*]  table[x= ndof, y=err_l2, col sep = comma] {Data/⁨Interior_line_singularity⁩/MNS_0p5/L2_error_DPG_ILS_gamma3p5_p2_MN_0p5.txt};
		\addplot [color = black,mark=square*] table[x= ndof, y=err_l2, col sep = comma] {Data/⁨Interior_line_singularity⁩/MNS_0p5/L2_error_DPG_ILS_gamma3p5_p3_MN_0p5.txt};
	    \addplot  [dashed,line width=1.5pt,mark=none, black,forget plot] table[x= ndof, y=exslp, col sep = comma] {Data/⁨Interior_line_singularity⁩/MNS_0p5/L2_error_DPG_ILS_gamma3p5_p1_MN_0p5.txt};
	    	    \addplot  [dashed,line width=1.5pt,mark=none, black,forget plot] table[x= ndof, y=exslp, col sep = comma] {Data/⁨Interior_line_singularity⁩/MNS_0p5/L2_error_DPG_ILS_gamma3p5_p2_MN_0p5.txt};
	    	    	    \addplot  [dashed,line width=1.5pt,mark=none, black,forget plot] table[x= ndof, y=exslp, col sep = comma] {Data/⁨Interior_line_singularity⁩/MNS_0p5/L2_error_DPG_ILS_gamma3p5_p3_MN_0p5.txt};
		\legend{$P =1$,$P =2$,$P =3$}
		\end{loglogaxis}
	\end{tikzpicture}
	\caption{}
\end{subfigure}
\begin{subfigure}[b]{0.5\textwidth}
\begin{tikzpicture}[scale=0.8]
		\begin{loglogaxis}[xmin=10,xmax=300, ymin=1e-7,ymax=1,xlabel=\large{$\sqrt[2]{ndof}$},ylabel=\large{$||U-U_h||_{E(\Omega)}$},grid=major,legend style={at={(1,1)},anchor=north east,font=\tiny,rounded corners=2pt}]
		\addplot [color = blue,mark=square*] table[x= ndof, y=EE, col sep = comma] {Data/⁨Interior_line_singularity⁩/MNS_0p5/EE_error_DPG_ILS_gamma_3p5_p1_MN_0p5.txt};
		\addplot [color = red,mark=square*] table[x= ndof, y=EE, col sep = comma] {Data/⁨Interior_line_singularity⁩/MNS_0p5/EE_error_DPG_ILS_gamma_3p5_p2_MN_0p5.txt};
			\addplot [color = black,mark=square*] table[x= ndof, y=EE, col sep = comma] {Data/⁨Interior_line_singularity⁩/MNS_0p5/EE_error_DPG_ILS_gamma_3p5_p3_MN_0p5.txt};
		\legend{$P =1$,$P =2$,$P =3$}
		\end{loglogaxis}
\end{tikzpicture}
\caption{}
\end{subfigure}	
\caption{Convergence plots for different polynomial order of $L^2$ error in $u_h$ and Energy norm at $\gamma = 3.5$ using scaled V norm.} \label{convergence_ils}
\end{figure}
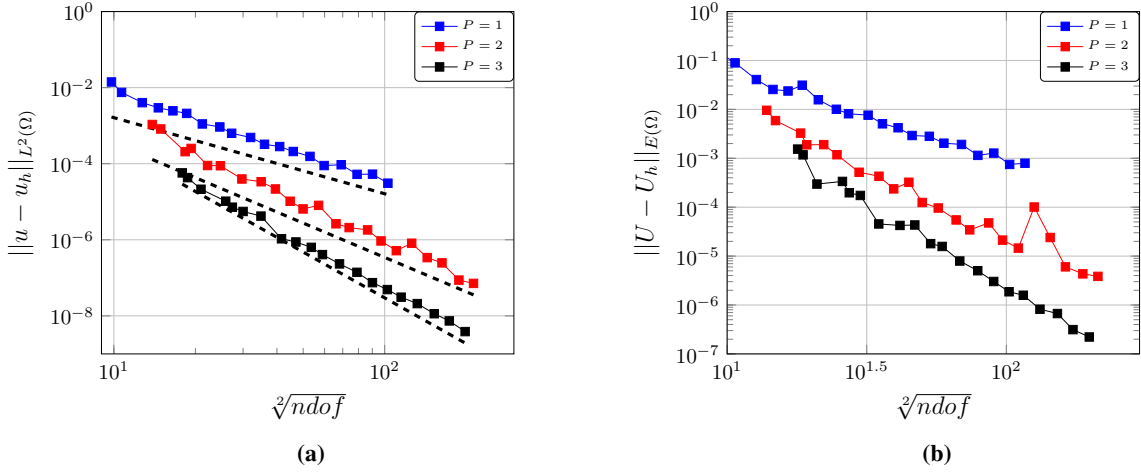

\begin{equation}
\begin{aligned}
-{\nabla}^2 u &= s(\mathbf{x}) \quad && \mathbf{x}\in \Omega = {(0,1)}^2 \\
u &= g_D && \mathbf{x} \in \partial \Omega
\end{aligned}
\end{equation}
Source term $s(\mathbf{x})$ is chosen in such a way that the exact solution is given by:
\begin{equation}
u(\mathbf{x}) = \begin{cases} \cos(\pi (y - 0.5)) & \qquad  x \leq\Theta y + 0.5 \\
\cos(\pi (y - 0.5)) + {(x-\Theta y - 0.5)}^{\gamma}2^{\gamma} & \qquad x > \Theta y + 0.5.
\end{cases}
\end{equation}
The boundary condition $g_D$ is set to be the exact solution. 
The regularity of the solution depends upon the parameter $\gamma$ as the solution lies in fractional Sobolev space  $H^{(\gamma + 0.5 -\epsilon)} \, \forall \epsilon > 0$.  Consequently, the optimal order of convergence of solution $u$ in $L^2$ norm is 
\begin{equation}
min(p+1,\gamma + 0.5 -\epsilon) \label{opt_conv_order_ils}
\end{equation}

where $p$ is the order of approximation. ~\Cref{convergence_ils} shows the convergence plots in $L^2$ norm of the primary field variable $u$ and the energy norm induced by the scaled V norm. The primary objective of performing this numerical experiment is to test the optimality of the meshes being produced by the proposed methodology. Since, the proposed methodology directly extracts information from the inbuilt estimator, optimality means that the mesh adapted for a given parameter $\gamma$ should be sub-optimal for a different value of $\gamma$. We will be demonstrating this with two different values of $\gamma$ $(\gamma = 2.0 \, \text{and}  \,\gamma = 3.3)$.  

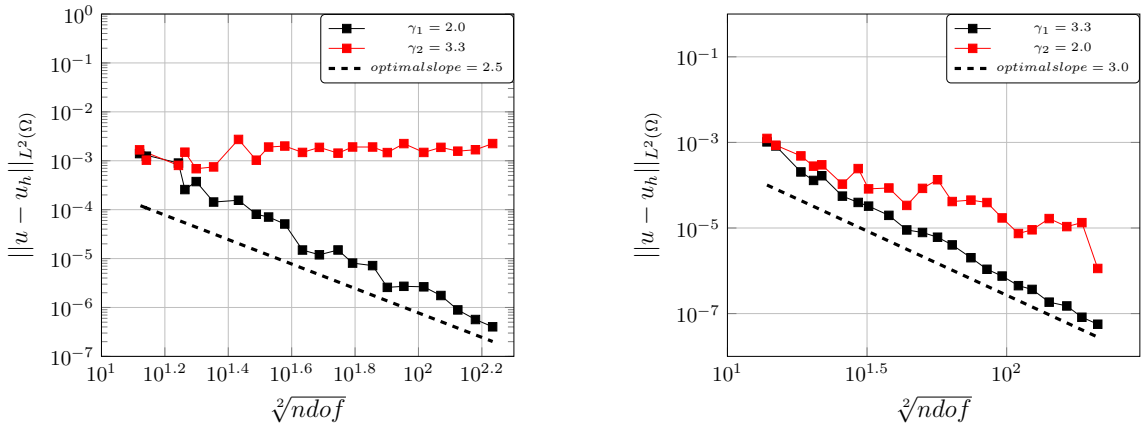
\begin{figure}[H]
\begin{subfigure}[b]{0.5\textwidth}
	\begin{tikzpicture}[scale=0.8]
		\begin{loglogaxis}[xmin=10,xmax=200, ymin=1e-7,ymax=1,xlabel=\large{$\sqrt[2]{ndof}$},ylabel=\large{$||u-u_h||_{L^{2}(\Omega)}$},grid=major,legend style={at={(1,1)},anchor=north east,font=\tiny,rounded corners=2pt}]
		\addplot [color = black,mark=square*] table[x= ndof, y=err_l2, col sep = comma] {Data/⁨Interior_line_singularity⁩/MNS_0p5/L2_error_DPG_ILS_casea_gamma1_p2_MN_0p5.txt};
		\addplot [color = red,mark=square*] table[x= ndof, y=err_l2, col sep = comma] {Data/⁨Interior_line_singularity⁩/MNS_0p5/L2_error_DPG_ILS_casea_gamma2_p2_MN_0p5.txt};
	    \addplot  [dashed,line width=1.5pt,mark=none, black] table[x= ndof, y=exslp, col sep = comma] {Data/⁨Interior_line_singularity⁩/MNS_0p5/L2_error_DPG_ILS_casea_gamma1_p2_MN_0p5.txt};
		\legend{$\gamma_1  = 2.0$,$\gamma_2 = 3.3$,$optimal slope =  2.5$}
		\end{loglogaxis}
	\end{tikzpicture}
%
\end{subfigure}
\begin{subfigure}[b]{0.5\textwidth}
	\begin{tikzpicture}[scale=0.8]
		\begin{loglogaxis}[xmin=10,xmax=300, ymin=1e-8,ymax=1,xlabel=\large{$\sqrt[2]{ndof}$},ylabel=\large{$||u-u_h||_{L^{2}(\Omega)}$},grid=major,legend style={at={(1,1)},anchor=north east,font=\tiny,rounded corners=2pt}]
		\addplot [color = black,mark=square*] table[x= ndof, y=err_l2, col sep = comma] {Data/⁨Interior_line_singularity⁩/MNS_0p5/L2_error_DPG_ILS_caseb_gamma1_p2_MN_0p5.txt};
		\addplot [color = red,mark=square*] table[x= ndof, y=err_l2, col sep = comma] {Data/⁨Interior_line_singularity⁩/MNS_0p5/L2_error_DPG_ILS_caseb_gamma2_p2_MN_0p5.txt};
	    \addplot  [dashed,line width=1.5pt,mark=none, black] table[x= ndof, y=exslp, col sep = comma] {Data/⁨Interior_line_singularity⁩/MNS_0p5/L2_error_DPG_ILS_caseb_gamma1_p2_MN_0p5.txt};
		\legend{$\gamma_1  = 3.3$,$\gamma_2 = 2.0$,$optimal slope =  3.0$}
		\end{loglogaxis}
	\end{tikzpicture}
%
\end{subfigure}

\caption{Convergence plots for optimal $\gamma_1$ and suboptimal $\gamma_2$ in $L^2$ norm of $u$ using scaled V norm.} \label{optimal_L2_norm}
\end{figure}

\begin{figure}[H]
\begin{subfigure}[b]{0.5\textwidth}
\begin{tikzpicture}[scale=0.8]
		\begin{loglogaxis}[xmin=10,xmax=200, ymin=1e-4,ymax=1,xlabel=\large{$\sqrt[2]{ndof}$},ylabel=\large{$||U-U_h||_{E(\Omega)}$},grid=major,legend style={at={(1,1)},anchor=north east,font=\tiny,rounded corners=2pt}]
		\addplot [color = black,mark=square*] table[x= ndof, y=EE, col sep = comma] {Data/⁨Interior_line_singularity⁩/MNS_0p5/EE_error_DPG_ILS_casea_gamma1_p2_MN_0p5.txt};
		\addplot [color = red,mark=square*] table[x= ndof, y=EE, col sep = comma] {Data/⁨Interior_line_singularity⁩/MNS_0p5/EE_error_DPG_ILS_casea_gamma2_p2_MN_0p5.txt};
		\legend{$\gamma_1  = 2.0$,$\gamma_2 = 3.3$}
		\end{loglogaxis}
	\end{tikzpicture}
\end{subfigure}
\begin{subfigure}[b]{0.5\textwidth}
\begin{tikzpicture}[scale=0.8]
		\begin{loglogaxis}[xmin=10,xmax=300, ymin=1e-6,ymax=1e-1,xlabel=\large{$\sqrt[2]{ndof}$},ylabel=\large{$||U-U_h||_{E(\Omega)}$},grid=major,legend style={at={(1,1)},anchor=north east,font=\tiny,rounded corners=2pt}]
		\addplot [color = black,mark=square*] table[x= ndof, y=EE, col sep = comma] {Data/⁨Interior_line_singularity⁩/MNS_0p5/EE_error_DPG_ILS_caseb_gamma1_p2_MN_0p5.txt};
		\addplot [color = red,mark=square*] table[x= ndof, y=EE, col sep = comma] {Data/⁨Interior_line_singularity⁩/MNS_0p5/EE_error_DPG_ILS_caseb_gamma2_p2_MN_0p5.txt};
		\legend{$\gamma_1  = 3.3$,$\gamma_2 = 2.0$}
		\end{loglogaxis}
	\end{tikzpicture}
\end{subfigure}
\caption{Convergence plots for optimal $\gamma_1$ and suboptimal $\gamma_2$ in Energy norm using scaled V norm.} \label{optimal_energy_norm}
\end{figure}
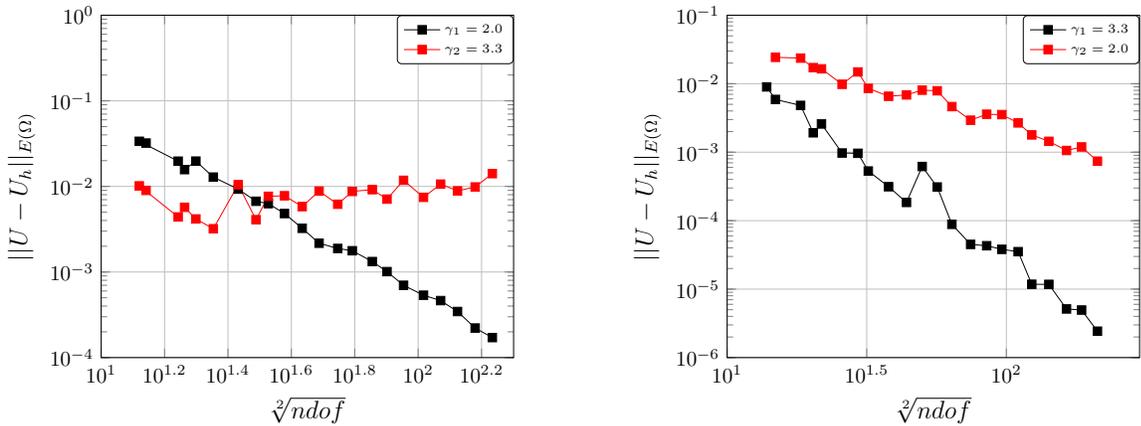

\begin{figure}[H]
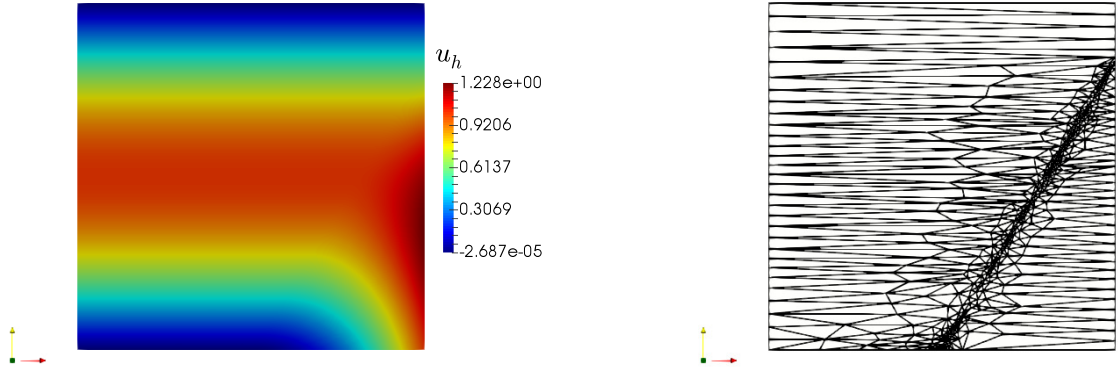

\begin{subfigure}[b]{0.5\textwidth}
\includegraphics[scale=0.33]{Data/⁨Interior_line_singularity⁩/MNS_0p5/sol_ils_gamma_2_856-eps-converted-to.pdf}
\end{subfigure}
\hspace{0.8cm}
\begin{subfigure}[b]{0.5\textwidth}
\includegraphics[scale=0.33]{Data/⁨Interior_line_singularity⁩/MNS_0p5/mesh_ils_gamma_2_856-eps-converted-to.pdf}
\end{subfigure}
\caption{Solution and Mesh after 15 adaptations $(\gamma = 2.0,Ne = 856, P = 2)$ using scaled V norm.}\label{ILS_a}
\end{figure}

\begin{figure}[H]
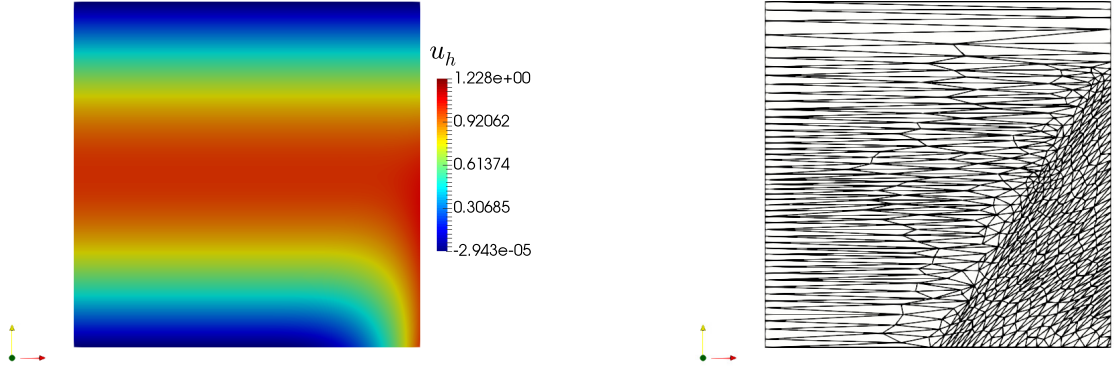

\begin{subfigure}[b]{0.5\textwidth}
\includegraphics[scale=0.25]{Data/⁨Interior_line_singularity⁩/MNS_0p5/sol_ils_gamma_2_925-eps-converted-to.pdf}
\end{subfigure}
\hspace{0.8cm}
\begin{subfigure}[b]{0.5\textwidth}
\includegraphics[scale=0.25]{Data/⁨Interior_line_singularity⁩/MNS_0p5/mesh_ils_gamma_2_925-eps-converted-to.pdf}
\end{subfigure}
\caption{Solution and Mesh after 15 adaptations $(\gamma = 3.3,Ne = 925, P = 2)$ using scaled V norm.} \label{ILS_b}
\end{figure}

To perform this test, we start with a isotropic mesh containing $32$ elements and we increase $N$ by $30\%$ in each adaptation. The meshes are initially adapted for $\gamma_1$ and all the meshes in this adaptation run are saved. We, then solve the poisson problem on these saved meshes with $\gamma_2$. In ~\cref{optimal_L2_norm}, it can be  observed that the optimal convergence order for primary field variable $u$ as given by ~\cref{opt_conv_order_ils} is only achieved in case of $\gamma_1$. This sub-optimal behaviour can also be seen in the convergence plots for the energy error (~\cref{optimal_energy_norm}). This clearly demonstrates the strong dependence of mesh on the problem at hand which is achieved by directly utilizing the inbuilt error estimator.  In ~\cref{ILS_a} and ~\cref{ILS_b}, we have presented the solution for different $\gamma$ and corresponding adapted mesh. 

\subsection{L-shaped Domain}
\begin{equation}
\begin{aligned}
-{\nabla}^2 u &= s(\mathbf{x}) \quad && \mathbf{x} \in \Omega = [-1,1]^2 \setminus [0,1] \times [-1,0] \\
u &= g_D && \mathbf{x} \in \partial \Omega
\end{aligned}
\end{equation}
Source term $s(\mathbf{x})$ is chosen in such a way that the exact solution is given by:
\begin{equation}
u(\mathbf{x}) = r^{\frac{2}{3}}sin\left(\frac{2}{3}\theta\right) \quad where \quad \theta = tan^{-1}\left(\frac{y}{x}\right) \quad \text{and} \quad r = \sqrt{x^2 + y^2} 
\end{equation}

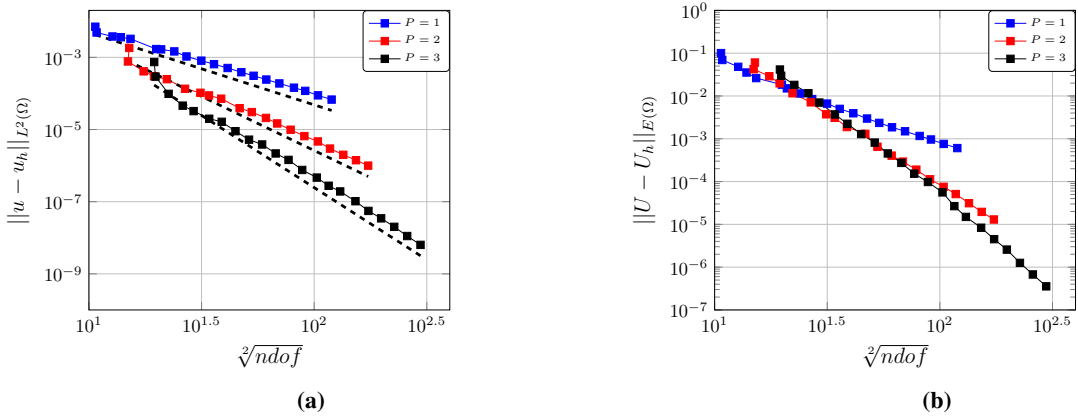
\begin{figure}[H]
\begin{subfigure}[b]{0.5\textwidth}
\begin{tikzpicture}[scale=0.7]
		\begin{loglogaxis}[xmin=10,xmax=400, ymin=1e-10,ymax=0.02,xlabel=\large{$\sqrt[2]{ndof}$},ylabel=\large{$||u-u_h||_{L^{2}(\Omega)}$},grid=major,legend style={at={(1,1)},anchor=north east,font=\tiny,rounded corners=2pt}]
		\addplot[color = blue,mark=square*] table[x= ndof, y=err_l2, col sep = comma] {Data/L_shaped⁩/L2_error_Lshaped_Domain_MNS_0p5_p1.txt};
		\addplot [color = red,mark=square*] table[x= ndof, y=err_l2, col sep = comma] {Data/L_shaped⁩/L2_error_Lshaped_Domain_MNS_0p5_p2.txt};
		\addplot [color = black,mark=square*] table[x= ndof, y=err_l2, col sep = comma] {Data/L_shaped⁩/L2_error_Lshaped_Domain_MNS_0p5_p3.txt};
	    \addplot  [dashed,line width=1.5pt,mark=none, black,forget plot] table[x= ndof, y=exslp, col sep = comma] {Data/L_shaped⁩/L2_error_Lshaped_Domain_MNS_0p5_p1.txt};
	    	    \addplot  [dashed,line width=1.5pt,mark=none, black,forget plot] table[x= ndof, y=exslp, col sep = comma] {Data/L_shaped⁩/L2_error_Lshaped_Domain_MNS_0p5_p2.txt};
	    	    	    \addplot  [dashed,line width=1.5pt,mark=none, black,forget plot] table[x= ndof, y=exslp, col sep = comma] {Data/L_shaped⁩/L2_error_Lshaped_Domain_MNS_0p5_p3.txt};
		\legend{$P =1$,$P =2$,$P =3$}
		\end{loglogaxis}
	\end{tikzpicture}
	\caption{}
\end{subfigure}
\begin{subfigure}[b]{0.5\textwidth}
\begin{tikzpicture}[scale=0.7]
		\begin{loglogaxis}[xmin=10,xmax=400, ymin=1e-7,ymax=1,xlabel=\large{$\sqrt[2]{ndof}$},ylabel=\large{$||U-U_h||_{E(\Omega)}$},grid=major,legend style={at={(1,1)},anchor=north east,font=\tiny,rounded corners=2pt}]
		\addplot[color = blue,mark=square*]  table[x= ndof, y=EE, col sep = comma] {Data/L_shaped⁩/EE_error_Lshaped_Domain_MNS_0p5_p1.txt};
		\addplot [color = red,mark=square*] table[x= ndof, y=EE, col sep = comma]  {Data/L_shaped⁩/EE_error_Lshaped_Domain_MNS_0p5_p2.txt};
			\addplot [color = black,mark=square*] table[x= ndof, y=EE, col sep = comma]  {Data/L_shaped⁩/EE_error_Lshaped_Domain_MNS_0p5_p3.txt};
		\legend{$P =1$,$P =2$,$P =3$}
		\end{loglogaxis}
\end{tikzpicture}
\caption{}
\end{subfigure}	
\caption{Convergence plots of (a) $L^2$ error in $u_h$ and (b) Energy norm using scaled V norm} \label{convergence_Lshaped_domain_MNS_0p5}
\end{figure}

\begin{figure}[H]
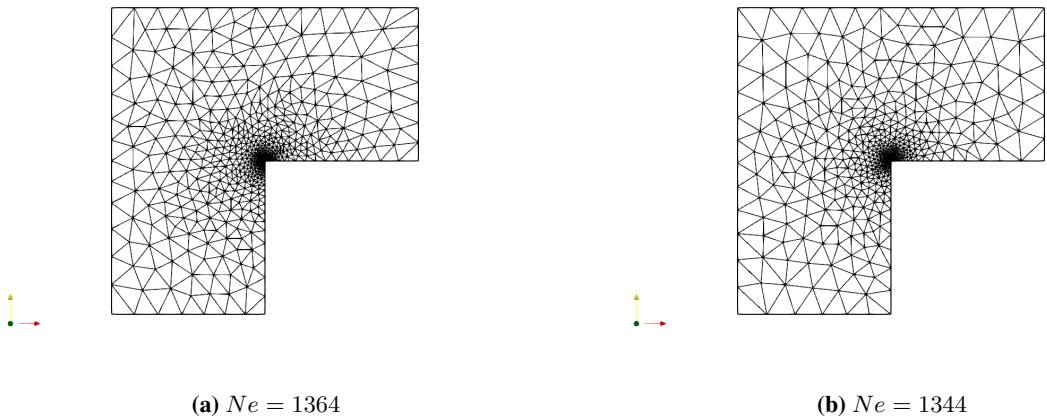

\begin{subfigure}[b]{0.5\textwidth}
\includegraphics[scale=0.2]{Data/L_shaped⁩/Lshapeddomain_p2_15thadaptedmesh_1364el-eps-converted-to.pdf}
\caption{$Ne  = 1364$}
\end{subfigure}
\begin{subfigure}[b]{0.5\textwidth}
\includegraphics[scale=0.2]{Data/L_shaped⁩/Lshapeddomain_p3_15thadaptedmesh_1344el-eps-converted-to.pdf}
\caption{$Ne  = 1344$}
\end{subfigure}
\caption{Adapted mesh for (a) $P =2$ and  (b) $P =3$} \label{Lshaped_domain_adapted_mesh}
\end{figure}

Boundary condition $g_D$ is the exact solution. L-shaped domain with a $r^{\alpha}$ type solution exhibits corner singularity. For the  present L-shaped domain, $\alpha = \frac{2.0}{3.0}$. Through this numerical experiment, we wanted to demonstrate the behaviour of the proposed methodology in a presence of a singularity.  Despite having a singular solution, it can be noticed that $O(h^{p+1})$ is achieved for $h$ adaptivity.  We have presented the convergence in $L^2$ and energy norm in ~\cref{convergence_Lshaped_domain_MNS_0p5}.
The exponential grading of meshes is qualitatively visible from ~\cref{Lshaped_domain_adapted_mesh}. In \cite{Yano2012} Theorem 4.4, the expression for the optimal grading of mesh elements for degree $p$ polynomial approximation of $u$ has been presented. The same analysis can be done with a focus on $\sigma$ such that the optimal mesh can minimize $L^2$ error in $\sigma$. In this case the mesh grading comes out to be,
\begin{equation}
h(r) = C r^\beta
\end{equation}
where $C$ is a constant independent of $r$ and 
\begin{equation}
\beta = 1 - \frac{\alpha}{p + 2} \label{theor_exponent}
\end{equation}

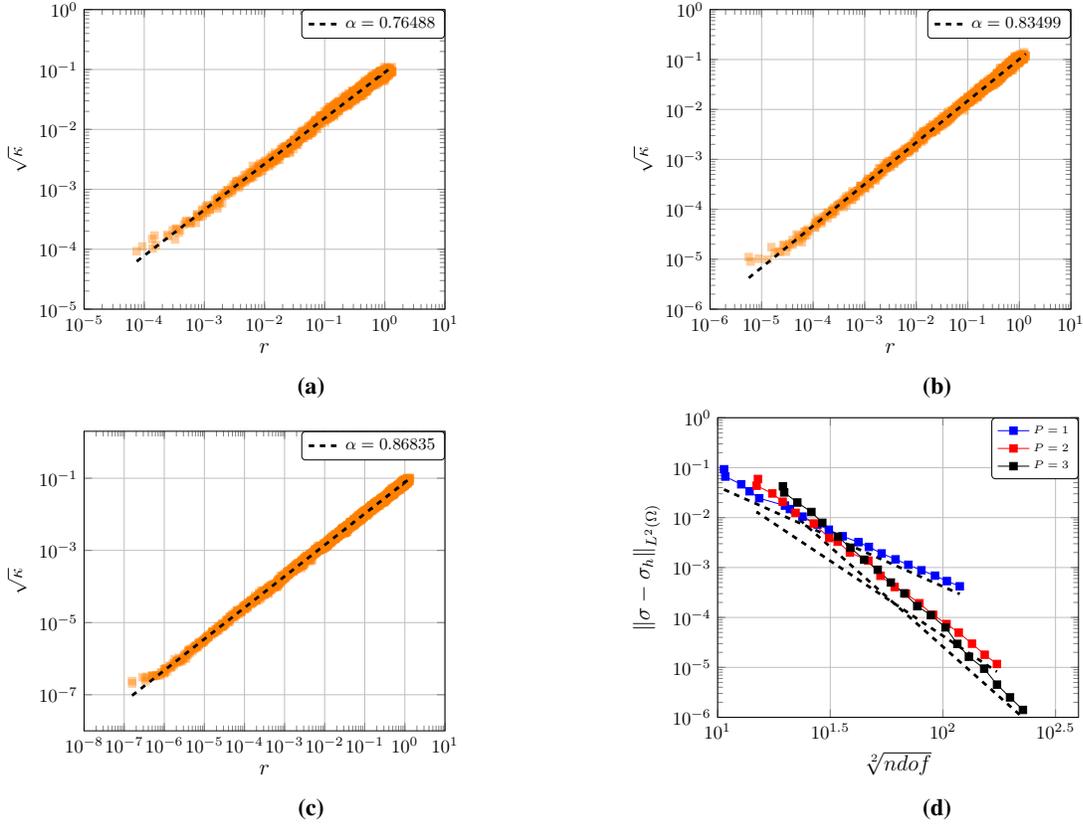
\begin{figure}[H]
\begin{subfigure}[b]{0.5\textwidth}
\begin{tikzpicture}[scale=0.7]
		\begin{loglogaxis}[clip mode=individual,xmin=1e-5,xmax=10, ymin=1e-5,ymax=1,xlabel=\large{$r$},ylabel=$\sqrt{\kappa}$,grid=major,legend style={at={(1,1)},anchor=north east,font=\small,rounded corners=2pt}]
		\addplot[only marks,orange,mark=square*,opacity=0.4,forget plot] table[x= r, y=h, col sep = comma] {Data/L_shaped⁩/cellgrading_Lshaped_p1.txt};
		\addplot [dashed,line width=1.5pt,mark=none, black] table[x= r, y=ext, col sep = comma] {Data/L_shaped⁩/cellgrading_Lshaped_p1.txt};
		\legend{$\alpha = 0.76488$}
		\end{loglogaxis}
	\end{tikzpicture}	
	\caption{}
\end{subfigure}
\begin{subfigure}[b]{0.5\textwidth}
\begin{tikzpicture}[scale=0.7]
		\begin{loglogaxis}[clip mode=individual,xmin=1e-6,xmax=10, ymin=1e-6,ymax=1,xlabel=\large{$r$},ylabel=$\sqrt{\kappa}$,grid=major,legend style={at={(1,1)},anchor=north east,font=\small,rounded corners=2pt}]
		\addplot[only marks,orange,mark=square*,opacity=0.4,forget plot] table[x= r, y=h, col sep = comma] {Data/L_shaped⁩/cellgrading_Lshaped_p2.txt};
		\addplot [dashed,line width=1.5pt,mark=none, black] table[x= r, y=ext, col sep = comma] {Data/L_shaped⁩/cellgrading_Lshaped_p2.txt};
			\legend{$\alpha = 0.83499$}
		\end{loglogaxis}
	\end{tikzpicture}
	\caption{}
\end{subfigure}

\begin{subfigure}[b]{0.5\textwidth}
\begin{tikzpicture}[scale=0.7]
		\begin{loglogaxis}[clip mode=individual,xmin=1e-8,xmax=10, ymin=1e-8,ymax=2,xlabel=\large{$r$},ylabel=$\sqrt{\kappa}$,grid=major,legend style={at={(1,1)},anchor=north east,font=\small,rounded corners=2pt}]
		\addplot[only marks,orange,mark=square*,opacity=0.4,forget plot] table[x= r, y=h, col sep = comma] {Data/L_shaped⁩/cellgrading_Lshaped_p3.txt};
		\addplot [dashed,line width=1.5pt,mark=none, black] table[x= r, y=ext, col sep = comma] {Data/L_shaped⁩/cellgrading_Lshaped_p3.txt};
		\legend{$\alpha = 0.86835$}
		\end{loglogaxis}
	\end{tikzpicture}	
	\caption{}
\end{subfigure}	
\begin{subfigure}[b]{0.5\textwidth}
\begin{tikzpicture}[scale=0.7]
		\begin{loglogaxis}[xmin=10,xmax=400, ymin=1e-6,ymax=1,xlabel=\large{$\sqrt[2]{ndof}$},ylabel=\large{${\Vert \sigma- \sigma_h \Vert}_{L^{2}(\Omega)}$},grid=major,legend style={at={(1,1)},anchor=north east,font=\tiny,rounded corners=2pt}]
		\addplot[color = blue,mark=square*] table[x= ndof, y=net_sigma_err, col sep = comma] {Data/L_shaped⁩/L2_error_Lshaped_Domain_MNS_0p5_p1_sigma.txt};
		\addplot [color = red,mark=square*] table[x= ndof, y=net_sigma_err, col sep = comma] {Data/L_shaped⁩/L2_error_Lshaped_Domain_MNS_0p5_p2_sigma.txt};
		\addplot [color = black,mark=square*] table[x= ndof, y=net_sigma_err, col sep = comma] {Data/L_shaped⁩/L2_error_Lshaped_Domain_MNS_0p5_p3_sigma.txt};
	    \addplot  [dashed,line width=1.5pt,mark=none, black,forget plot] table[x= ndof, y=exslp, col sep = comma] {Data/L_shaped⁩/L2_error_Lshaped_Domain_MNS_0p5_p1_sigma.txt};
	    	    \addplot  [dashed,line width=1.5pt,mark=none, black,forget plot] table[x= ndof, y=exslp, col sep = comma] {Data/L_shaped⁩/L2_error_Lshaped_Domain_MNS_0p5_p2_sigma.txt};
	    	    	    \addplot  [dashed,line width=1.5pt,mark=none, black,forget plot] table[x= ndof, y=exslp, col sep = comma] {Data/L_shaped⁩/L2_error_Lshaped_Domain_MNS_0p5_p3_sigma.txt};
		\legend{$P =1$,$P =2$,$P =3$}
		\end{loglogaxis}
	\end{tikzpicture}
	\caption{}
	\end{subfigure}
\caption{Grading of meshes in L-shaped domain for (a) $P =1$, (b) $P =2$, (c) $P =3$ and (d) convergence in $\sigma$ .} \label{exponentialgrading}
\end{figure}

Since the energy error is equivalent to composite $L^2$ error and the grading in $\sigma$ has a higher exponent than $u$, one can infer that the grading of the mesh is driven by $\sigma$ rather than $u$.  In ~\cref{exponentialgrading}, we present the variation of $\sqrt{\kappa}$ which is a measure of size of an element with respect to its distance from the singularity for different polynomial order using log-log axis. In order to obtain the approximation of the expoenent $\beta$, we did a linear fit to the data and the slopes of the linear fit agrees with theoretical exponent mentioned in ~\cref{theor_exponent}.

\begin{figure}[H]
\begin{subfigure}[b]{0.5\textwidth}
\begin{tikzpicture}[scale=0.7]
		\begin{loglogaxis}[xmin=1,xmax=5e6, ymin=1e1,ymax=1e14,xlabel=\large{$h_{min}^{-1}$},ylabel=\large{Condition number},grid=major,legend style={at={(1,1)},anchor=north east,font=\tiny,rounded corners=2pt}]
		\addplot[color = blue,mark=square*] table[x= hinv, y=cnd_no, col sep = comma] {Data/L_shaped⁩/DLS_cond_number_Lshaped_p2_15adap.txt};
		\addplot [color = red,mark=square*] table[x= hinv, y=cnd_no, col sep = comma] {Data/L_shaped⁩/DPG_cond_number_Lshaped_p2_15adap.txt};

	    \addplot  [dashed,line width=1.5pt,mark=none, black,forget plot] table[x= hinv, y=exslp, col sep = comma] {Data/L_shaped⁩/DLS_cond_number_Lshaped_p2_15adap.txt};
	    	    \addplot  [dashed,line width=1.5pt,mark=none, black,forget plot] table[x= hinv, y=exslp, col sep = comma] {Data/L_shaped⁩/DPG_cond_number_Lshaped_p2_15adap.txt};

		\legend{$DLS$,$DPG$}
		\end{loglogaxis}
	\end{tikzpicture}
	\caption{}
\end{subfigure}
\begin{subfigure}[b]{0.5\textwidth}
\begin{tikzpicture}[scale=0.7]
		\begin{loglogaxis}[xmin=1,xmax=1e8, ymin=1e1,ymax=1e16,xlabel=\large{$h_{min}^{-1}$},ylabel=\large{Condition number},grid=major,legend style={at={(1,1)},anchor=north east,font=\tiny,rounded corners=2pt}]
				\addplot[color = blue,mark=square*] table[x= hinv, y=cnd_no, col sep = comma] {Data/L_shaped⁩/DLS_cond_number_Lshaped_p3_15adap.txt};
		\addplot [color = red,mark=square*] table[x= hinv, y=cnd_no, col sep = comma] {Data/L_shaped⁩/DPG_cond_number_Lshaped_p3_15adap.txt};

	    \addplot  [dashed,line width=1.5pt,mark=none, black,forget plot] table[x= hinv, y=exslp, col sep = comma] {Data/L_shaped⁩/DLS_cond_number_Lshaped_p3_15adap.txt};
	    	    \addplot  [dashed,line width=1.5pt,mark=none, black,forget plot] table[x= hinv, y=exslp, col sep = comma] {Data/L_shaped⁩/DPG_cond_number_Lshaped_p3_15adap.txt};
		\legend{$DLS$,$DPG$}
		\end{loglogaxis}
	\end{tikzpicture}
	\caption{}
\end{subfigure}

\caption{Comparison of condition number between DLS and DPG (normal equations) for (a) $P =2$ and (b) $P =3$.} \label{Conditon_compare_DLS_DPG}
\end{figure}
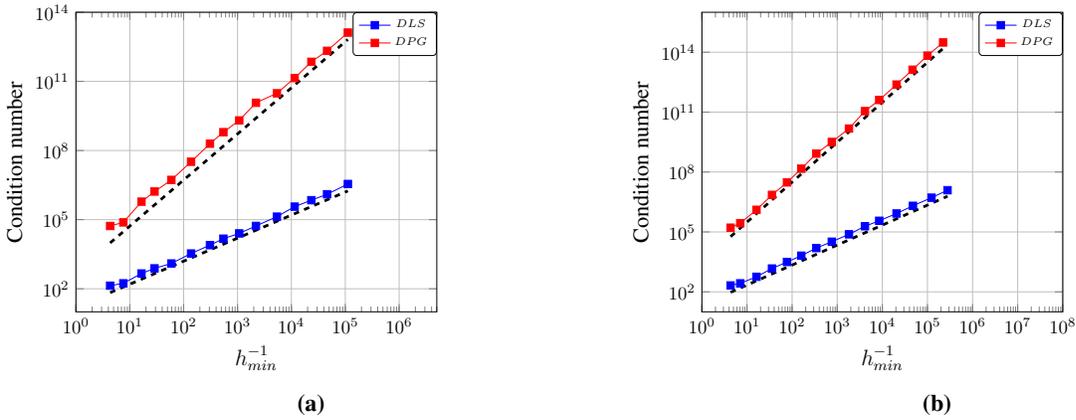

In ~\cref{Discretization}, we have mentioned about the importance of the discrete least square implementation. Here  in ~\cref{Conditon_compare_DLS_DPG}, we present quantitative result showing the growth of condition number for adapted meshes using normal equations and DLS implementation for two different polynomial orders. As mentioned earlier, condition number for the normal equations has quadratic growth rate whereas DLS results in linear growth. This has significant implications as using normal equations might limit the extent to which one can adapt the mesh especially when there is presence of strong features in solution.